\documentclass[11pt]{article}
\usepackage{amssymb,amsmath}
\usepackage{graphicx}
\usepackage{verbatim}
\usepackage{tabbert} 

\newcommand{\hu}{\hat{u}}
\newcommand{\hT}{\hat{T}}
\newcommand{\hS}{\hat{S}}
\newcommand{\hv}{\hat{v}}
\newcommand{\tu}{\tilde{u}}
\newcommand{\tT}{\tilde{T}}
\newcommand{\tS}{\tilde{S}}
\newcommand{\vu}{\check{u}}
\newcommand{\vT}{\check{T}}
\newcommand{\vv}{\check{v}}
\newcommand{\vsi}{\check{\sigma}} 
\newcommand{\vat}{\vartheta}

\allowdisplaybreaks[2]

\title{Asymptotics and numerical efficiency of the Allen-Cahn model
  for phase interfaces with low energy in solids}

\author{
Hans-Dieter Alber\addtocounter{footnote}{1}\footnote{alber@mathematik.tu-darmstadt.de}
\\
{\small Fachbereich Mathematik, Technische Universit\"at Darmstadt} \\ 
{\small Schlossgartenstr. 7, 64289 Darmstadt, Germany}
}

\date{}
\begin{document}
\maketitle
\begin{abstract}
  We study how the propagation speed of interfaces in the Allen-Cahn
  phase field model for phase transformations in solids consisting of
  the elasticity equations and the Allen-Cahn equation depends on two
  parameters of the model. The two parameters control the interface
  energy and the interface width but change also the interface speed.
  To this end we derive an asymptotic expansion of second order for
  the interface speed, called the kinetic relation, and prove that it
  is uniformly valid in both parameters. As a consequence we show that
  the model error is proportional to the interface width divided by
  the interface energy.  We conclude that simulations of interfaces
  with low interface energy based on this model require a very small
  interface width, implying a large numerical effort. Effective
  simulations thus need adaptive mesh refinement or other advanced
  techniques.

  This version of the paper contains the proofs of Theorem~4.5 and
  Lemma~5.8, which are omitted in the version published in Continuum
  Mechanics and Thermodynamics.

\end{abstract}
\begin{center}
  \parbox{.87\textwidth}{ {\em Key words:} Allen-Cahn phase field
    model for solids, asymptotic expansion, propagation speed of phase
    interfaces, kinetic relation, model error,
    numerical efficiency  }
\end{center}
\section{Introduction}\label{S1}

In this paper we study how the propagation speed of interfaces in the
Allen-Cahn phase field model for phase transformations in elastic
solids depends on two parameters of the model. The model consists of
the partial differential equations of linear elasticity coupled to the
standard Allen-Cahn phase field equation. The two parameters, which we
denote by $\mu$ and $\la$, control the interface energy and the
interface width, but variation of these parameters also changes the
interface speed, or more precisely, the form of the relation, which
determines the interface speed as a function of the stress field and
the curvature of the interface. In sharp interface models this
relation is called kinetic relation. We use this notion also for phase
field models. Our goal is therefore to determine the kinetic relation
for the Allen-Cahn model and the dependence of it on the two model
parameters. To this end we must derive an asymptotic expansion for the
propagation speed of the interface and prove an error estimate for
this asymptotic expansion, which holds uniformly in both parameters.
Our results have consequences for the efficiency of the Allen-Cahn
model in numerical simulations of interfaces with small interface
energy. These consequences are also discussed.

Let $\Om \subseteq \R^3$ be a bounded open set with a sufficiently
smooth boundary $\pa\Om$. The points of $\Om$ represent the material
points of a solid elastic body. The unknown functions in the model are
the displacement $u(t,x) \in \R$ of the material point $x$ at time
$t$, the Cauchy stress tensor $T(t,x)\in \ES^3$, where $\ES^3$ denotes
the set of all symmetric $3\times 3$-matrices, and the order parameter
$S(t,x)\in \R$. These unknowns must satisfy the model equations
\begin{eqnarray}
-\div_x\, T &=& {\sf b}, \label{E1.1}
\\
T &=& D\big(\ve(\na_x u ) - \ov{\ve}S \big),\label{E1.2}
\\
\pa_t S &=& - \frac{c}{(\mu\la)^{1/2}} \Big(\pa_S {\sf W} \big(\ve(\na_x
  u ), S \big) + \frac{1}{\mu^{1/2}} \hat{\psi}'(S) -
  \mu^{1/2} \la \Da_x S \Big) \label{E1.3} 
\end{eqnarray}
in the domain $[0,\infty)\ti \Om$. The boundary and initial conditions are
\begin{alignat}{2}
u(t,x) &= {\sf U}(t,x), &\qquad& (t,x) \in [0,\infty)\ti \pa\Om,  
  \label{E1.4}
\\
\pa_{n_{\pa\Om}} S(t,x) &= 0, && (t,x) \in [0,\infty)
  \ti \pa\Om,  \label{E1.4a}
\\ 
S(0,x) &= {\sf S}(x), && x \in \Om.  \label{E1.5}
\end{alignat}
Here ${\sf b}(t,x)\in \R^3$, ${\sf U}(t,x) \in \R^3$, ${\sf S}(t,x)
\in \R$ denote given data, the volume force, boundary displacement and
initial data. $\pa_{n_{\pa\Om}}$ denotes the derivative in
direction of the unit normal vector $n_{\pa\Om}$ to the boundary.  The
deformation gradient $\na_x u(t,x)$ is the $3\ti 3$--matrix of first
order partial derivatives of $u$ with respect to the components $x_k$
of $x$, and the strain tensor
\[
\ve(\na_x u) = \frac{1}{2}\big(\na_x u+(\na_x u)^T\big)
\]
is the symmetric part of the deformation gradient, where $(\na_x u)^T$
denotes the transpose matrix. The elasticity tensor $D:\ES^3 \to
\ES^3$ is a linear symmetric, positive definite mapping,
$\ov{\ve}\in\ES^3$ is a given constant matrix, the transformation
strain, and $\mu>0$ and $\lambda>0$ are parameters. The elastic energy
is given by  
\begin{equation}\label{E1.6}
{\sf W}\big(\ve(\na_x u),S\big)= \frac{1}{2}\Big(D\big(\ve(\na_x
 u) - \ov{\ve} S \big)\Big):\big(\ve(\na_x u) - \ov{\ve} S\big), 
\end{equation}
with the matrix scalar product $A:B=\sum_{i,j}a_{ij}b_{ij}$. Using
\eq{1.2}, we obtain for the derivative  
\begin{equation}\label{E1.7}
\pa_S {\sf W}(\ve,S)=-\ov{\ve}:D\big(\ve(\na_x u)-\ov{\ve}S\big)=-\ov{\ve}:T.
\end{equation}
$c>0$ is a given constant and $\hat{\psi}:\R\to[0,\infty)$ is a double
well potential satisfying  
\[
\hat{\psi}(0) = \hat{\psi}(1)=0, \qquad \hat{\psi}(\zeta)>0 \mbox{ for
  } \zeta \not= 0,1.
\]
The precise assumptions on $\hat{\psi}$, which we need in our
investigations, are stated in \reft{2.3}. This completes the
formulation of the model. 

\eq{1.1} and \eq{1.2} are the equations of linear elasticity theory.
This subsystem is coupled to the Allen-Cahn equation \eq{1.3}, which
governs the evolution of the order parameter $S$.  The system \eq{1.1}
-- \eq{1.3} satisfies the second law of thermodynamics. More
precisely, the Clausius-Duhem inequality is satisfied with the free
energy
\begin{equation}\label{E1.9}
\psi_{\mu\la}^*(\ve,S)={\sf W}(\ve,S)+ \frac{1}{\mu^{1/2}}
  \hat{\psi}(S) + \frac{\mu^{1/2}\lambda}{2} |\na_x S|^2.    
\end{equation}
From this expression we see that the parameter $\lambda$ determines
the energy density of the phase interface. We assume that the
parameters $\mu$ and $\la$ vary in intervals $(0,\mu_0]$ and
$(0,\la_0]$, respectively, with $\mu_0 > 0$ and $\la_0 > 0$ chosen
sufficiently small. The scaling $\frac{c}{(\mu\lambda)^{1/2}}$ on the
right hand side of \eq{1.3} is necessary for otherwise the propagation
speed of the diffuse interface would tend to zero for $\mu \ra 0$ or
$\la \ra 0$.

We give now a slightly sketchy overview of the main results in this
article. The precise definitions and statements are given in
Section~\ref{S2}, and in particular in Section~\ref{S2.4}. 

We denote solutions of the Allen-Cahn equation by $(u_{\rm AC},T_{\rm
  AC},S_{\rm AC})$, to distinguish them from aproximate solutions,
which we construct later. The values $S_{\rm AC}(t,x) \approx 0$ or
$S_{\rm AC}(t,x) \approx 1$ indicate that at the point $x \in \Om$ at
time $t$ the crystal structure of the material of the solid belongs to
phase~1 or to phase~2, respectively.  The set of all $x \in \Om$ with
$0 < S_{\rm AC}(t,x) < 1$ is the region of the diffuse interface at
time $t$.  The level set 
\begin{equation}\label{E1.1.9}
\Gm_{\rm AC}(t) = \big\{ x \in \Om \bigm| S_{\rm AC}(t,x) = \frac12
  \big\} 
\end{equation}
belongs to this region. For $x \in \Gm_{\rm
  AC}(t)$ we denote by $s_{\rm AC}(t,x) \in \R$ the normal speed of
this level set at $x$, and we call $s_{\rm AC}$ the speed of the
diffuse interface. For this speed we derive an expression of the
form
\begin{equation}\label{E1.1.10}
s_{\rm AC}(t,x) = (s_{00} + \la^{1/2}s_{01}) + \mu^{1/2} ( s_{10} +
 \la^{1/2} s_{11} ) + |\ln \mu|^3 \mu\, s_\infty^{(\mu\la)}.  
\end{equation}
We call this expression the kinetic relation of the Allen-Cahn model.
It is the central result of this paper. The remainder term
$s_\infty^{(\mu\la)}$ depends on $\mu$ and $\la$, but in
Section~\ref{S6} we prove that there exist numbers $\mu_0 > 0$ and
$\la_0 > 0$ and a constant $C_{\cal E}$ such that the $L^2$--norm
satisfies
\begin{equation}\label{E1.1.11}
\| s^{(\mu\la)}_\infty(t) \|_{L^2_{\Gm_{\rm AC}(t)}} \leq C_{\cal E} 
\end{equation}
for all $0 < \mu \leq \mu_0$ and all $0 < \la \leq \la_0$, where
$s^{(\mu\la)}_\infty(t)$ denotes the function $x \mapsto
s^{(\mu\la)}_\infty(t,x):\Gm_{\rm AC}(t) \ra \R$. Therefore
\eq{1.1.10} is an asymptotic expansion for the propagation speed of
the diffuse interface, which is uniformly valid with respect to the
parameters $\mu$ and $\la$. For sufficiently small $\mu$ the leading
term $s_0(t,x) = s_{00} + \la^{1/2} s_{01}$ and the second term
$\mu^{1/2}s_1(t,x) = \mu^{1/2} ( s_{10} + \la^{1/2} s_{11} )$ dominate
over the remainder term $|\ln \mu|^3 \mu\, s_\infty^{(\mu\la)}$. We
can therefore read off the behavior of the Allen-Cahn model with
respect to the parameters $\mu$ and $\la$ from the first two terms in
\eq{1.1.10}.

The terms $s_{00} + \la^{1/2}s_{01}$ and $s_{10} + \la^{1/2}s_{11}$
are explicitly given in \reft{2.3}. We restrict ourselves here to
state the form of the leading term. We assume that $0 \leq t_1 < t_2 <
\infty$ are given times. We study the propagation of the interface for
$t$ varying in the interval $[t_1,t_2]$. For $t$ from this interval
the leading term is
\begin{equation}\label{E1.1.12}
s_0(t,x) = s_{00} + \la^{1/2} s_{01} = \frac{c}{c_1} \Big(
  - \ov{\ve}:\langle \hat{T} \rangle(t,x) + \la^{1/2} c_1 \ka_\Gm(t,x)
  \Big), 
\end{equation}
where $c$ is the mobility constant from \eq{1.3}, $c_1 = \int_0^1
\sqrt{2\hat{\psi}(\vartheta)}\, d\vartheta$ is computed from the
double well potential, $\ka_\Gm(t,x)$ is twice the mean curvature of
the surface $\Gm_{\rm AC}(t)$ at the point $x$, and $\hT$ is the
stress field in the solution $(t,x) \mapsto
\big(\hu(t,x),\hT(t,x)\big)$ of the transmission problem
\begin{eqnarray}
-\div_x \hat{T} &=& {\sf b},  \label{E1.1.13}
\\
\hat{T} &=& D\big( \ve(\na_x \hat{u}) - \ov{\ve} \hat{S}\big),
  \label{E1.1.14}   
\\
{[\hat{u}]} &=& 0, \label{E1.1.15} 
\\
{[\hat{T}]}n &=& 0,\label{E1.1.16} 
\\
\hu(t)\rain{\pa\Om} &=& {\sf U}(t),\label{E1.1.17} 
\end{eqnarray}
Here $(t,x) \mapsto \hS(t,x)$ is a function, which takes only the
values $0$ or $1$ and jumps across the interface  
\[
\Gm_{\rm AC} = \{ (t,x) \in [t_1,t_2] \ti \Om \mid x \in \Gm_{\rm
  AC}(t) \}. 
\]
$[\hu]$ and $[\hT]$ are the jumps of the functions $\hu$ and $\hT$
across $\Gm_{\rm AC}$, and ${\sf b}$, ${\sf U}$ are the volume force
and boundary data from equations \eq{1.1} and \eq{1.4}. The equations
\eq{1.1.13}, \eq{1.1.14} hold in the domain $\big([t_1,t_2] \ti \Om
\big) \setminus \Gm_{\rm AC}$. For every fixed $t$ the problem
\eq{1.1.13} -- \eq{1.1.17} is an elliptic transmission problem for the
function $x \mapsto \big(\hu(t,x),\hT(t,x)\big)$ in the domain $\Om$.
Therefore $t$ can be considered to be a parameter in this problem.
Finally, the expression
\[
\langle \hT \rangle = \frac12 \big(\hT^{(+)} + \hT^{(-)}\big)
\]
in equation\eq{1.1.12} denotes the mean value of the values $\hT^{(+)}$
and $\hT^{(-)}$ on both sides of the interface $\Gm_{\rm AC}$. 

The terms $s_{10}$ and $s_{11}$ are determined by a more complicated
transmission problem, for which the transmission conditions are also
posed on the interfae $\Gm_{\rm AC}$, and by a coupled system of
ordinary differential equations for two functions $S_0$ and $S_1$,
which are needed in the construction of an asymptotic solution of the
Allen-Cahn model \eq{1.1} -- \eq{1.4a}. The coefficients of the second
transmission problem and of the system of ordinary differential
equations depend on the solution $(\hu,\hT)$ of \eq{1.1.13} --
\eq{1.1.17}. Both transmission problems together can thus be
considered to be a larger transmission problem, which is recursively
solvable.

The derivation of the kinetic relation \eq{1.1.10} is based on the
construction of an asymptotic solution
$(u^{(\mu\la)},T^{(\mu\la)},S^{(\mu\la)})$ for the Allen-Cahn
model. From this asymptotic solution it is seen that the width of the
diffuse interface in the Allen-Cahn model is proportional to the
parameter 
\begin{equation}\label{E1.1.18}
B = (\mu\la)^{1/2}.
\end{equation}
We call $B$ the interface width parameter. As will be explained in
Section~\ref{S2.4}, the interface energy density is proportional to
the parameter
\begin{equation}\label{E1.1.19}
E = \la^{1/2}.
\end{equation}
We call $E$ the interface energy parameter. 

The kinetic relation \eq{1.1.10} and the equation \eq{1.1.18} together
have consequences for the efficiency of numerical simulations of
interfaces with low interface energy density, which we sketch here. A
precise discussion is given in Section~\ref{S2.4}. 

The explicit expressions in \reft{2.3} show that the second term
$\mu^{1/2}(s_{10} + \la^{1/2} s_{11})$ in \eq{1.1.10} is of a very
special form. We therefore argue that this term does not have a
physical meaning, only the leading term $s_{00} + \la^{1/2} s_{01}$ is
physically relevant. This means that in \eq{1.1.10} the term
\[
{\cal E}^{(\mu\la)} = \mu^{1/2} ( s_{10} + \la^{1/2} s_{11} ) +
|\ln \mu|^3 \mu\, s_\infty^{(\mu\la)}
\]
is a mathematical error term, which in a precise numerical simulation
of the evolution of the interface must be made small by choosing
$\mu^{1/2}$ small enough. Therefore we call ${\cal E}^{(\mu\la)}$ the
model error and $F = \mu^{1/2}$ the error parameter. By \eq{1.1.18},
the interface width is proportional to the error parameter $F$. This
means that the interface width depends on the size of the model error.
The smaller the model error is, which we want to allow, the smaller
the interface width must be chosen. The total error in a numerical
simulation consists of the model error and the numerical error. In
order to make the numerical error small, the grid spacing must be
chosen small enough to resolve the transition of the order parameter
from $0$ to $1$ across the diffuse interface. When the interface width
is small, we must therefore choose the grid spacing small, which means
that the numerical effort is high.

From \eq{1.1.18} and \eq{1.1.19} we see that for constant values of
the error parameter $F$ the interface width is proportional to the
interface energy density parameter $E$. Thus, when we want to
precisely simulate an interface with small interface energy density,
we must choose small values for $E$ and $F$, hence the interface width
$B = EF$ becomes very small. As a consequence, also the grid spacing
must be chosen very small, which means that numerical simulations of
interfaces with low interface energy based on the Allen-Cahn model are
not efficient. Of course, the efficiency can be improved by using
adaptive mesh refinement and other advanced numerical techniques, but
still it would be advantageous if such tools could be avoided.

Often the Allen-Cahn model is formulated using the parameters $E$ and
$F$ instead of $\mu$ and $\la$. It might therefore be helpful to
shortly discuss the form, which our results take when this formulation
is used. With these parameters the Allen-Cahn equation \eq{1.3} is  
\[
\pa_t S = - \frac{c}{B} \Big(\pa_S {\sf W} \big(\ve(\na_x
  u ), S \big) + E \big(\frac{\hat{\psi}'(S)}{B} -
  B \Da_x S \big) \Big),  
\]
the free energy \eq{1.9} becomes 
\[
\psi_{EB}^*(\ve,S)={\sf W}(\ve,S)+ E \Big( \frac{ \hat{\psi}(S)}{B}
   + \frac{B}{2} |\na_x S|^2 \Big),  
\]
and the kinetic relation \eq{1.1.10} takes the form 
\[
s_{\rm AC}(t,x) = (s_{00} + E s_{01}) + \frac{B}{E} ( s_{10} +
 E s_{11} ) + \Big|2 \ln \Big(\frac{B}{E}\Big) \Big|^3
 \Big(\frac{B}{E}\Big)^2 s_\infty^{(BE)}.  
\]
From this equation we see that the model error 
\[
{\cal E}^{(BE)} = \frac{B}{E} ( s_{10} + E s_{11} ) + \Big|2 \ln
  \Big(\frac{B}{E}\Big) \Big|^3 \Big(\frac{B}{E}\Big)^2
  s_\infty^{(BE)}
\]
is governed by the ratio $\frac{B}{E}$. When one reduces the interface
energy density $E$ and one wants to keep the model error constant, one
must reduce the interface width $B$ by the same proportion. In a
simulation of an interface without interface energy, the total model
error is the sum ${\cal E}^{(BE)}_{\rm total} = E s_{01} + {\cal
  E}^{(BE)}$. To reduce the error in such a simulation we must reduce
$E$ and the fraction $\frac{B}{E}$, hence $B$ must be reduced faster
than $E$.

The paper is organized as follows. The main results are presented in
Section~\ref{S2}, where we first state the transmission problems and
the system of ordinary differential equations, whose solutions are
needed to compute the coefficients $s_{00},\ldots,s_{11}$ in
\eq{1.1.10}. These coefficients are explicitly given in \reft{2.3}.
Moreover, in this theorem we also state properties of the asymptotic
solution $(u^{(\mu\la)},T^{(\mu\la)},S^{(\mu\la)})$, which is
constructed in later sections. In particular, we state the scaling law
\eq{1.1.18} for the width of the diffuse interface. These properties
are needed in Section~\ref{S2.4}, where we precisely discuss the model
error and the numerical efficiency. The estimate \eq{1.1.11}, which is
the most important mathematical result of this paper, is stated in
\reft{2.8}.

Sections~\ref{S3} -- \ref{S5} contain the proof of \reft{2.3}. In
Section~\ref{S3} we construct the approximate solution $(u^{(\mu\la)},
T^{(\mu\la)}, S^{(\mu\la)})$. That is, we state the inner and outer
expansions which define the function $(u^{(\mu\la)}, T^{(\mu\la)},
S^{(\mu\la)})$. In these asymptotic expansions functions appear, which
are obtained as solutions of systems of algebraic and differential
equations. These systems are also stated in Section~\ref{S3}. The
system for the outer expansion can be readily solved, and the solution
of the system of ordinary differential equations for the inner
expansion is more involved and is discussed in Section~\ref{S4}. In
two equations of this system a linear differential operator appears
with kernel different from $\{0\}$. In order that these differential
equations be solvable the right hand sides must satisfy orthogonality
conditions. The right hand sides contain the coefficients $s_{00},
\ldots,s_{11}$ of the kinetic relation \eq{1.1.10}. The orthogonality
conditions dictate the form of these coefficients; therefrom the
equation \eq{1.1.10} results. In Section~\ref{S5} we verify that
$(u^{(\mu\la)}, T^{(\mu\la)}, S^{(\mu\la)})$ is really an asymptotic
solution of the model equations \eq{1.1} -- \eq{1.4a} and prove the
necessary estimates.

In Section~\ref{S6} we prove the estimate \eq{1.1.11}. The proof uses
the residue, with which the function $(u^{(\mu\la)}, T^{(\mu\la)},
S^{(\mu\la)})$ satisfies the equations \eq{1.1} -- \eq{1.3}. The main
difficulty in the proof is that though we want to prove that
$s_\infty^{\mu\la}$ is bounded uniformly with respect to $\mu$ and
$\la$, the residue term itself is not bounded for $\la \ra 0$, but
instead behaves like $\la^{-1/2}$.


In the bibliography of \cite{JElast2012} we gave many references to
the literature on existence, uniqueness and asymptotics for models
containing the Allen-Cahn and Cahn-Hilliard equations. We refer the
reader to that bibliography and discuss here only some publications,
which are of interest in the construction of asymptotic solutions.

We believe that for the model \eq{1.1} -- \eq{1.3} an asymptotic
solution was constructed and used to identify the associated sharp
interface problem for the first time in \cite{FG94}, following earlier
such investigations for other phase field models. For example, in
\cite{Cagi86-1} these investigations were carried out for a model from
solidification theory, which consists ot the Allen-Cahn equation
coupled to the heat equation.

The considerations in \cite{Cagi86-1,FG94} are formal, since it is not
shown that the asymptotic solution converges to an exact solution of
the model equations for $\mu \ra 0$. Under the assumption that the
associated sharp interface problems have smooth solutions, this was
proved in \cite{MotSchatz95} for the Allen-Cahn equation, in
\cite{ABCh94} for the Cahn-Hilliard equation, in \cite{CagiCh98} for
the model from solidification theory and in \cite{AbelsSchau2014} for
a model consisting of the Cahn-Hilliard equation coupled with the
elasticity equations. The proofs use variants of a spectral estimate
derived in \cite{Chen94}. For the model from solidification theory the
associated sharp interface model is the Mullins-Sekerka model with
surface tension.

In \cite{CCO05} an asymptotic solution for the Cahn-Hilliard equation
has recently been constructed with a method different from the one
used in \cite{ABCh94}, and which is similar to our method.

In \cite{KarRap98} the numerical efficiency of simulations 
based on the phase field model consisting of the Allen-Cahn equation
coupled to the heat equation is studied. It is shown that for suitably
chosen coefficients of the model the second order term in an
asymptotic expansion of the solution vanishes. By arguments similar to
the ones we gave in the above discussion it is seen that this improves
the numerical efficiency of the model. This result has been improved and
generalized in \cite{Alm99,ChenCagi06,GarckeStin06}. A similar idea is
also present in \cite{FifePen95}.

Since the construction of asymptotic solutions is based on sharp
interface problems, a rigorous analysis of these problems is of
special interest. Of particular interest is the Hele-Shaw problem with
surface tension, since this is the sharp interface problem associated
with the Cahn-Hilliard equation. Existence, uniqueness, and regularity
of classical solutions of this problem have been investigated in
\cite{ChenEx96,EschSim96,EschSim97}. In \cite{EschSim98} it is shown
that if the initial data are close to a sphere then a classical
solution exists and converges to spheres. Existence of solutions to
the Mullins-Sekerka problem mentioned above has been shown in
\cite{Luckhaus90}.

The model \eq{1.1} -- \eq{1.5} describes the evolution of phase
transitions in a solid when temperature effects are negligible.  This
model is the prototype of a large class of models obtained by
extensions and generalizations of the model, which are used in the
engineering sciences to simulate the behavior of complex and
functional materials. From the very large literature in this field we
cite here only
\cite{Bhattacharya03,SchradeMuellXuGross07,XuSchradeMuellGrossRoedel10,ZhangBhat2006,ZuoGenKleinSteinXu14}.

\section{The kinetic relation}\label{S2}

\subsection{Notations}\label{S2.1}

For given fixed times $0 \leq t_1 < t_2 < \infty$ let 
\[
Q = [t_1,t_2] \ti \Om \subseteq \R^4.
\]
The construction of the asymptotic solution $(u^{(\mu\la)},
T^{(\mu\la)}, S^{(\mu\la)})$ is based on a surface
$\Gm^{(\mu\la)}(t)$, which for $t_1 \leq t \leq t_2$ moves in $\Om$
and which will be the level set 
\[
\Gm^{(\mu\la)}(t) = \big\{ x \in \Om \bigm| S^{(\mu\la)}(t,x) =
  \frac12 \big\}. 
\]
We set
\begin{equation}\label{E2.1.1}
\Gm^{(\mu\la)} = \{ (t,x) \in Q \mid x \in \Gm^{(\mu\la)}(t) \}.
\end{equation}
To simplify the notation we often drop the index $\la$ or both indices
$\mu$ and $\la$ and write $\Gm^{(\mu)}(t)$ and $\Gm^{(\mu)}$ or simply
$\Gm(t)$ and $\Gm$. Similarly, we often write $(u^{(\mu)}, T^{(\mu)},
S^{(\mu)})$ for the asymptotic solution and use the same convention
also in other notations. Both indices are specified if the dependence
on $\la$ becomes important.

The precise definition of the family $\{ \Gm^{(\mu\la)}(t) \}_{t_1
  \leq t \leq t_2}$ is given in the nect section, and in
Section~\ref{S2.4} we associate $\Gm^{(\mu\la)}(t)$ with the level set
$\Gm^{(\mu\la)}_{\rm AC}(t) = \Gm_{\rm AC}(t)$ introduced in
\eq{1.1.9}. To introduce notations we assume here that $\Gm$ is a
known, orientable, three dimensional $C^k$--manifold with $k \geq 1$
sufficiently large embedded in $Q$ such that $\Gm(t)$ is a regular two
dimensional surface in $\Om$ for every $t\in [t_1,t_2]$. Let
\begin{equation}\label{E2.1}
n:\Gm \to \R^3
\end{equation}
be a continuous vector field such that $n(t,x) \in \R^3$ is a unit
normal vector to $\Gm(t)$ at $x \in \Gm(t)$, for every $t \in
[t_1,t_2]$. For $\da > 0$ and $t \in [t_1,t_2]$ define the sets    
\begin{equation}\label{E2.2}
{\cal U}_\da(t) = \{x \in \Om \mid {\rm dist}(x,\Gm(t)) < \da \}\quad
\mbox{and} \quad 
{\cal U}_\da = \{ (t,x) \in Q \mid x \in {\cal U}_\da(t) \}.  
\end{equation}
We assume that there is $\da > 0$ such that ${\cal U}_\da \subseteq
Q$. Since $\Gm$ is a regular $C^1$--manifold in $Q$, 
then $\da$ can be chosen sufficiently small such that for all $t \in
[t_1,t_2]$ the mapping  
\begin{equation}\label{E2.3}
(\eta,\xi) \mapsto x(t,\eta,\xi) = \eta + \xi n(t,\eta):\Gm(t) \ti
 (-\da,\da) \to {\cal U}_\da(t) 
\end{equation} 
is bijective. We say that this mapping defines new coordinates
$(\eta,\xi)$ in ${\cal U}_\da(t)$ and $(t,\eta,\xi)$ in ${\cal
  U}_\da$. If no confusion is püossible we switch freely between the
coordinates $(t,x)$ and $(t,\eta,\xi)$. In particular, if $(t,x)
\mapsto w(t,x)$ is a function defined on ${\cal U}_\da$ we write
$w(t,\eta,\xi)$ for $w\big(t,x(t,\eta,\xi)\big)$, as usual.

We use the standard convention and denote for a function $w$ defined
on a subset $U$ of $Q$ by $w(t)$ the function $x \mapsto w(t,x)$,
which is defined on the set $\{ x \mid (t,x) \in U \} \subseteq \R^3$.

If $w$ is a function defined on ${\cal U}_\da(t) \setminus \Gm(t)$, we
set for $\eta \in\Gm(t)$ 
\begin{eqnarray*}
w^{(\pm)} (\eta) &=& \lim_{\genfrac{}{}{0pt}{2}{\xi\to 0}{\xi >
    0}} w \big( \eta \pm \xi n(t,\eta)\big), 
\\ 
(\pa_n^i w)^{(+)} (\eta) &=& \lim_{\genfrac{}{}{0pt}{2}{\xi\to 0}{\xi >
   0} } \frac{\pa^i}{\pa\xi^i} w \big( \eta + \xi
   n(t,\eta)\big), \quad i \in \N,
\\
(\pa_n^i w)^{(-)} (\eta) &=& \lim_{\genfrac{}{}{0pt}{2}{\xi\to
    0}{\xi < 0} } \frac{\pa^i}{\pa\xi^i} w \big( \eta +
    \xi n(t,\eta)\big),\quad i \in \N,
\\ 
{[w]} (\eta) &=& w^{(t)} (\eta) - w^{(-)}(\eta),
\\[1ex]
{[\pa_n^i w]}(\eta) &=& (\pa_n^i w)^{(+)}(\eta) - (\pa_n^i
  w)^{(-)}(\eta),  
\\
\lan w\ran (\eta) &=& \ha \big( w^{(+)} (\eta) + w^{(-)} (\eta) \big),
\end{eqnarray*}
provided that the one-sided limits in these equations
exist. If $w$ is defined on ${\cal U}_\da \setminus \Gm$, we set 
\[
w^{(\pm)} (t,\eta) = \big(w^{(\pm)} (t)\big)(\eta), \qquad (\pa_n^i
w)^{(\pm)} (t,\eta) = \big( (\pa_n^i w)^{(\pm)} (t) \big)(\eta), 
\]
and define $[w](t,\eta)$, $\lan w\ran(t,\eta)$, $[\pa_n^i w](t,\eta)$
as above. Let $\tau_1(\eta)$, $\tau_2(\eta)\in\R^3$ be two orthogonal
unit vectors to $\Gm(t)$ at $\eta \in\Gm(t)$. For functions $w:\Gm(t)
\to\R$, $W:\Gm(t)\to\R^3$ we define the surface gradients by
\begin{eqnarray}
\label{E2.4}
\na_\Gm w &=& (\pa_{\tau_1} w)\tau_1 + (\pa_{\tau_2} w)\tau_2,
\\
\label{E2.5} 
\na_\Gm W &=& (\pa_{\tau_1} W)\otimes\tau_1 + (\pa_{\tau_2}W) \otimes
  \tau_2,  	
\end{eqnarray}
where for vectors $c,d \in \R^3$ a $3\times 3$-matrix is defined by
\[
c\otimes d = (c_i d_j)_{i,j=1,2,3}.
\]
With \eq{2.4}, \eq{2.5} we have for functions $w:{\cal U}_\da(t) \ra
\R$ and $W:{\cal U}_\da(t) \ra \R^3$ at $\eta \in \Gm(t)$ the
decompositions  
\begin{eqnarray}
\na_x w &=& (\pa_n w) n + \na_\Gm w, \label{E2.6}
\\
\na_x W &=& (\pa_n W)\otimes n + \na_\Gm W, \label{E2.7} 
\end{eqnarray}
where $n = n(t,\eta)$ is the unit normal vector to $\Gm(t)$.

The normal speed of the family of surfaces $t \mapsto \Gm(t)$ is of
fundamental importance in this paper. Therefore we give a precise
definition.

\begin{tdefi}\label{D2.1}
Let $m(t,\eta) = \big( m'(t,\eta),m''(t,\eta) \big) \in \R \ti \R^3$
be a normal vector to $\Gm$ at $(t,\eta) \in \Gm$. The normal speed
of the family of surfaces $t \mapsto \Gm(t)$ at $\eta \in \Gm(t)$ is
defined by
\begin{equation}\label{E2.normspeed1}
s(t,\eta) = \frac{-m'(t,\eta)}{m''(t,\eta) \cdot n(t,\eta)}, 
\end{equation}
with the unit normal vector $n(t,\eta) \in \R^3$ to $\Gm(t)$.
\end{tdefi}
Note that with this definition the speed is measured positive in the
direction of the normal vector field $n$.  Since $m''(t,\eta) \in
\R^3$ is a normal vector to $\Gm(t)$, the denominator in
\eq{2.normspeed1} is different from zero. 

If $\om = (\om',\om'') \in \R \ti \R^3$ is a tangential vector to
$\Gm$ at $(t,\eta)$ with $\om' \neq 0$, then with the unit normal
$n(t,\eta) \in \R^3$ to $\Gm(t)$ the vector $(-\om'' \cdot n,\om' n)$ is
a normal vector to $\Gm$ at $(t,\eta)$, hence \eq{2.normspeed1}
implies that the normal speed at $\eta \in \Gm(t)$ is given by
\begin{equation}\label{E2.normspeed2}
s(t,\eta) = \frac{n \cdot \om'' }{\om' n\cdot n} = \frac{n \cdot \om''
  }{\om'}. 
\end{equation}
For later use we prove the following

\begin{lem}\label{L2.normspeed}
Let $x \in {\cal U}_\da(t_0)$ be
a point having the representation $x = \eta + n(t,\eta) \xi$
in the $(\eta,\xi)$--coordinates, where $\eta = \eta(t,x) \in \Gm(t)$
and $\xi = \xi(t,x)$. Then the normal speed satisfies 
\begin{equation}\label{E2.normspeed3}
s(t_0,\eta) = n(t_0,\eta) \cdot \pa_t \eta(t_0,x) = - \pa_t\,
  \xi(t_0,x). 
\end{equation}
The tangential component of the vector $\pa_t \eta(t_0,x) \in \R^3$ to
the surface $\Gm(t_0)$ is equal to $-\xi \pa_t n\big( t_0,\eta(t_0,x)
\big)$.  
\end{lem}
{\bf Proof:} By definition of ${\cal U}_\da$, there is a
neighborhood $U$ of $t_0$ in $[t_1,t_2]$ such that $\{x\} \ti U
\subseteq {\cal U}_\da$, which implies that $x$ has the representation
\[
x = \eta(t,x) + \xi(t,x) n\big( t,\eta(t,x) \big).
\]
for all $t \in U$. We differentiate this equation and obtain 
\begin{equation}\label{E2.patetadeco}
0=\pa_t x=n \,\pa_t \xi + \xi\, \pa_t n + \pa_t \eta.
\end{equation}
From $0=\pa_t 1 = \pa_t |n|^2 = 2n \cdot \pa_t n$ we see that $\pa_t
n$ is tangential to $\Gm(t)$, hence \eq{2.patetadeco} implies that the
tangential component of $\pa_t \eta$ is equal to $- \xi \pa_t n
$. Multiplication of \eq{2.patetadeco} with $n$ yields 
\begin{equation}\label{E2.normspeed4}
\pa_t \xi= - n \cdot \pa_t \eta.
\end{equation}
Since $\pa_t \big(t, \eta(t,x) \big) = \big(1, \pa_t \eta(t,x)\big)$ 
is a tangential vector to $\Gm$, it follows from \eq{2.normspeed2}
that $s = \frac{n \cdot \pa_t \eta}{1} = n \cdot \pa_t \eta$, which
together with \eq{2.normspeed4} implies \eq{2.normspeed3}. \qed

\subsection{The evolution problem for the level set
  $\boldsymbol{\Gm^{(\mu)}}$}\label{S2.2}

The level set $\Gm = \Gm^{(\mu\la)}$ of $S^{(\mu\la)}$ defined in
\eq{2.1.1} is determined by an evolution problem for the family of
surfaces $t \mapsto \Gm(t)$. To state this evolution problem let
${\cal N}$ be the operator, which assigns the normal speed to the
family $t \mapsto \Gm(t)$, i.e.
\[
s(t,x) = {\cal N}(\Gm)(t,x),
\] 
with $s(t,x) = s^{(\mu)}(t,x)$ defined by \eq{2.normspeed1}. The
evolution problem is given by
\begin{equation}\label{E2.evolution0}
{\cal N}(\Gm)(t) = {\cal K}^{(\mu)}\big(\Gm (t)\big),\qquad t_1 \leq t
 \leq t_2\,, 
\end{equation}
where ${\cal K}^{(\mu)}$ is the non-local evolution operator, which
has the form 
\begin{equation}\label{E2.evolution1}
{\cal K}^{(\mu)}\big(\Gm(t)\big)(x) = s_0 \big(\hT,\ka_\Gm,\la^{1/2}
  \big)(t,x) + \mu^{1/2} s_1 \big(\hu,\hT,\vT,S_0,S_1,\la^{1/2}
  \big)(t,x),       
\end{equation}
for $x \in \Gm(t)$. Here $(\hu,\hT,\vu,\vT,S_0,S_1)$ is the solution
of a transmission-boundary value problem for a coupled system of
elliptic partial differential equations and ordinary differential
equations, which can be solved recursively. $\ka_\Gm(t,x)$ denotes
twice the mean curvature of the surface $\Gm(t)$ at $x \in \Gm(t)$.
With the principle curvatures $\kappa_1$, $\kappa_2$ of $\Gm(t)$ at $x
\in \Gm(t)$ we thus have
\[
\kappa_\Gm (t,x)= \kappa_1(t,x) + \kappa_2(t,x). 
\]
The transmission condition is posed on $\Gm(t)$. Therefore the
functions $\hu$, $\hT$, $\vu$, $\vT$ and $S_1$ depend on $\Gm(t)$. We
first state and discuss the transmission-boundary value problem. The
precise form of the functions $s_0$ and $s_1$ is given in \reft{2.3}
following below.

Let $\hS:~Q \setminus \Gm \to \{ 0,1\}$ be a piecewise constant
function, which only takes the values $0$ and $1$ with a jump across
$\Gm$. The sets 
\begin{align*}
\gm &= \{(t,x) \in Q \setminus \Gm  \mid \hS(t,x) = 0 \},&  
\gm(t) &= \{ x \in \Om \setminus \Gm(t)  \mid (t,x) \in \gm \}, 
\\
\gm' &= \{(t,x) \in Q \setminus \Gm  \mid \hS(t,x) = 1 \},&   
\gm'(t) &= \{ x \in \Om \setminus \Gm(t)  \mid (t,x) \in \gm' \}
\end{align*}
yield partitions $Q = \gm \cup \Gm \cup \gm'$ and $\Om = \gm(t) \cup
\Gm(t) \cup \gm'(t)$ of $Q$ and $\Om$, respectively. If $x$ belongs to
$\gm(t)$ or $\gm'(t)$, then the crystal structure at the material
point $x$ at time $t$ belongs to phase~$1$ or phase~$2$, respectively.
We assume that the normal vector field $n$ given in \eq{2.1} is such
that the vector $n(t,x)$ points into the set $\gm'(t)$ for every $x
\in \Gm(t)$.

The transmission-boundary value problem can be separated into two
transmission-boundary value problems for the elasticity equations and
a boundary value problem for a coupled system of two ordinary
differential equations. To state the complete problem we fix $t \in
[t_1,t_2]$ and assume that $\Gm(t)$ is known. In the first
transmission-boundary problem the unknowns are the displacement $x
\mapsto \hat{u}(t,x)\in\R^3$ and the stress tensor $x \mapsto
\hat{T}(t,x)\in \ES^3$, which must satisfy the equations
\begin{eqnarray}
\label{E2.8}
-\div_x \hat{T} &=& {\sf b},
\\
\label{E2.9}
\hat{T} &=& D\big( \ve(\na_x \hat{u}) - \ov{\ve} \hat{S}\big),
\\
\label{E2.10}
{[\hat{u}]} &=& 0,
\\
\label{E2.11}
{[\hat{T}]}n &=& 0,
\\
\label{E2.12}
\hu(t)\rain{\pa\Om} &=& {\sf U}(t),	
\end{eqnarray}
with ${\sf b}$ and ${\sf U}$ given in \eq{1.1} and \eq{1.4}. In the
second transmission-boundary problem the unknowns are the
displacement $x \mapsto \vu(t,x)\in \R^3$ and the stress tensor $x
\mapsto \vT(t,x)\in \ES^3$, and the problem is
\begin{eqnarray}
\label{E2.13}
-\div_x \vT &=& 0,
\\
\label{E2.14}
\vT &=& D\Big(\ve(\na_x \vu) -  \ov{\ve}\, \frac{
   \hat{T}:\ov{\ve} }{\hat{\psi}''(\hS)} \Big),  
\\	
\label{E2.15} 
[\vu] &=& 0,
\\
\label{E2.16}
{[\vT]} n &=& 0,
\\
\label{E2.17}	
\vu(t)\rain{\pa\Om} &=& 0.
\end{eqnarray}
The equations \eq{2.8}, \eq{2.9} and \eq{2.13}, \eq{2.14} must hold on
the set $\Om \setminus \Gm(t)$, whereas the equations \eq{2.10},
\eq{2.11} and \eq{2.15}, \eq{2.16} are posed on $\Gm(t)$.

In the boundary value problem for the ordinary differential equations
the unknowns are $S_0:\R \to \R$, $S_1: \Gm \ti \R \to \R$ and $s_0:
\Gm \ra \R$. We use the notations $S_1'(t,\eta,\zeta) = \pa_\zeta
S_1(t,\eta,\zeta)$, $S_1''(t,\eta,\zeta) = \pa^2_\zeta
S_1(t,\eta,\zeta)$. In this problem not only $t$, but also $\eta \in
\Gm(t)$ is a parameter. For all $\zeta \in \R$ and all values of the
parameter $\eta \in \Gm(t)$ the unknowns must satisfy the coupled
ordinary differential equations
\begin{eqnarray}
\hat{\psi}'\big(S_0(\zeta)\big) - S_0''(\zeta) &=& 0,  
  \label{E2.18}
\\
\hat{\psi}'' \big(S_0(\zeta)\big) S_1(t,\eta,\zeta) -
  S_1''(t,\eta,\zeta) &=& F_1(t,\eta,\zeta),
  \label{E2.19} 
\end{eqnarray}
and the boundary conditions  
\begin{align} 
S_0(0)=\frac{1}{2},\quad \lim_{\zeta \to -\infty} S_0(\zeta)
  &= 0, \quad  \lim_{\zeta \to \infty} S_0(\zeta) = 1,
  \label{E2.20}
\\
\lim_{\zeta \to -\infty} S_1(t,\eta,\zeta) &=
  \frac{\overline{\ve}:\hat{T}^{(-)}(t,\eta)}{\hat{\psi}''(0)}, 
  \label{E2.21} 
\\   
\lim_{\zeta \to +\infty} S_1(t,\eta,\zeta) &=
  \frac{\overline{\ve}:\hat{T}^{(+)}(t,\eta)}{\hat{\psi}''(1)}, 
  \label{E2.22}
\\
S_1(t,\eta,0) &= 0, \label{E2.23}  
\end{align}
with the right hand side of \eq{2.19} given by 
\begin{equation}\label{E2.24}
\begin{split}
F_1(t,\eta,\zeta)= \ov{\ve}: \Big(&[\hT] (t,\eta) S_0(\zeta) + 
   \hT^{(-)}(t,\eta) \Big) 
\\ &
 + \Big(\frac{s_0(t,\eta)}{c} - \la^{1/2} \ka_\Gm (t,\eta) \Big)
  S'_0(\zeta),  
\end{split}
\end{equation}
where the constant $c > 0$ is given in \eq{1.3}. 

The linear elliptic system \eq{2.8}, \eq{2.9} differs from the
standard elasticity system only by the term $-D\ov{\ve} \hat{S}$. This
term is known since $\Gm(t)$ is given. Under suitable regularity
assumptions for the given functions ${\sf b}$ and ${\sf U}$ and very
mild assumptions on the regularity of the interface $\Gm(t)$ the
problem has a unique weak solution $(\hu,\hT)$. This can be proved by
standard methods from functional analysis. Of course, the regularity
of the solution depends on the regularity of ${\sf b, \ U}$ and
$\Gm(t)$.

After insertion of the stress tensor $\hT$ from this solution into
\eq{2.14}, the equations \eq{2.13} -- \eq{2.17} form a
transmission-boundary value problem of the same type as \eq{2.8} --
\eq{2.12}, with unique solution $(\vu,\vT)$ determined by the same
methods.

We also insert $\hT$ into \eq{2.21}, \eq{2.22} and \eq{2.24}, which
determines the right hand side of the differential equation \eq{2.19}
and the boundary conditions \eq{2.21}, \eq{2.22} posed at $\pm
\infty$. The nonlinear differential equation \eq{2.18} has a unique
solution $S_0$ satisfying the boundary conditions \eq{2.20}. By
insertion of $S_0$ into \eq{2.19} and \eq{2.24}, equation \eq{2.19}
becomes a linear differential equation for $S_1$, however with an
additional unknown function $s_0$ in the right hand side. This
function is constant with respect to $\zeta$. We sketch here the
procedure used to determine $s_0$. This procedure is standard in
investigations of the asymptotics of phase field models:

The second order differential operator $\big(\hat{\psi}''(S_0) -
\pa_\zeta^2\big)$ is selfadjoint in the Hilbert space $L^2(\R)$ with a
one dimensional kernel spanned by the function $S_0'$. This is seen by
differentiating the equation \eq{2.18}. From functional analysis we
thus know that for $F_1 \in L^2(\R)$ the differential equation
$\big(\hat{\psi}''(S_0) - \pa_\zeta^2\big)w = F_1$ has a solution 
$w \in L^2(\R)$ if and only if the orthogonality condition
\begin{equation}\label{E2.25}
\int_{-\infty}^\infty F_1(t,\eta,\zeta) S_0'(\zeta)\,d\zeta = 0
\end{equation}
holds. It turns out that though the function $F_1$ defined in
\eq{2.24} does not in general belong to $L^2(\R)$ and the solution
$S_1(t,\eta,\cdot)$ is not sought in $L^2(\R)$, which is seen from the
boundary conditions \eq{2.21}, \eq{2.22}, the orthogonality condition
\eq{2.25} is sufficient for the solution $S_1$ to exist.  Comparison
with \eq{2.24} shows that \eq{2.25} can be satisfied by choosing the
constant $s_0(t,\eta)$ suitably. This defines the function $s_0:\Gm
\ra \R$ uniquely. Since $F_1$ depends on $\hT$, $\ka_\Gm$ and
$\la^{1/2}$, it follows that also $s_0$ is a function of these
variables:
\[
s_0(t,\eta) = s_0\big(\hT,\ka_\Gm,\la^{1/2} \big)(t,x).
\]
The explicit expression for $s_0$ obtained in this way is stated below
in \eq{2.32}.  In fact, $s_0(\hT,\ka_\Gm,\la^{1/2})$ is the first term
on the right hand side in the expression \eq{2.evolution1} for ${\cal
  K}^{(\mu)}\big(\Gm(t) \big)$.

The procedure sketched here is discussed precisely in
Section~\ref{S4.2} when we determine the second term $s_1$ in 
\eq{2.evolution1}, which is obtained from a similar, but more
complicated boundary value problem.

\subsection{The asymptotic solution and the kinetic
  relation}\label{S2.3}

To state the properties of the asymptotic solution and the kinetic
relation in \reft{2.3}, we introduce some definitions.

We need in our investigations that the second derivatives
$\hat{\psi}''(0)$ and $\hat{\psi}''(1)$ of the double well potential
at the minima $0$ and $1$ are positive, and we set 
\[
a = \min\Big\{ \sqrt{\hat{\psi}''(0)},\sqrt{\hat{\psi}''(1)} \Big\}.
\]
Depending on the parameters $\la$ and $\mu$, we partition $Q$ into the inner
neighborhood $Q_{\rm inn}^{(\mu\la)}$ of $\Gm$, into the matching region
$Q_{\rm match}^{(\mu\la)}$ and into the outer region 
$Q_{\rm out}^{(\mu\la)}$. These sets are defined by   
\begin{equation}\label{E2.25a}
\begin{split}
Q_{\rm inn}^{(\mu\la)} &= \Big\{ (t,\eta,\xi) \in {\cal U}_\da  \Bigm|
  |\xi| < \frac32 \frac{ (\mu\la)^{1/2} |\ln\mu|}{a} \Big\}, 
\\
Q_{\rm match}^{(\mu\la)} &= \Big\{ (t,\eta,\xi) \in {\cal U}_\da \Bigm| 
  \frac32 \frac{ (\mu\la)^{1/2} |\ln\mu|}{a} \leq |\xi| \leq 
  \frac{3 (\mu\la)^{1/2} |\ln\mu|}{a} \Big\}, 
\\[1ex] 
Q_{\rm out}^{(\mu\la)} &= Q \setminus \big( Q_{\rm inn}^{(\mu\la)} \cup
  Q_{\rm match}^{(\mu\la)} \big). 
\end{split}
\end{equation}
We always assume that the parameters $\la$ and $\mu$ satisfy $0 < \la
\leq \la_0$ and $0 < \mu \leq \mu_0$, where $\la_0$, $\mu_0$ are fixed
constants satisfying
\[
\mu_0 \leq e^{-2}, \qquad \frac{3 (\mu_0\la_0)^{1/2} |\ln \mu_0|}{a}
  < \da. 
\]
The first condition is imposed for purely technical reasons and
guarantees that the function $\mu \mapsto \mu^{1/2} |\ln \mu|$ is
increasing, the second condition guarantees that $Q_{\rm
  inn}^{(\mu\la)},Q_{\rm match}^{(\mu\la)} \subset {\cal U}_\da$ and
that $Q_{\rm out}^{(\mu\la)} \cap {\cal U}_\da $ is a nonempty,
relatively open subset of ${\cal U}_\da$.

By \eq{2.10}, the function $\hat{u}:Q \to \R^3$ is continuous at every
point $(t,\eta) \in \Gamma$, but the first and higher derivatives of
$\hat{u}$ in the direction of the normal vector $n(t,x)$ can jump
across $\Gamma$. For these jumps we write 
\begin{eqnarray}
u^\ast(t,\eta) &=& [\pa_n \hat{u}](t,\eta),\label{E2.26}
\\
a^\ast (t,\eta) &=& [\pa_n^2 \hat{u}](t,\eta). \label{E2.27}
\end{eqnarray}
We also set
\begin{equation}\label{E2.28}
c_1 = \int_0^1 \sqrt{2\hat{\psi}(\vartheta)}\, d\vartheta. 
\end{equation}

\begin{theo}\label{T2.3} Suppose that the double well potential
$\hat{\psi} \in C^5(\R)$ satisfies  
\begin{equation}\label{E2.29}
\begin{split}
\hat{\psi}(r)>0,\quad \mbox{for } 0 < r < 1,   
\\
\hat{\psi}(r)=\hat{\psi}'(r)= 0,\quad \mbox{for } r=0,1,
 \\   
a = \min\Big\{ \sqrt{\hat{\psi}''(0)},\sqrt{\hat{\psi}''(1)} \Big\} 
 >  0. 
\end{split}
\end{equation}
Moreover, suppose that $\hat{\psi}$ satisfies the symmetry condition 
\begin{equation}\label{E2.30}
\hat{\psi}(\frac12 - \zeta) = \hat{\psi}(\frac12 + \zeta),\quad \zeta
  \in \R.  
\end{equation}
Assume that there is a solution $\Gm$ of the evolution problem
\eq{2.evolution0}, \eq{2.evolution1} with $s_0 =
s_0(\hT,\ka_\Gm,\la^{1/2} ):\Gm \ra \R$ given by
\begin{equation}\label{E2.32}  
s_0 = \frac{c}{c_1} \Big(
  - \ov{\ve}:\langle \hat{T} \rangle + \la^{1/2} c_1 \ka_\Gm \Big),
\end{equation}
and with $s_1 = s_1(\hu,\hT,\vT,S_0,S_1,\la^{1/2}):\Gm \ra \R$
defined by 
\begin{equation}\label{E2.32s1}
s_1 = s_{10} + \la^{1/2} s_{11} = s_{10}(\hT,\vT,S_0,S_1) + \la^{1/2}
  s_{11} (\hu,S_0), 
\end{equation}
where 
\begin{eqnarray}
s_{10} &=& \frac{c}{c_1} \Biggl( - \ov{\ve}:\langle
  \vT \rangle + \ov{\ve}: [\hT] \Big( \Big\langle \frac{\ov{\ve}:\hT }{
  \hat{\psi}''(\hS)} \Big\rangle -  \int_{-\infty}^\infty S_1 S'_0\,
  d\zeta \Big) 
\nn \\ 
&& \hspace{8ex} \mbox{} 
  + \frac{1}{c_1} \ov{\ve}: \langle \hT \rangle \int_{-\infty}^\infty
  S_1' S_0' \,d\zeta + \frac12 \int_{-\infty}^\infty
  \hat{\psi}'''(S_0) S_1^2 S_0' \,d\zeta \Biggr), \label{E2.33}
\\
s_{11} &=& - \frac{c}{c_1}\, \ov{\ve}:D \ve(a^* \otimes n + \na_\Gm u^*) 
  \int_{-\infty}^\infty S_0(\zeta) S_0(-\zeta)\, d\zeta. \label{E2.34}
\end{eqnarray}
In \eq{2.32} and \eq{2.33}, \eq{2.34} we have $S_0 = S_0(\zeta)$ and
$S_1 = S_1(t,\eta,\zeta)$, for all other functions the argument is
$(t,\eta)$. The positive constant $c$ is defined in \eq{1.3}. The
notations $[\cdot]$ and $\langle \cdot \rangle$ are introduced in
Section~\ref{S2.1}. In particular, we have 
\[
\Big\langle \frac{ \ov{\ve}:\hT } { \hat{\psi}''(\hS)} \Big\rangle 
= \frac12 \Big(\frac{\ov{\ve}:\hT^{(+)}} {\hat{\psi}''(1)} +
  \frac{\ov{\ve}:\hT^{(-)}} {\hat{\psi}''(0)} \Big). 
\] 
With these functions the normal speed $s(t,\eta)$ of $\Gm(t)$ at $\eta
\in \Gm(t)$ is thus given by
\begin{equation}\label{E2.31}
s(t,\eta) = s_0 (t,\eta) + \mu^{1/2}s_1(t,\eta,\la^{1/2}) = s_0
  (t,\eta) + \mu^{1/2} \big( s_{10}(t,\eta) + \la^{1/2} s_{11}(t,\eta)
  \big).  
\end{equation}
We assume moreover that the solution $\Gm$ is a $C^5$--manifold and
that the functions $\hu$ and $\vu$ defined by the evolution problem
satisfy $\hu \in C^4(\gm \cup \gm',\R^3)$, $\vu \in C^3(\gm \cup
\gm',\R^3)$ and that $\hu$ has $C^4$--extensions, $\vu$ has
$C^3$--extensions from $\gm$ to $\gm \cup \Gm$ and from $\gm'$ to
$\gm' \cup \Gm$. For the given right hand side of \eq{1.1} we assume 
that ${\sf b} \in C^1(\ov{Q})$.

Under these assumptions there is an approximate solution
$(u^{(\mu)},T^{(\mu)},S^{(\mu)})$ of the Allen-Cahn model \eq{1.1} --
\eq{1.4a}, for which $\Gm$ is the level set 
\begin{equation}\label{E2.36}
\Gm = \Big\{ (t,x) \in Q \bigm| S^{(\mu)}(t,x) = \frac12  \Big\},
\end{equation}
and which satisfies the equations 
\begin{gather}
-\div_x T^{(\mu)} = {\sf b} + f_1^{(\mu\la)}, \label{E2.36a}  
 \\
T^{(\mu)} = D\big(\ve( \na_x u^{(\mu)}) - \ov{\ve}S^{(\mu)}\big), 
  \label{E2.36b}  
\\
\pa_t S^{(\mu)} + \frac{c}{(\mu\la)^\frac12} \Big( \pa_S {\sf W} 
  \big( \ve(\na_x u^{(\mu)}), S^{(\mu)} \big) +
  \frac{1}{\mu^\frac12} \hat{\psi}'(S^{(\mu)}) - \mu^\frac12\la\Da_x
  S^{(\mu)}\Big) = f_2^{(\mu\la)}, \label{E2.36c}
\\
u^{(\mu)}(t,x) = {\sf U} (t, x),
   \qquad (t,x) \in [t_1,t_2]\ti \pa\Om, \label{E2.36d} 
\\
\pa_{n_{\pa\Om}} S^{(\mu)}(t,x) = f_3^{(\mu\la)}, \qquad (t,x) \in
  [t_1,t_2] \ti \pa\Om,  \label{E2.36e}
\end{gather}
where to the right hand sides $f_1^{(\mu\la)}$,\ldots, $f_3^{(\mu\la)}$ 
there exist nonnegative constants $K_1,\ldots, K_5$ such that for all $\mu \in
(0,\mu_0]$ and all $\la \in (0,\la_0]$   
\begin{eqnarray}
\| f_1^{(\mu\la)} \|_{L^\infty(Q_{\rm inn}^{(\mu\la)} \cup
  Q_{\rm match}^{(\mu\la)})} &\leq& |\ln \mu|^2\, \Big( \frac{\mu}{\la}
\Big)^\frac12 K_1\, , \label{E2.37a}
\\ 
\| f_1^{(\mu\la)} \|_{L^\infty(Q_{\rm out}^{(\mu\la)})} &\leq&
\mu^\frac32 K_2\,, \label{E2.37b} 
\\
\|  f_2^{(\mu\la)}  \|_{L^\infty(Q_{\rm inn}^{(\mu\la)} \cup
  Q_{\rm match}^{(\mu\la)})} &\leq& |\ln \mu|^2 \Big( \frac{\mu}{\la}
  \Big)^\frac12 K_3\,, 
  \label{E2.38a}
\\
\|  f_2^{(\mu\la)} \|_{L^\infty(Q^{(\mu\la)}_{\rm out})} &\leq&
  \frac{\mu}{\la^{1/2}} K_4\,,  
  \label{E2.38b}  
\\
\|  f_3^{(\mu\la)} \|_{L^\infty(\pa\Om)} &\leq& \mu^\frac12 K_5\,,
  \label{E2.38c}
\end{eqnarray}
In the neighborhood $Q_{\rm inn}^{(\mu\la)}$ of $\Gm$ the order
parameter in the approximate solution is of the form  
\begin{equation}\label{E2.39}
S^{(\mu)}(t,x)=S_0\Big(\frac{\xi}{(\mu\lambda)^{1/2}}\Big)+
 \mu^{1/2} S_1 \Big(t,\eta, \frac{\xi}{(\mu\lambda)^{1/2}}
 \Big) + \mu S_2 \Big(t,\eta,
 \frac{\xi}{(\mu\lambda)^{1/2}}\Big),   
\end{equation}
where the monotonically increasing transition profile $S_0: \R \to \R$
and the function  $S_1: \Gm \ti \R \ra \R$ are given as solution of
the coupled problem \eq{2.18} -- \eq{2.24}, and where $S_2: \Gm \ti \R
\ra \R$ satisfyies $S_2(t,\eta,0) = 0$ and  
\begin{equation}\label{E2.40}
|S_2(t,\eta,\zeta)| \leq C (1 + |\zeta|),\qquad \mbox{for }
(t,\eta,\zeta) \in \Gm \ti \R,   
\end{equation}
with a constant $C$ independent of $(t,\eta,\zeta)$. 
\end{theo}
We mention that the positive constant $c$ in \eq{1.3} does not play a
major role in the analysis and could be replaced by $1$. We refrain
from replacing it to show how $c$ appears in the kinetic relation.

The {\bf proof} of this theorem forms the content of Sections~\ref{S3}
-- \ref{S5}. We remark that the symmetry assumption \eq{2.30} for the
double well potential $\hat{\psi}$ serves to simplify the computations
in the derivation of the asymptotic solution. Without this assumption
the term $s_1$ in the kinetic relation \eq{2.31} would contain other
terms in addition to the terms $s_{10}$ and $s_{11}$ given in
\eq{2.33} an \eq{2.34}. 

The regularity properties of $\Gm$ and of $\hu$, $\vu$ are of course
not independent, since $\hu$ and $\vu$ are solutions of the elliptic
transmission problems \eq{2.8} -- \eq{2.12} and \eq{2.13} --
\eq{2.17}, respectively. Therefore the regularity theory of elliptic
equations shows that $\hu$ and $\vu$ automatically have the
differentiability properties assumed in the theorem if the manifold
$\Gm$ and the right hand side ${\sf b}$ are sufficiently smooth.
\\[1ex]
Since by definition of $Q^{(\mu\la)}_{\rm inn}$ and $Q^{(\mu\la)}_{\rm
  match}$ in \eq{2.25a} we have
\[
{\rm meas}(Q^{(\mu\la)}_{\rm inn} \cup Q^{(\mu\la)}_{\rm match}) \leq
C_3 (\mu\la)^{1/2} |\ln \mu |,
\]
we immediately obtain from \eq{2.37a} -- \eq{2.38b} the following

\begin{coro}\label{C2.2}
There are constants $K_6$, $K_7$ such that for all $0 < \mu \leq
\mu_0$ and all $0 < \la \leq \la_0$  
\begin{eqnarray}
\| f_1^{(\mu\la)}\|_{L^1(Q)} &\leq& |\ln \mu|^3 \mu K_6 \,,
  \label{E2.37}    
\\[1ex] 
\Big\| f_2^{(\mu\la)} \|_{L^1(Q)} &\leq& \frac{|\ln \mu|^3 \,
    \mu}{\la^{1/2}} K_7 \,. \label{E2.38}
\end{eqnarray}
\end{coro}
The leading term $s_0$ given in \eq{2.32} can be written in a
more common and more general form. To give this form, we need a result
on the jump of the Eshelby tensor. The Eshelby tensor to the solution
$(\hu,\hT)$ of the transmission problem \eq{2.8} -- \eq{2.12} is
defined by   
\begin{equation}\label{E2.45} 
\hat{C} (\na_x \hat{u}, \hat{S})= \psi_\mu \big(\ve(\na_x \hat{u}),
\hat{S}\big) I-(I+\na_x \hat{u})^T \hat{T}, 
\end{equation}
where $I \in \ES^3$ is the unit matrix and where  
\begin{equation}\label{E2.46}
\psi_\mu (\ve,S)= {\sf W}(\ve, S)+ \frac{1}{\mu^{1/2}}
\hat{\psi} (S) 
\end{equation}
is that part of the free energy $\psi^*_\mu$ defined in \eq{1.9}
without gradient term. The last term on the right hand side of
\eq{2.45} is a matrix product. We use the standard convention to
denote the matrix product of two matrices $A \in \R^{k \times m}$ and
$B \in \R^{m \times \ell}$ by $A B \in \R^{k \ti \ell}$.

\begin{lem}\label{L2.3}
Let $(\hat{u}, \hat{T})$ be the solution of the transmission problem
\eq{2.8} -- \eq{2.12} and let $n$ be a unit normal vector field to
$\Gm(t)$. Then the jump $[\hat{C}]$ of the Eshelby tensor to
$(\hu,\hT)$ across $\Gamma$ satisfies 
\begin{equation}\label{E2.44} 
n \cdot [\hat{C}]n = \frac{1}{\mu^{1/2}}\, [\hat{\psi}(\hS)] -
        \ov\ve:\langle \hat{T} \rangle.  
\end{equation}

\end{lem} 
This result is known \cite{AK90}. In \cite{CMT2011} it is stated as
equation~(3.4) and proved on pages~154, 155.

\begin{coro}\label{C2.4}
The leading term $s_0$ of the kinetic relation defined in \eq{2.32}
satisfies 
\begin{equation}\label{E2.55}
 s_0 = \frac{c}{c_1} \Big( n \cdot[\hat{C}]n + \lambda^{1/2} c_1 
 \kappa_\Gamma\Big). 
\end{equation} 
\end{coro}
This corollary follows immediately from \eq{2.44}, since by assumption
\eq{2.29} we have $[\hat{\psi}(\hat{S})] = \hat{\psi}(1)-
\hat{\psi}(0)=0$, which implies that 
$n \cdot[\hat{C}]n=-\ov\ve : \langle \hat{T}\rangle$. 
\qed

\subsection{Consequences for numerical simulations}\label{S2.4}

In this section we discuss the consequences of \reft{2.3} for
numerical simulations of interfaces with small interface energy.

In many functional materials the phase interfaces consist only of a
few atomic layers. For interfaces with such small width mathematical
models with sharp interface are appropriate. We therefore base the
following considerations on the hypothesis that the propagation speed
of the interface in the sharp interface model is a good approximation
to the propagation speed of the interface in the real material. The
model error of the Allen-Cahn model is then the difference of the
propagation speed of the sharp interface and the propagation speed of
the diffuse interface in the phase field model. The parameters $\mu$
and $\la$ in the Allen-Cahn model should be chosen such that this
model error is small and such that numerical simulations based on the
Allen-Cahn model are effective.

To make this precise we must first determine the sharp interface model
to be used. The model consists of the transmission problem \eq{2.8} --
\eq{2.12} combined with a kinetic relation. To find this relation, one
proceeds in the usual way and uses that by the second law of
thermodynamics the Clausius-Duhem inequality
\[
\pa_t \psi_{\rm sharp} + \div_x\, q_{\rm sharp} \leq \hu_t \cdot {\sf
  b} 
\]
must be satisfied to impose restrictions on the form of the kinetic
relation.  Here $\psi_{\rm sharp}$ denotes the free energy in the
sharp interface problem and $q_{\rm sharp}$ is the flux of the free
energy. We use the standard free energy and flux
\begin{align}
\psi_{\rm sharp}\big(\ve(\na_x \hat{u}), \hat{S}\big) &=  {\sf W}
\big(\ve(\na_x \hat{u}), \hat{S}\big) + \lambda^{1/2} c_1 
 \int_{\Gamma(t)}\,{\rm d} \sigma, \label{E2.56}
\\
q_{\rm sharp} ( \hT, \hS ) &= - \hT \cdot \hu_t \,. \nn 
\end{align}
The last term on the right hand side of \eq{2.56} is the interface
energy, hence $\la^{1/2} c_1$ is the interface energy density. It is
well known that if $\big(\hu(t),\hT(t)\big)$ is a solution of the
transmission problem \eq{2.8} -- \eq{2.12} at time $t$ and if the
interface $\Gm(t)$ in this problem moves with the given normal speed
$s_{\rm sharp}(t,x)$ at $x \in \Gm(t)$, then the Clausius-Duhem
inequality holds if and only if the inequality
\begin{equation}\label{E2.cdequiv}
s_{\rm sharp}(t,x) \Big( n(t,x) \cdot
  [\hat{C}](t,x)n(t,x) + \la^{1/2} c_1  \ka_\Gm(t,x) \Big) \geq 0
\end{equation}
is satisfied at every point $x \in \Gm(t)$. A proof of this well known
result is given in \cite{Al00}, however only for the case where $\la =
0$ in \eq{2.56}. The proof can be readily generalized to the case $\la
> 0$.

A simple linear kinetic relation, for which \eq{2.cdequiv} obviously
holds, is 
\begin{equation}\label{E2.kinrelsharp}
s_{\rm sharp} = \frac{c}{c_1}\, \big( n \cdot [\hat{C}]n + \la^{1/2}
  c_1  \ka_\Gm \big). 
\end{equation}
The sharp interface problem thus consists of the transmission
problem \eq{2.8} -- \eq{2.12} combined with the kinetic relation
\eq{2.kinrelsharp}. For this problem the Clausius-Duhem inequality is
satisfied.

We can now define the model error. To this end let $\big(u^{(\mu)},
T^{(\mu)}, S^{(\mu)} \big)$ be the asymptotic solution in the domain
$Q = [t_1,t_2] \ti \Om$ constructed in \reft{2.3}, where by \eq{2.36},
the manifold $\Gm$ is the level set $\{ S^{(\mu)}=\frac12 \}$. 
Let $\hat{t}\in [t_1,t_2]$ be a fixed time 
and let$\big( u_{\rm AC}^{(\mu)}, T_{\rm AC}^{(\mu)}, S_{\rm
  AC}^{(\mu)} \big)$ be the exact solution of the Allen-Cahn model
\eq{1.1} -- \eq{1.3} in the domain $[\hat{t},t_2] \ti \Om$, which
satisfies the boundary and initial conditions
\begin{eqnarray} 
u^{(\mu)}_{\rm AC}(t,x) &=& {\sf U}(t,x), \qquad (t,x) \in [\hat{t},t_2]
  \ti \pa\Om, \label{E2.BC1AC}
\\
\pa_{n_{\pa\Om}} S^{(\mu)}_{\rm AC}(t,x) &=& f_3^{(\mu\la)}(t,x), \qquad
 (t,x) \in [\hat{t},t_2] \ti \pa\Om, \label{E2.BC2AC}  
\\
S_{\rm AC}^{(\mu)}(\hat{t},x) &=& S^{(\mu)} (\hat{t},x),  \quad x
  \in \Om, \label{E2.ICAC}
\end{eqnarray}
where $f_3^{(\mu\la)}$ is the right hand side of \eq{2.36e}. 
The level set of the order parameter $S_{\rm AC}^{(\mu)}$ is
denoted by   
\begin{equation}\label{E2.GmAC} 
\Gamma_{\rm AC}  = \Big\{ (t,x) \in Q \Bigm|
  S_{\rm AC}^{(\mu)}(t,x)=\frac{1}{2}\Big\}.
\end{equation}
Let $\Gm_{\rm sharp} \subseteq Q$ be the sharp interface in the
solution of the sharp interface problem \eq{2.8} -- \eq{2.12},
\eq{2.kinrelsharp}, which satisfies the initial condition
\begin{equation}\label{E2.ICsharp} 
\Gm_{\rm sharp}(\hat{t}) = \Gm(\hat{t}).
\end{equation}
The normal speeds of the different surfaces are 
\[ 
s = s^{(\mu\la)} = \EN\big(\Gamma^{(\mu\la)}\big), \qquad   
  s_{\rm AC} = s_{\rm AC}^{(\mu\la)}= \EN\big(\Gamma_{\rm
  AC}^{(\mu\la)}\big), \qquad s_{\rm sharp} = \EN\big(\Gm_{\rm sharp}
\big), 
\]
where $\EN$ is the normal speed operator introduced at the beginning
of Section~\ref{S2.2}. Of course, $s_{\rm sharp}$ is given by
\eq{2.kinrelsharp}. Note that the functions $s^{(\mu\la)}(\hat{t})$,
$s_{\rm AC}^{(\mu\la)}(\hat{t})$, $s_{\rm sharp}(\hat{t})$ are defined
on the same set, since the initial condition \eq{2.ICAC} and
\eq{2.ICsharp} together imply $\Gm_{\rm AC}(\hat{t}) = \Gm(\hat{t}) =
\Gm_{\rm sharp}(\hat{t})$.

\begin{tdefi}\label{D2.7}
We call the function ${\cal E}= {\cal
  E}^{(\mu\la)}(\hat{t}):\Gm(\hat{t}) \ra \R$ defined by  
\begin{equation}\label{E2.modelerror}
{\cal E} = s_{\rm AC}(\hat{t}) - s_{\rm sharp}(\hat{t})
\end{equation}
the model error of the Allen-Cahn model at time $\hat{t}$ to the
parameters $\mu$ and $\la$.   
\end{tdefi}
We can now discuss the choice of the parameters $\mu$ and $\la$. 
Since \eq{2.kinrelsharp} coincides with the leading term
$s_0$ in the asymptotic expansion \eq{2.31} of the kinetic relation of
the Allen-Cahn model, which is seen from \eq{2.55}, we have
\begin{equation}\label{E2.ssharp}
s_{\rm sharp} = s_0.
\end{equation}
Therefore \eq{2.31} yields
\begin{equation}\label{E2.decomperror}
{\cal E} = s_{\rm AC} - s_{\rm sharp} = s_{\rm AC} - s_0 = ( s_{\rm
  AC} - s ) + ( s - s_0 ) =  ( s_{\rm AC} - s )  + \mu^{1/2}(s_{10} +
  \la^{1/2} s_{11}) .  
\end{equation}
The difference $ s_{\rm AC} - s$ between
the propagation speeds of the exact solution and the asymptotic
solution tends to zero for $\mu \ra 0$ faster than the term
$\mu^{1/2}s_{10}$, and the convergence is uniform with respect to
$\la$. This is the basic result, which allows to discuss the optimal
choice of $\mu$ and $\la$. The precise result is

\begin{theo}\label{T2.8}
There is a constant $C_{\cal E} > 0$ such that for all $0 < \mu \leq
\mu_0$ and all $0 < \la \leq \la_0$ we have the estimate   
\begin{equation}\label{E2.assumpA}
\|s_{\rm AC}(\hat{t}) - s(\hat{t})\|_{L^2(\Gm(\hat{t}))}
 \leq C_{\cal E} | \ln \mu |^3 \mu.  
\end{equation}
\end{theo}
The {\bf proof} of this theorem is given in Section~\ref{S6}.
\\[1ex]
\eq{2.decomperror} and \eq{2.assumpA} together yield 
\begin{equation}\label{E2.esterror}
\| {\cal E}^{(\mu\la)} \|_{L^2(\Gm(\hat{t}))} \leq C
  \mu^{1/2},  
\end{equation} 
with a constant $C$, which can be chosen independently of $\la$. By this
inequality, $\mu^{1/2}$ controls the model error. Therefore we write
$F = \mu^{1/2}$ and call $F$ the error parameter. Moreover, since
$\la^{1/2} c_1$ is the interface energy density, we call $E =
\la^{1/2}$ the interface energy parameter. Also, since by \eq{2.39}
the interface width is proportional to $(\mu\la)^{1/2}$, we call $B =
(\mu\la)^{1/2}$ the interface width parameter. These three parameters
and the propagation speed $s_{\rm AC}$ are connected by the
fundamental relations 
\begin{gather}
B = EF, \label{E2.fundeqnAC1}
\\
s_{\rm AC} = \frac{c}{c_1} n \cdot [\hat{C}] n + c \ka_\Gm E + {\cal
  E}[E,F],  \label{E2.fundeqnAC2} 
\\
\|{\cal E}[E,F]\|_{L^2(\Gm(\hat{t}))} \leq C F,
  \label{E2.fundeqnAC3}  
\end{gather}
where we use the notation ${\cal E}[E,F] = {\cal E}^{(\mu\la)}$. The
first equation is an immediate consequence of the definition of the
parameters, the second is obtained by insertion of \eq{2.kinrelsharp}
into \eq{2.modelerror}, and the last inequality is just a restatement
of \eq{2.esterror}.

Now assume that we want to use a phase field model to numerically
simulate the propagation of a phase interface. In such a simulation
the numerical effort is proportional to $h^{-p}$, where $h$ denotes
the grid spacing and where the power $p > 1$ depends on whether we
want to simulate a problem in $2$--d or in $3$--d and it depends on
the numerical scheme we use. In order for the simulation to be
precise, we must guarantee that the model error and the numerical
error are small. To make the numerical error small, we must choose the
grid spacing $h$ small enough to resolve the transition of the order
parameter across the interface, which means that we must choose $h <
B$, hence we have $h^{-p} > B^{-p}$. Therefore we see that the
numerical effort of a simulation based on a phase field model is
measured by the number $B^{-p}$. We call the number
\[
e_{\rm num} = B^{-p}  
\]
the parameter of numerical effort. For a simulation based on the
Allen-Cahn model we see from \eq{2.fundeqnAC1} that the numerical
effort is  
\[
e_{\rm num} = (EF)^{-p}.
\]
Assume that the interface, which we want to simulate with the
Allen-Cahn model, has very small interface energy density. Such
interfaces are common in metallic or functional materials. For such
materials the interface energy parameter $E$ is small. To make the
model error small, we must also choose the error parameter $F$ small,
which means that the numerical effort parameter $e_{\rm num} =
(EF)^{-p}$ is very large as a product of two large numbers $E^{-p}$
and $F^{-p}$.

To be more specific, we consider an interface without interface
energy, which means that the free energy $\psi_{\rm sharp}$ does not
contain the last term on the right hand side of \eq{2.56}. From
\eq{2.kinrelsharp} we see that the propagation speed of the sharp
interface with zero interface energy density is 
\[
s_{\rm sharp} = \frac{c}{c_1} n \cdot [\hat{C}]n.
\]
From this
equation and from \eq{2.fundeqnAC2} we see that in this case the total
model error, which we denote by ${\cal E}_{\rm total}$, is
\[
{\cal E}_{\rm total} = s_{\rm AC} - s_{\rm sharp} = c\ka_\Gm E +
  {\cal E}[E,F] .  
\] 
This means that the term $c\ka_\Gm E$ is now part of the total model
error. This term does not vanish identically, since we cannot set $\la
= 0$ in the Allen-Cahn equation \eq{1.3}. Instead the values of $\la$
and of $E = \la^{1/2}$ must be positive.

If we prescribe the $L^2$-norm ${\cal E}_{L^2} = \|{\cal E}_{\rm
  total}\|_{L^2(\Gm(\hat{t}))}$ of the total model
error, we must therefore choose the parameters $E$ and $F$ such that
\begin{eqnarray}
 c \|\ka_\Gm\|_{L^2(\Gm(\hat{t}))} E + \|{\cal E}[E,F]\|_{L^2(\Gm(\hat{t}))} &\leq&
 {\cal E}_{L^2},  \label{E2.opti1}  
\\
EF &\overset{!}{=}& \max, \label{E2.opti2}
\end{eqnarray}
where the second condition is imposed by the requirement to make the
numerical effort $e_{\rm num} = (EF)^{-p}$ as small as possible. To
discuss this optimization problem, we assume first that the term
$s_{10}$ in the asymptotic expansion \eq{2.31} of the kinetic relation
of the Allen-Cahn model is not identically equal to zero. In this case
we conclude from \eq{2.decomperror} and \eq{2.assumpA} by the inverse
triangle inequality that for sufficiently small $\la^{1/2} = E$ and
for sufficiently small $\mu^{1/2} = F$
\begin{multline*}
\|{\cal E}[E,F]\|_{L^2(\Gm(\hat{t}))} = \| \mu^{1/2} s_{10} +
  (\mu\la)^{1/2} s_{11} + (s_{\rm AC} -s) \|_{L^2(\Gm(\hat{t}))} 
\\
\geq \mu^{1/2} \|s_{10}\|_{L^2(\Gm(\hat{t}))} - (\mu\la)^{1/2}
  \|s_{11}\|_{L^2(\Gm(\hat{t}))} - \|s_{\rm AC} -s
  \|_{L^2(\Gm(\hat{t}))} 
\\
\geq \mu^{1/2} \big( \|s_{10} \|_{L^2(\Gm(\hat{t}))} - \la^{1/2}
 \|s_{11}\|_{L^2(\Gm(\hat{t}))} - C_{\cal E} |\ln \mu|^3 \mu^{1/2}
 \big) 
\\ 
\geq \mu^{1/2} \Big( \| s_{10} \|_{L^2(\Gm(\hat{t}))} - \frac12
  \|s_{10} \|_{L^2(\Gm(\hat{t}))} \Big) = \frac12
  \|s_{10}\|_{L^2(\Gm(\hat{t}))} F. 
\end{multline*}
This inequality and \eq{2.opti1} imply that the solution $(E,F)$ of
the optimization problem \eq{2.opti1}, \eq{2.opti2} satisfies
\[
F \leq \frac{2}{\|s_{10}\|_{L^2(\Gm(\hat{t}))}}  \|{\cal E}[E,F]\|_{L^2(\Gm(\hat{t}))}
  \leq \frac{2}{\|s_{10}\|_{L^2(\Gm(\hat{t}))}}\, {\cal E}_{L^2} 
\qquad \mbox{and} \qquad E \leq \frac{1}{c \|\ka_\Gm\|_{L^2(\Gm(\hat{t}))}}\,
  {\cal E}_{L^2} .
\]
From this result we obtain
\begin{coro}\label{C2.5}
Let ${\cal E}_{\rm max}$ denote the total model error of the Allen-Cahn
model in the simulation of an interface without interface energy.
If the term $s_{10}$ in the asymptotic expansion \eq{2.31} of the
kinetic relation of the Allen-Cahn model is not identically equal to
zero, then the interface width $B$ satisfies
\begin{equation}\label{E2.withErr}
B = EF \leq \frac{2}{c \|s_{10}\|_{L^2(\Gm(\hat{t}))}\|\ka_\Gm\|_{L^2(\Gm(\hat{t}))}}\,
{\cal E}_{L^2}^2\,.  
\end{equation}
In a numerical simulation of an interface without interface energy the
parameter of numerical effort satisfies
\begin{equation}\label{E2.numeff}
e_{\rm num} \geq \left(
  \frac{c  \|s_{10}\|_{L^2(\Gm(\hat{t}))}  
  \|\ka_\Gm\|_{L^2(\Gm(\hat{t}))} }{ 2\,{\cal E}_{L^2}^2} \right)^p 
\end{equation} 
with a power $p > 1$ depending on the space dimension and the
numerical method used. 
\end{coro}
The interface width thus decreases with the square of the model
error. Since the time step in a simulation must be decreased when the
grid spacing $h$ in $x$--direction is decreased, the number $p$ can be
larger than $4$ in a three dimensional simulation. From \eq{2.numeff}
we thus see that the numerical effort grows very rapidely when the
required accuracy is increased. The Allen-Cahn model is therefore
ineffective when used to accurately simulate interfaces with low
interface energy.

If the term $s_{10}$ vanishes identically, then the same
considerations show that instead of \eq{2.withErr} and \eq{2.numeff}
we would have $B = O({\cal E}_{L^2}^{3/2})$ and $e_{\rm num} \geq
C {\cal E}_{L^2}^{-\frac32 p}$. The numerical effort would still
grow fast when the required accuracy is increased, though less fast
than for $s_{10} \neq 0$. However, a close investigation of the terms
in the definition \eq{2.33} of $s_{10}$, which we do not present here,
shows that only in very exceptional situations one can expect that
$s_{10}$ vanishes identically.

In \refc{2.5} we assumed that the mesh is globally refined.  Of
course, one can improve the effectivity of simulations by using
local mesh refinement in the neighborhood of the interface. We do not
discuss this question of numerical analysis here, but \refc{2.5} in
fact shows that adaptive mesh refinement and other advanced numerical
techniques are needed to make precise simulations of interfaces with
small energy based on the Allen-Cahn model effective.

\paragraph{Comparison to the hybrid phase field model} With \refc{2.5}
we can refine the comparison given in \cite{JElast2012} of the
Allen-Cahn model and another phase field model, which we call the
hybrid model. The hybrid model was introduced and discussed in
\cite{Al00,Al04,AlPZ06,AlPZ07,JElast2012}. By formal construction of
asymptotic solutions we showed in \cite{JElast2012} the following
result:
\\[1ex]
Let $B_{\rm AC}({\cal E}_{L^2})$ and $B_{\rm hyb}({\cal E}_{L^2})$ be
the interface widths in the Allen-Cahn model and the 
hybrid model, respectively, which result when the model parameters are
adjusted to model an interface without interface energy with the total
model error ${\cal E}_{L^2}$. Then we have for ${\cal E}_{L^2}
\ra 0$ that
\[
B_{\rm hyb}({\cal E}_{L^2}) = O({\cal E}_{L^2}), \qquad
  B_{AC}({\cal E}_{L^2}) = o(1)O({\cal E}_{L^2}) = o({\cal
  E}_{L^2}).   
\]
The Landau symbol $o(1)$ denotes terms, which tend to zero for ${\cal
  E}_{\rm max} \ra 0$. This result was obtained under the assumption
that estimates corresponding to \eq{2.assumpA} hold true for both
models, without proving this assumption.

To achieve a prescribed small value ${\cal E}_{L^2}$ of the total
model error we must therefore choose the interface width in the
Allen-Cahn model smaller than in the hybrid model. Consequently, the
hybrid model is numerically more effective, but how much more depends
on the rate of decay of the $o({\cal E}_{L^2})$ term in the result
for the Allen-Cahn model. In \cite{JElast2012} we could not determine
this decay rate, since the asymptotic solution constructed in
\cite{JElast2012} for the Allen-Cahn model was only of first order.

\refc{2.5} yields this decay rate. From the result for the hybrid
model and from \refc{2.5} we thus obtain for the parameters $e_{\rm
  num}^{\rm hyb}$ and $e_{\rm num}^{\rm AC}$ of the hybrid model and
the Allen-Cahn model, respectively, that
\[
e_{\rm num}^{\rm hyb} \leq C {\cal E}_{L^2}^{-p}, \qquad e_{\rm
  num}^{\rm AC} \geq C {\cal E}_{L^2}^{-2p}, 
\]
which shows that when the prescribed error ${\cal E}_{L^2}$ is
small, the hybrid model can be quite considerably more effectice in
numerical simulations of interfaces with low interface energy or no
interface energy than the Allen-Cahn model.


\subsection{The jump of solutions of the transmission problems
  }\label{S2.5}

In this section we prove some results on
the jumps of the solutions $(\hu,\hT)$ and $(\vu,\vT)$ of the
transmission problems \eq{2.8} -- \eq{2.12} and \eq{2.13} --
\eq{2.17}, which we need in the following sections.

We define a scalar product $\alpha:_D \beta$ on $\ES^3$ by $\alpha:_D
\beta=\alpha:(D\beta)$, for $\alpha, \beta \in \ES^3$. For a unit
vector $n \in \R^3$ let a linear subspace of $\ES^3$ be given by 
\begin{equation}\label{E2.41}
\ES_n^3= \Big\{\frac{1}{2} (\om \otimes n+n\otimes \om) \,\big\vert\,
\om \in \R^3\Big\}, 
\end{equation}
let $P_n:\ES^3 \to \ES^3$ be the projector onto $\ES_n^3$, which is
orthogonal with respect to the scalar product $\alpha:_D \beta$ and
let $Q_n=I-P_n$.

\begin{lem}\label{L2.6}
Let $\om^* \in \R^3$ be a vector. This vector satisfies $\Big( D\big(
\ve(\om^* \otimes n) - \ov{\ve} \big) \Big) n = 0$ if and only if
$\ve(\om^* \otimes n) = P_n \ov{\ve}$ holds. 
\end{lem}
This lemma is proved in \cite[Lemma 2.2]{CMT2011}.

\begin{lem}\label{L2.7}
Let $(\hat{u}, \hat{T})$ be a solution of the transmission problem
\eq{2.8} -- \eq{2.12}. Assume that $\hu$ is continuous in $Q$ and
that the limits $(\na_x \hu)^{(\pm)}$ exist and define continuous
extensions of $\na_x \hu$ from the set $\gm'$ to $\gm' \cup \Gm$ and
from the set $\gm$ to $\gm \cup \Gm$, respectively. Then we have
\begin{align}
&[\ve(\na_x \hat{u})] =\ve(u^* \otimes n), &&
    [\hat{T}]=D\big(\ve(u^*\otimes n)- \ov{\ve}\big),
    \label{E2.42}\\ 
&[\ve(\na_x \hat{u})] = P_n \ov\ve, && [\hat{T}]=-D
     Q_n \ov\ve, \label{E2.43} 
\end{align}
\end{lem} 
{\bf Proof:} Equation \eq{2.42} is proved in \cite[Lemma
2.2]{JElast2012}, \eq{2.43} is proved in \cite{CMT2011}. For
completeness we give the short proofs here.

Since by assumption $\hu$ is continuous across $\Gm$ and since $\na_x
\hu$ has continuous extensions from both sides of $\Gm$ onto $\Gm$,
the surface gradients $(\na_\Gm \hu)^{(+)}$ and $(\na_\Gm \hu)^{(-)}$
on both sides of $\Gm$ coincide, hence $[\na_\Gm \hu] = 0$. Using the
decomposition \eq{2.7} and the definition \eq{2.26} of $u^*$
we therefore obtain  
\begin{equation}\label{E2.47}
[\na_x \hu] = [(\pa_n \hu) \otimes n + \na_\Gm \hu] = [(\pa_n \hu)
\otimes n] + [\na_\Gm \hu] = [\pa_n \hu] \otimes n = u^* \otimes n. 
\end{equation}
Thus, by \eq{2.9},
\[
D\big( \ve(u^* \otimes n) - \ov{\ve} \big) = D\big( [\ve(\na_x \hu)] -
\ov{\ve} [\hS] \big) = \big[ D\big( \ve (\na_x \hu) - \ov{\ve} \hS
  \big) \big] = [\hT].
\]
This proves \eq{2.42}. From \eq{2.11} and \eq{2.42} we infer that 
\[
0 = [\hT]n = \Big( D \big( \ve(u^* \otimes n) - \ov{\ve} \big) \Big) n,
\]
so that $[\ve(\na_x \hu)] = \ve(u^* \otimes n) = P_n \ov{\ve}$, by
\refl{2.6}. Therefore we find 
\[
[\hT] = D\big( [\ve(\na_x \hu)] - \ov{\ve} \big) = D(P_n \ov{\ve} -
\ov{\ve}) = - D Q_n \ov{\ve},
\]
which proves \eq{2.43}.  \qed

\begin{lem}\label{L2.8}
Let $(\hu,\hT)$ and $(\vu,\vT)$ be solutions of the transmission
problems \eq{2.8} -- \eq{2.12} and \eq{2.13} -- \eq{2.16},
respectively. Assume that $\hu$ and $\vu$ are continuous in $Q$ and
that the limits $(\na_x \hu)^{(\pm)}$ and $(\na_x \vu)^{(\pm)}$ exist
and define continuous extensions of $\na_x \hu$ and of $\na_x \vu$
from the set $\gm'$ to $\gm' \cup \Gm$ and from the set $\gm$ to $\gm
\cup \Gm$, respectively. Then we have
\begin{equation}\label{E3.jumpvu}
[\pa_n \vu] = \big[ \frac{\ov{\ve}:\hT}{\hat{\psi}''(\hS)}
     \big] u^*, \qquad  [\na_x \vu ] = \big[
     \frac{\ov{\ve}:\hT}{\hat{\psi}''(\hS)} \Big] u^* \otimes n. 
\end{equation}
\end{lem}
{\bf Proof:} The decomposition \eq{2.7} yields 
\begin{equation}\label{E3.jumppavu}
[\na_x \vu] = [(\pa_n \vu) \otimes n + \na_\Gm \vu] = [(\pa_n \vu)
\otimes n] + [\na_\Gm \vu] = [\pa_n \vu] \otimes n. 
\end{equation}
From this equation and from \eq{2.14}, \eq{2.16} we infer 
\[
0 = [\vT]n = \Big( D \big( [\ve(\na_x \vu)] -  \big[
  \frac{\ov{\ve}:\hT}{\hat{\psi}''(\hS)} \big] \ov{\ve} \big) \Big) n
  = \Big( D \big( \ve\big( [\pa_n \vu] \otimes n \big) - \big[
  \frac{\ov{\ve}:\hT}{\hat{\psi}''(\hS)} \big] \ov{\ve} \big) \Big)
  n.
\]
Thus, \refl{2.6}, the linearity of the projector $P_n$ and \eq{2.42},
\eq{2.43} imply
\[
\ve\big( [\pa_n \vu] \otimes n \big) = \big[ \frac{\ov{\ve}:\hT}
  {\hat{\psi}''(\hS)} \big] P_n \ov{\ve} = \big[ \frac{\ov{\ve}:\hT}
  {\hat{\psi}''(\hS)} \big] \ve(u^* \otimes n) = \ve\Big( \big[
  \frac{\ov{\ve}:\hT} {\hat{\psi}''(\hS)} \big] u^* \otimes n \Big),      
\]
whence
\[
\Big(  [\pa_n \vu] -  \big[ \frac{\ov{\ve}:\hT} {\hat{\psi}''(\hS)}
  \big]  u^* \Big) \otimes n + n \otimes  \Big(  [\pa_n \vu] -  \big[
  \frac{\ov{\ve}:\hT} {\hat{\psi}''(\hS)} \big]  u^* \Big) = 0.
\]
We multiply this equation from the right with $n$ and obtain
\[
\Big( [\pa_n \vu] -  \big[ \frac{\ov{\ve}:\hT} {\hat{\psi}''(\hS)}
  \big]  u^* \Big) + n  \Big(  [\pa_n \vu] -  \big[
  \frac{\ov{\ve}:\hT} {\hat{\psi}''(\hS)} \big]  u^* \Big)\cdot n = 0,
\]
which means that $[\pa_n \vu] -  \big[ \frac{\ov{\ve}:\hT}
{\hat{\psi}''(\hS)} \big]  u^* $ is a multiple of $n$. Scalar
multiplication of the last equation with $n$ yields $ \big( [\pa_n
\vu] -  \big[ \frac{\ov{\ve}:\hT} {\hat{\psi}''(\hS)} \big]  u^* \big)
\cdot n =0$, from which we now conclude that the first equation in
\eq{3.jumpvu} holds. The second equation is obtained from
\eq{3.jumppavu}. \qed

\section{The asymptotic solution}\label{S3}

This section forms the first part of the proof \reft{2.3}. We state
in this section the form of the asymptotic solution
$(u^{(\mu)},T^{(\mu)},S^{(\mu)})$. The proof continues in
Section~\ref{S4}, where we study properties of the functions
$S_0,\ldots,S_2$ appearing in the asymptotic solution. In
Section~\ref{S5} we use these properties to show that
$(u^{(\mu)},T^{(\mu)},S^{(\mu)})$ is an asymptotic solution by
verifying that the estimates \eq{2.37a} --\eq{2.38b} hold. This
concludes the proof of \reft{2.3}.

\subsection{Notations}\label{S3.1}

Before we can start with the construction of the asymptotic solution
we must introduce more definitions and notations. In particular, we
must introduce parallel manifolds to the manifold $\Gm$ and we must
extend the definition of the surface gradients for functions defined
on $\Gm$, which are given in Section~\ref{S2.1}, to functions defined
on the parallel manifolds.  These definitions and notations are
needed throughout the remaining sections.

Let $\da > 0$ be the number from \eq{2.2}. For $\xi$ satisfying
$-\delta < \xi < \delta$ 
\[
\Gm_\xi = \{ (t,\eta + n(t,\eta)\xi) \mid (t,\eta) \in \Gm \} 
\]
is a three dimensional parallel manifold of $\Gm$ embedded in ${\mathcal
  U}_\da$, and 
\[
\Gm_\xi(t) = \{ x \in \Om \mid (t,x) \in \Gm_\xi \}
\]
is a two-dimensional parallel surface of $\Gm(t)$ embedded in
${\mathcal U}_\da(t)$. Let $\tau_1,\ \tau_2 \in \R^3$ be two
orthogonal unit vectors tangent to $\Gm_{\xi}(t)$ at $x \in
\Gm_{\xi}(t)$. For functions $w:\Gm_\xi(t) \rightarrow \R$, $W:
\Gm_\xi(t) \rightarrow \R^3$ and $\hat{W}: \Gm_\xi(t) \rightarrow
\R^{3\ti 3}$ we define the surface gradient and the surface
divergence on $\Gm_\xi(t)$ by 
\begin{eqnarray} 
\na_{\Gm_\xi} w &=& (\pa_{\tau_1} w)\tau_1 + (\pa_{\tau_2}
   w)\tau_2 , \label{Esurfgrad-intro}
\\[1ex]  
\na_{\Gm_\xi} W &=& (\pa_{\tau_1}W)\otimes\tau_1 + (\pa_{\tau_2}W)
  \otimes \tau_2 , \label{Esurfgradmat-intro}
\\
\div_{\Gm_\xi} W & = & \tau_1 \cdot \pa_{\tau_1} W + \tau_2 \cdot
  \pa_{\tau_2} W = \sum_{i=1}^2 \tau_i \cdot (\na_{\Gm_\xi} W)\tau_i,  
  \label{Esurfdiv} 
\\  
\div_{\Gm_\xi} \hat{W} & = & (\pa_{\tau_1} \hat{W}) \tau_1 + (\pa_{\tau_2}
  \hat{W}) \tau_2 . \label{Esurfdivmat}    
\end{eqnarray}
Clearly, with $\na_\Gm$ defined in \eq{2.4} and \eq{2.5} we have
$\na_{\Gm_0}=\na_\Gm$. For brevity we write $\div_\Gm = \div_{\Gm_0}$.
If $w$, $W$, $\hat{W}$ are defined on $\Gm_\xi$, we define
$\na_{\Gm_\xi} w: \Gm_\xi \mapsto \R^3$, $\na_{\Gm_\xi} W: \Gm_\xi
\mapsto \R^{3\times3}$, $\div_{\Gm_\xi} W: \Gm_\xi \mapsto \R$,
$\div_{\Gm_\xi} \hat{W}: \Gm_\xi \mapsto \R^3$ by applying the
operators $\na_{\Gm_\xi}$ and $\div_{\Gm_\xi}$ to the restrictions
$w\rain{\Gm_\xi(t)}$, $W\rain{\Gm_\xi(t)}$, $\hat{W}\rain{\Gm_\xi(t)}$
for every $t$. With these definitions we have the splittings 
\begin{eqnarray} 
\na_x w(t,x) &=& \pa_\xi w(t,\eta,\xi)\, n(t,\eta) +
  \na_{\Gm_\xi} w(t,\eta,\xi),  \label{Eskalgradsplit}
\\ 
\na_x W(t,x) &=& \pa_\xi W(t,\eta,\xi)\otimes n(t,\eta) + \na_{\Gm_\xi}
  W(t,\eta,\xi),  \label{Egradsplit}
\\
\div_x \hat{W}(t,x) &=& \big(\pa_\xi \hat{W}(t,\eta,\xi)\big) n(t,\eta) +
  \div_{\Gm_\xi} \hat{W}(t,\eta,\xi), \label{Edivsplit}  
\end{eqnarray}
where, as usual, $W(t,\eta,\xi) = W\big(t,\eta + n(t,\eta) \xi\big)$. 

The operators $\na_\Gm$ and $\div_\Gm$ can be applied to functions
defined on subsets of $\Gm$. In contrast, the operator $\na_\eta$
introduced next can be applied to functions defined on $\Gm \ti J$,
where $J \subseteq \R$ is an interval. For $w:\Gm \ti J \ra \R$,
$W:\Gm \ti J \ra \R^3$, $\hat{W}:\Gm \ti J \ra \R^{3\ti 3}$ consider
the functions $\eta \mapsto w_{t,\xi}(\eta) = w(t,\eta,\xi)$, $\eta
\mapsto W_{t,\xi}(\eta) = W(t,\eta,\xi)$, $\eta \mapsto
\hat{W}_{t,\xi}(\eta) = \hat{W}(t,\eta,\xi)$, which are defined on
$\Gm(t)$. To these functions the operators $\na_\Gm$ and $\div_{\Gm}$
can be applied. We set  
\begin{eqnarray}
\na_\eta w(t,\eta,\xi) &=& \na_\Gm w_{t,\xi}(\eta) \in
  \R^3, \label{Enaeta} \\
\na_\eta W(t,\eta,\xi) &=& \na_\Gm W_{t,\xi}(\eta) \in \R^{3\ti 3},
\label{EnaetaW} \\
\div_\eta W(t,\eta,\xi) &=& \div_\Gm W_{t,\xi}(\eta) \in \R, 
\label{Ediveta} \\ 
\div_\eta \hat{W}(t,\eta,\xi) &=& \div_\Gm \hat{W}_{t,\xi}(\eta) \in
  \R^3. \label{EdivetahW}  
\end{eqnarray}
If $W$ is defined on ${\cal U}_\da$, then $(t,\eta,\xi) \ra
W(t,\eta,\xi) = W(t,\eta + n(t,\eta) \xi)$ is defined on $\Gm\ti
(-\da,\da)$. Consequently, the gradient $\na_\eta W$ is
defined. The connection between $\na_\eta W$ and $\na_{\Gm_\xi} W =
\na_{\Gm_\xi} W\rain{\Gm_\xi}$ is given by the chain rule, which
yields  
\begin{equation}\label{Eetatrans}
\na_\eta W (t,\eta,\xi) = \big(\na_{\Gm_\xi} W(t,\eta + n(t,\eta) \xi)\big)
 \big( I + \xi \, \na_\eta n(t,\eta) \big). 
\end{equation}
In particular, we have $\na_\eta W(t,\eta,0) = \na_\Gm W(t,\eta)$. Similar
formulas and relations hold for $\na_\eta w$, $\div_\eta W$, $\div_\eta
\hat{W}$. If $W:{\cal U}_\da \ra \R^3$ is constant on all
the lines normal to $\Gm(t)$, for all $t$, we have $W(t,\eta,\xi) =
W(t,\eta)$. For such functions we sometimes interchangeably use the notations
$\na_\eta W$ and $\na_\Gm W$. Similarly, we interchangeably use the notations
$\na_\eta w$ and $\na_\Gm w$, $\div_\eta W$ and $\div_\Gm W$, $\div_\eta
\hat{W}$ and $\div_\Gm \hat{W}$ if $w$ and $\hat{W}$ are independent of
$\xi$.   

Note that by \eq{etatrans} we have for $x \in \Gm_\xi(t)$ that 
\begin{equation}\label{Esurfgradtrans} 
\na_{\Gm_\xi} W(t,x) = \big(\na_\eta W(t,\eta,\xi)\big) A(t,\eta,\xi),
\end{equation}
where $A(t,\eta,\xi) \in \R^{3\ti3}$ is the inverse of the linear mapping
$\big( I + \xi \na_\eta n(t,\eta)\big):\R^3 \ra \R^3 $. From the mean value
theorem we obtain the expansion 
\begin{equation}\label{EAexpansion}
A(t,\eta,\xi) = I + \xi\, R_A(t,\eta,\xi)  
\end{equation}
where the remainder term $R_A(t,\eta,\xi) \in \R^{3\ti 3}$ is bounded
when $(t,\eta,\xi)$ varies in $\Gm \ti (-\da,\da)$. Insertion into
\eq{surfgradtrans} yields
\begin{equation}\label{Esurfgraddeco}
\na_{\Gm_\xi} W(t,x) = \na_\eta W(t,\eta,\xi)\, \big(I +
  \xi R_A(t,\eta,\xi) \big). 
\end{equation}
For $w :{\cal U}_\da \to \R$ we consider $\na_{\Gm_\xi} w$ and
$\na_\eta w$ to be column vectors. For such $w$ the equation
corresponding to \eq{surfgraddeco} is
\begin{equation}\label{EnaGmxiw}
\na_{\Gm_\xi} w(t,x) = A^T(t,\eta,\xi) \na_\eta w(t,\eta,\xi) = \big(I + 
  \xi\, R_A^T(t,\eta,\xi) \big)\, \na_\eta w(t,\eta,\xi) .   
\end{equation}
Furthermore, \eq{surfdiv}, \eq{surfgraddeco} and \eq{diveta} together yield
for $W:{\cal U}_\da \ra \R^3$ that   
\begin{equation}\label{EdivGmxiW}
\div_{\Gm_\xi} W = \sum_{i=1}^2 \tau_i \cdot \big((\na_\eta W)(I +
  \xi R_A)\tau_i\big) = \div_\eta W + \xi\, \div_{\Gm,\xi} W,
\end{equation}
with the remainder term  
\begin{equation}\label{Edivxi}
\div_{\Gm,\xi} W(t,\eta,\xi) = \sum_{i=1}^2 \tau_i \cdot \big( (
\na_\eta W ) R_A \tau_i \big) = \sum_{i=1}^2 \tau_i \cdot \big( (
  \na_\Gm W_{t,\xi} ) R_A \tau_i \big),   
\end{equation}
and this equation implies for $\hat{W}:{\cal U}_\da \ra \R^{3\ti3}$
that  
\begin{equation}\label{Esurfdivmatdeco}
\div_{\Gm_\xi} \hat{W}(t,\eta,\xi) = \div_\eta \hat{W} +  \xi\,
  \div_{\Gm,\xi} \hat{W},   
\end{equation}
where $\div_{\Gm,\xi} \hat{W} = \sum_{i,j=1}^2 (\pa_{\tau_j}
\hat{W}_{t,\xi})\, \tau_i (\tau_j \cdot R_A \tau_i)$. 
The terms $\div_{\Gm,\xi} W$ and $\div_{\Gm,\xi} \hat{W}$ are bounded when
$(t,\eta,\xi)$ varies in $\Gm \ti (-\da,\da)$.  

For functions $w$ with values in $\R$ we define the second gradients
$\na^2_{\Gm_\xi} w$, $\na^2_\eta w$ by applying the operators
$\na_{\Gm_\xi}$, $\na_\eta$ to the vector functions $\na_{\Gm_\xi} w$,
$\na_\eta w$. For $W$ with values in $\R^3$ we define second gradients
$\na^2_{\Gm_\xi} W$, $\na^2_\eta W$ by applying these operators to the
rows of $\na_{\Gm_\xi} W$, $\na_\eta W$. We remark that 
\[
\Da_{\Gm_\xi} w = \div_{\Gm_\xi} \na_{\Gm\xi}\, w
\]
is the surface Laplacian.

\begin{tdefi}\label{D3.1}
Let $I \subseteq \R$ be an interval. For $k,m \in \N_0$ and $p =
1,3$ we define the space 
\begin{multline*}
C^k \big( I, C^m (\Gm , \R^p)\big) \\
= \{ (t,\eta,\xi) \ra
 w(t,\eta,\xi): \Gm \ti I \ra \R^p \mid \pa^\ell_\xi
 \pa^i_t \na^j_\eta w \in C(\Gm \ti I), \ \ell \leq k, \ i+j \leq m
   \}.  
\end{multline*}

\end{tdefi}

\subsection{Construction of the asymptotic solution}\label{S3.2}

We start with the construction of the asymptotic solution $(u^{(\mu)},
T^{(\mu)}, S^{(\mu)})$. We assume that the hypotheses of Theorem
\ref{T2.3} are satisfied. In particular, we assume that there is a
sufficiently smooth solution $\Gm=\Gm^{(\mu)}$ of the evolution
problem \eq{2.evolution0}, \eq{2.evolution1}, with $s_0
(\hT,\ka_\Gm,\la^{1/2})$, $s_1 (\hu,\hT,\vT,S_0,S_1,\la^{1/2})$
defined in \eq{2.32} -- \eq{2.34}. By this assumption, the function
$(\hat{u}, \hat{T}, \vu, \vT, S_0, S_1)$ is known as a solution of the
transmission-boundary value problem \eq{2.8} -- \eq{2.24}. We use the
notation
\begin{equation}\label{E3.1}
1^+(r)=
\begin{cases}
1, & r >0\\
0, & r \leq 0
\end{cases},
\qquad 1^-(r)=1-1^+(r), 
\qquad r^\pm = r\,1^\pm(r). 
\end{equation}
Let $\phi \in C_0^\infty((-2,2))$ be a function satisfying $0 \leq
\phi(r) \leq 1$ for all $r \in \R$ and $\phi(r) = 1$ for $|r| \leq
1$. With the constant $a$ from \eq{2.29} we define a function
$\phi_{\mu\la}:Q \ra [0,1]$ by 
\begin{equation}\label{E3.2}
\begin{split}
\phi_{\mu \la} (t,x) &= \phi_{\mu\la}(t,\eta,\xi) = \phi \Big(
\frac{2a \xi}{3 (\mu\la)^{1/2} | \ln \mu| } \Big), \quad \mbox{for }
(t,x) \in {\cal U}_\da, \\
\phi_{\mu \la} (t,x) &= 0, \qquad \mbox{otherwise}.
\end{split}
\end{equation}
By \eq{2.25a}, $\phi_{\mu\la}$ is equal to $1$ in 
$Q_{\rm inn}^{(\mu\la)}$, transits smoothly from $1$ to $0$ in $Q_{\rm
  match}^{(\mu\la)}$, and vanishes in $Q_{\rm out}^{(\mu\la)}$. With this
function the asymptotic solution is defined by
\begin{align}
u^{(\mu)}(t,x) &= u_1^{(\mu)} (t,x)\, \phi_{\mu\la}(t,x) +
  u_2^{(\mu)}(t,x)\,  \big( 1-\phi_{\mu\la} (t,x) \big) ,
\label{E3.3}\\
S^{(\mu)}(t,x) &= S_1^{(\mu)} (t,x)\, \phi_{\mu\la}(t,x) +
  S_2^{(\mu)}(t,x)\,  \big( 1-\phi_{\mu\la} (t,x) \big) ,
\label{E3.4}\\
T^{(\mu)}(t,x) &= D \Big(\ve\big(\na_x u^{(\mu)}(t,x)\big)-
\ov\ve S^{(\mu)}(t,x)\Big), \label{E3.5} 
\end{align}
where $u_1^{(\mu)}$, $S_1^{(\mu)}$ are components of the inner
expansion $\big(u_1^{(\mu)}, T_1^{(\mu)}, S_1^{(\mu)}\big)$ defined in
${\cal U}_\da$, and $u_2^{(\mu)}$, $S_2^{(\mu)}$ are components of the
outer expansion $\big(u_2^{(\mu)}, T_2^{(\mu)}, S_2^{(\mu)} \big)$
defined in $Q \setminus \Gamma$. The function
$\big(u^{(\mu)},T^{(\mu)},S^{(\mu)} \big)$ is equal to the inner
expansion $\big(u_1^{(\mu)},T_1^{(\mu)},S_1^{(\mu)} \big)$ in the
region $Q_{\rm inn}^{(\mu\la)}$ and equal to the outer expansion
$\big(u_2^{(\mu)},T_2^{(\mu)},S_2^{(\mu)} \big)$ in the region $Q_{\rm
  out}^{(\mu\la)}$. In the region $Q_{\rm match}^{(\mu\la)}$ both
expansions are matched.

\paragraph{The outer expansion}

The outer expansion is defined as follows. With the solutions
$(\hat{u}, \hat{T})$ of the transmission problem \eq{2.8} -- \eq{2.12}
and $(\vu,\vT)$ of the transmission problem \eq{2.13} -- \eq{2.17} we
set for $(t,x) \in Q \setminus \Gm$  
\begin{eqnarray}
u_2^{(\mu)} (t,x) &=& \hu(t,x)+ \mu^{1/2} \vu (t,x)+ \mu\,
  \tilde{u}(t,x), 
\label{E3.6}\\[1ex] 
S_2^{(\mu)}(t,x) &=& \hat{S}(t,x)+ \mu^{1/2} \tilde{S}_1(t,x)+ \mu
\tilde{S}_2(t,x) +\mu^{3/2} \tilde{S}_3(t,x), 
\label{E3.7}\\[1ex] 
T_2^{(\mu)}(t,x) &=& D \Big(\ve\big(\na_x u_2^{(\mu)}(t,x)\big)-
\ov\ve S_2^{(\mu)}(t,x)\Big). \label{E3.8} 
\end{eqnarray}
The functions $\tu$, $\tS_1, \dots, \tS_3$ and another unknown
function $\tT$ solve the system of algebraic and partial differential
equations 
\begin{eqnarray}
-\div_x \tT &=& 0,  \label{E3.9}
\\
\tT &=& D\big(\ve(\na_x \tu) - \ov{\ve} \tS_2 \big),
\label{E3.10}\\ 
- \hT:\ov{\ve} + \hat{\psi}'' (\hS) \tS_1 &=& 0, \label{E3.11}
\\  
- \vT:\ov{\ve} + \hat{\psi}'' (\hS) \tS_2 + \frac12 \hat{\psi}'''
 (\hS) \tS_1^2 &=& 0,  \label{E3.12}
\\ 
- \tT:\ov{\ve} + \hat{\psi}'' (\hS) \tS_3 + \hat{\psi}''' (\hS)
  \tS_1 \tS_2 + \frac16 \hat{\psi}^{(IV)} (\hS)\tS_1^3  \quad \nn
\\
\mbox{} + \frac{\la^{1/2}}{c} \pa_t \tS_1 - \la \Da_x \tS_1 &=& 0,
\label{E3.13}  
\end{eqnarray}
in the set $Q \setminus \Gamma$. Moreover, $\tu$ satisfies the
boundary conditions  
\begin{eqnarray}
\tu (t,x) &=& 0, \qquad (t,x) \in [t_1,t_2] \ti \pa\Om,  
  \label{E3.14} 
\\
\tu^{(-)}(t,\eta) &=& \la^{1/2} u^*(t,\eta) \int_{-\infty}^0 
  S_1(t,\eta,\zeta) - \frac{\ov{\ve} : \hT^{(-)}(t,\eta)}
  {\hat{\psi}''(1)}\, d\zeta, \quad (t,\eta) \in \Gm, 
  \label{E3.15} 
\\ 
\tu^{(+)}(t,\eta) &=& \la^{1/2} u^*(t,\eta) \int_0^\infty
  S_1(t,\eta,\zeta) - \frac{\ov{\ve} : \hT^{(+)}(t,\eta)}
  {\hat{\psi}''(1)} \,d\zeta \nn
\\
&& \mbox{} + \la\, a^*(t,\eta) \int_{-\infty}^\infty \Big(
  \int_{-\infty}^\zeta S_0(\vartheta) \, d\vartheta - \zeta^+ \Big)
  d\zeta,   \quad (t,\eta) \in \Gm. 
  \label{E3.16}
\end{eqnarray}
Since by assumption $\Gm$, $\hT$, $\vT$, $S_1$ are known from
the evolution problem, this system can be solved recursively. To see
this, note that \eq{3.11} yields 
\begin{equation}\label{E3.17}
\tilde{S}_1= \frac{\hat{T}:\ov{\ve}}{\hat{\psi}''(\hat{S})}.
\end{equation}
We insert this equation into \eq{3.12} and solve this equation for
$\tS_2$ to obtain 
\begin{equation}\label{E3.18}
\tS_2 = \frac{\vT : \ov{\ve}}{\hat{\psi}''(\hS)} -
 \frac{\hat{\psi}'''(\hS)}{2\hat{\psi}''(\hS)}
 \Big( \frac{\hT:\ov{\ve}}{\hat{\psi}''(\hS)} \Big)^2.  
\end{equation}
Using this function in \eq{3.10}, we can determine $\tu$ and $\tT$
from the boundary value problem \eq{3.9}, \eq{3.10}, \eq{3.14} --
\eq{3.16}. Finally, we can solve \eq{3.13} for $\tilde{S}_3$.

\paragraph{The inner expansion}

The inner expansion $(u_1^{(\mu)},T_1^{(\mu)},S_1^{(\mu)})$ is
essentially obtained by smoothing the jumps of the functions $\hu$,
$\vu$ and $\hS$ from the evolution problem for the surface $\Gm$.
Before we can define the inner expansion we must therefore study in
the next two lemmas the jumps of $\hu$ and $\vu$ across $\Gm$.

Let $u^*= [\pa_n \hat{u}]$ and $a^* = [\pa_n^2 \hat{u}]$ be the jumps
of derivatives of $\hu$ across $\Gm$. These functions are introduced
in \eq{2.26}, \eq{2.27}. For $(t,x) = (t,x(t,\eta,\xi)) \in {\cal
  U}_\da$ we decompose $\hu$ and $\vu$ in the form
\begin{eqnarray}
\hu(t,x) &=& u^*(t,\eta)\, \xi^+ + a^*(t,\eta) \frac12(\xi^+)^2 +
  \hat{v}(t,x),  
\label{E3.19} \\
\vu(t,x) &=&  u^*(t,\eta) \Big(\frac{\ov{\ve} :
  \hat{T}^{(+)}(t,\eta)}{\hat{\psi}''(1)}\, \xi^+ + \frac{\ov{\ve} :
  \hat{T}^{(-)}(t,\eta)}{\hat{\psi}''(0)}\, \xi^- \Big) +
  \vv(t,x), 
\label{E3.20}    
\end{eqnarray}
where $\xi^+$, $\xi^-$ are defined in \eq{3.1} and where the remainder
terms $\hv$ and $\vv$ are defined by \eq{3.19}, \eq{3.20}. The
decomposition \eq{3.19} is motivated by the fact that
\begin{equation}\label{E3.jumphv}
[\pa_n^i \hat{v}] = 0, \qquad i = 0,1,2,  
\end{equation}
which follows immediately from \eq{3.19}, \eq{2.10} and \eq{2.26},
\eq{2.27}. The first two terms on the right hand side of \eq{3.19}
thus serve to separate off the jumps of the first and second
derivatives of $\hu$ at $\Gm$.  Similarly, the normal derivatives of
first order of $\vv$ do not jump across $\Gm$.  More precisely, we
have the following result.

\begin{lem}\label{L3.2}
Let $(\hu,\hT)$ and $(\vu,\vT)$ be solutions of the transmission
problems \eq{2.8} -- \eq{2.12} and \eq{2.13} -- \eq{2.17},
respectively.  
\\
(i) $\vv$ defined in \eq{3.20} satisfies
\begin{equation}\label{E3.jumpvv}
[\pa_n^i \vv] = 0, \qquad i=0,1.  
\end{equation}
(ii) Assume that $\Gm$ is a $C^5$--manifold. Suppose that $\hu \in
C^4(\gm \cup \gm',\R^3)$, $\vu \in C^3(\gm \cup \gm',\R^3)$ and that
$\hu$ has $C^4$--extensions, $\vu$ has $C^3$--extensions from $\gm$ 
to $\gm \cup \Gm$ and from $\gm'$ to $\gm' \cup \Gm$. With the
function spaces introduced in \refd{3.1} we then have
\begin{eqnarray}
\hv &\in& C^2\big( (-\da,\da), C^2(\Gm)\big) \cap C^3\big(
  (-\da,0], C^1(\Gm)\big) \cap C^3\big( [0,\da), C^1(\Gm)\big), 
\label{E3.hvregu}\\
\vv &\in& C^1 \big( (-\da,\da), C^2(\Gm)\big) \cap C^3 \big( \Gm \ti
  (-\da,0] \big) \cap C^3 \big( \Gm \ti [0,\da) \big).  
\label{E3.vvregu}
\end{eqnarray}   
\end{lem}
{\bf Proof:} 
To prove \eq{3.jumpvv} note that by definition of $[\pa_n
w]$ in Section~\ref{S2.1} and by definiton of $\xi^\pm$ in \eq{3.1}
we have  
\[
\Big[ \pa_n \Big(\frac{\ov{\ve} :
  \hat{T}^{(+)}}{\hat{\psi}''(1)}\, \xi^+ + \frac{\ov{\ve} :
  \hat{T}^{(-)}}{\hat{\psi}''(0)}\, \xi^- \Big) u^*
  \Big] 
= \Big( \frac{\ov{\ve} :
  \hat{T}^{(+)}}{\hat{\psi}''(1)} - \frac{\ov{\ve} :
  \hat{T}^{(-)}}{\hat{\psi}''(0)} \Big) u^* = \Big[ \frac{\ov{\ve} :
  \hat{T}}{\hat{\psi}''(\hS} \Big] u^*.
\]
From this equation, from the first equation in \eq{3.jumpvu} and from
\eq{3.20} we obtain \eq{3.jumpvv} for $i=1$. For $i=0$ equation
\eq{3.jumpvv} is an immediate consequence of \eq{3.20} and
\eq{2.15}. This proves (i).  

(ii) Since $\Gm$ is a $C^5$--manifold, the coordinate mapping
$(t,\eta,\xi) \mapsto (t,x(t,\eta,\xi)) = \big(t,\eta + \xi\,
n(t,\eta) \big)$ and the inverse mapping $(t,x) \mapsto
\big(t,\eta(t,x),\xi(t,x)\big)$ are $C^4$. It follows from this
differentiability property of the coordinate mapping and from our
differentiability assumptions for $\hu$ that $(t,\eta,\xi) \mapsto
\hu(t,\eta,\xi)$ is $C^4$ in $\Gm \ti (-\da,0]$ and in $\Gm \ti
[0,\da)$, and that $(t,\eta) \mapsto u^*(t,\eta) = n(t,\eta) \cdot
[\na_x \hu](t,\eta)$ belongs to $C^3 (\Gm)$ and
$(t,\eta) \mapsto a^*(t,\eta) = n(t,\eta) \cdot [\pa_n \na_x
\hu](t,\eta)$ belongs to $C^2 (\Gm)$. Since by \eq{3.19} we have
\[
\hv(t,\eta,\xi) = \hu(t,\eta,\xi) - u^*(t,\eta) \xi^+ -
a^*(t,\eta) \frac12(\xi^+)^2,    
\]   
these properties imply that 
\[
\hv \in \bigcap_{m=0}^1 \Big( C^{2+m}\big( (-\da,0],C^{2-m} (\Gm)\big)
\cap C^{2+m}\big( [0,\da),C^{2-m}(\Gm)\big) \Big).
\] 
From this relation and from \eq{3.jumphv} we conclude
that \eq{3.hvregu} holds. Relation \eq{3.vvregu} is obtained in the
same way, using \eq{3.jumpvv} instead of \eq{3.jumphv}. \qed
\\[1ex]  
For brevity in notation we define 
\begin{gather}
\hat{\si}(\xi) = \hat{\si}(t,\eta,\xi) = \ov{\ve}: D\ve\big(\na_x
  \hv(t,x) \big), \qquad
\hat{\si}'(\xi) = \pa_\xi \hat{\si}(t,\eta,\xi), \label{E3.21}
\\
\check{\si}(\xi) = \check{\si}(t,\eta,\xi) =  \ov{\ve}: D\ve\big(\na_x
  \vv(t,x) \big). \label{E3.22}
\end{gather}
Later we need the following result, which shows how $\vsi(t,\eta,0)$
can be computed from the limit values of $\hT$ and $\vT$ at $\Gm$.

\begin{lem}\label{L3.3}
The function $\vsi$ defined in \eq{3.22} satisfies    
\begin{eqnarray}
\ov{\ve}:\vT^{(+)} &=& \vsi(0) + \ov{\ve}:[\hT]\,
  \frac{\ov{\ve}:\hT^{(+)}} {\hat{\psi}''(1)}, \label{E3.22a}
\\
\ov{\ve}:\vT^{(-)} &=& \vsi(0) + \ov{\ve}:[\hT]\,
  \frac{\ov{\ve}:\hT^{(-)}} {\hat{\psi}''(0)},  \label{E3.22b}
\\
\vsi(0) &=& \ov{\ve}: \langle \vT \rangle -  \ov{\ve}:[\hT] \Big\langle
  \frac{\ov{\ve}:\hT} {\hat{\psi}''(\hS)} \Big\rangle.  \label{E3.22c}
\end{eqnarray}

\end{lem}
{\bf Proof:} We apply the decomposition \eq{2.7} of the gradient to
the function 
$W(t,\eta,\xi) = u^*(t,\eta) \Big( \frac{\ov{\ve} :
  \hT^{(+)}(t,\eta)}{\hat{\psi}''(1)} \xi^+ + \frac{\ov{\ve} :
  \hT^{(-)}(t,\eta)}{\hat{\psi}''(0)} \xi^- \Big).$ This yields   
\[
(\na_x  W)^{(+)}(t,\eta) = \big((\pa_n W) \otimes n + \na_\Gm
  W\big)^{(+)} = u^* \otimes n\, \frac{\ov{\ve} :
    \hT^{(+)}}{\hat{\psi}''(1)}.   
\]
\eq{3.20} thus implies 
\[
(\na_x \vu)^{(+)} = u^* \otimes n\, \frac{\ov{\ve} :
 \hT^{(+)}}{\hat{\psi}''(1)} + \na_x \vv. 
\]
Insertion of these equations into \eq{2.14} yields 
\[
\vT^{(+)} = D\big( \ve(u^* \otimes n) - \ov{\ve} \big) \frac{\ov{\ve}
  : \hT^{(+)}}{\hat{\psi}''(1)} + D\ve(\na_x \vv).  
\]
We take the scalar product with $\ov{\ve}$ on both sides of this
equation and note \eq{2.42} and the definition of $\vsi$ in \eq{3.22}
to obtain \eq{3.22a}. Equation \eq{3.22b} is obtained in the same
way. To prove \eq{3.22c}, we add \eq{3.22a} and \eq{3.22b} and solve the
resulting equation for $\vsi(0)$. This proves the lemma. \qed 

\paragraph{Definition of the inner expansion} 
We can now construct the inner expansion
$(u_1^{(\mu)},T_1^{(\mu)},S_1^{(\mu)})$. With the remainder terms
$\hv$, $\vv$ introduced in \eq{3.19}, \eq{3.20} we set in the
neighborhood ${\cal U}_\da$ of $\Gm$
\begin{eqnarray}
u^{(\mu)}_1(t,x) &=&  (\mu\la)^{1/2} u_0
  \big(t,\eta,\frac{\xi}{(\mu\la)^{1/2}} \big) + \mu \la^{1/2} u_1
  \big(t,\eta,\frac{\xi}{(\mu\la)^{1/2}} \big) + \mu \la u_2
  \big(t,\eta,\frac{\xi}{(\mu\la)^{1/2}} \big) \nn \\  
 && \mbox{} + \hat{v}(t,x) + \mu^{1/2}\vv(t,x),  \label{E3.23}
\\
S^{(\mu)}_1(t,x) &=& S_0\big(\frac{\xi}{(\mu\la)^{1/2}}\big) + \mu^{1/2}
  S_1\big(t,\eta,\frac{\xi}{(\mu\la)^{1/2}}\big) + \mu
  S_2\big(t,\eta,\frac{\xi}{(\mu\la)^{1/2}}\big), \label{E3.24}
\\[1ex]
T^{(\mu)}_1(t,x) &=& D\Big(\ve\big(\na_x u^{(\mu)}_1(t,x)\big) -
\ov{\ve} S^{(\mu)}_1(t,x) \Big), \label{E3.25}
\end{eqnarray}
where by assumption $S_0$, $S_1$ are known from the evolution problem
for $\Gm$, and where the functions $u_0,\ldots ,u_2$ are defined by
\begin{eqnarray}
u_0(t,\eta,\zeta) &=& u^*(t,\eta) \int_{-\infty}^\zeta
  S_0(\vat)\,d\vat, \label{E3.26}
\\ 
u_1(t,\eta,\zeta) &=& u^*(t,\eta) \int_0^\zeta S_1(t,\eta,\vat) \, d
  \vat, \label{E3.27} 
\\
u_2 (t,\eta,\zeta) &=& a^*(t,\eta)  \int_{-\infty}^\zeta
\int_{-\infty}^{\vat} S_0(\vat_1)\,d\vat_1 d\vat, \label{E3.28} 
\end{eqnarray}
The function $S_2 = S_2(t,\eta,\zeta)$ together with another unknown
function $s_1 = s_1(t,\eta)$ solve a boundary value problem. To state
this boundary value problem let $\ka(t,\eta,\xi)$ denote twice the
mean curvature of the surface $\Gm_\xi(t)$ at $\eta \in \Gm_\xi(t)$.
With the notation introduced in Section~\ref{S2.2} we thus have
$\ka(t,\eta,0) = \ka_\Gm(t,\eta)$. We write $\ka'(0) = \pa_\xi
\ka(t,\eta,0)$. 

The boundary value problem for $S_2$ and $s_1$ consists of the
ordinary differential equation
\begin{equation}\label{E3.29} 
\hat{\psi}'' \big(S_0(\zeta)\big) S_2(t,\eta,\zeta) -
  S_2''(t,\eta,\zeta) = F_2(t,\eta,\zeta), 
\end{equation}
with the right hand side given by 
\begin{align}
F_2(t,\eta,\zeta) &= 
  \vsi(0) + \ov{\ve}: [\hT]\, S_1 
  - \frac{1}{c_1} \ov{\ve}:\langle \hT
  \rangle S_1' - \frac12 \hat{\psi}'''(S_0) S_1^2  
\nn \\ 
& \mbox{  } + \la^{1/2}\Big( \hat{\si}'(0) \zeta + 
  \ov{\ve} : D\ve(a^* \otimes n + \na_\Gm u^*) \int_{-\infty}^\zeta
  S_0(\vartheta)\, d\vartheta \Big)  
\nn \\
&\mbox{  } + \Big( \frac{s_1}{c} - \la \ka'(0) \zeta
  \Big) S_0'\,,  \label{E3.30}
\end{align}
and of boundary conditions. To formulate these boundary conditions, we
choose $\vp \in C^\infty(\R,[0,1])$ such that
\begin{equation}\label{E3.31} 
\vp(\zeta) = 
\begin{cases} 0, &  \zeta \leq 1, \\
1, & \zeta \geq 2, \end{cases}
\end{equation}
set 
\begin{equation}\label{E3.32}
\vp_+(\zeta) = \frac{\vp(\zeta)}{\hat{\psi}''(1)}, \qquad \vp_-(\zeta)
= \frac{\vp(-\zeta)}{\hat{\psi}''(0)},  
\end{equation}
and define
\begin{align}
\rho_2(t,\eta,\zeta) = \vp_-(\zeta) \Biggl( 
  \ov{\ve}:\vT^{(-)} -
  \frac{\hat{\psi}'''(0)}{2} & \Big( \frac{\ov{\ve} :
    \hT^{(-)}}{\hat{\psi}''(0)} \Big)^2 
  + \la^{1/2}\, \hat{\si}'(0) \zeta \Biggr) 
\nn \\
\mbox{} + \vp_+(\zeta) \Biggl( 
  \ov{\ve}:\vT^{(+)} -
  \frac{\hat{\psi}'''(1)}{2} & \Big( \frac{\ov{\ve} :
  \hT^{(+)}}{\hat{\psi}''(1)} \Big)^2  
  + \la^{1/2} \hat{\si}'(0) \zeta 
\nn \\
\mbox{} + 
  \la^{1/2}& \ov{\ve} : D\ve (a^* \otimes n + \na_\Gm u^*) \zeta^+   
\Biggr). \label{E3.33}
\end{align}
With this function the boundary conditions are 
\begin{gather}
S_2(t,\eta,0) = 0,  \label{E3.34} 
\\
\lim_{\zeta \ra \pm\infty} \big( S_2 (t,\eta,\zeta) -
\rho_2(t,\eta,\zeta) \big) = 0. \label{E3.35} 
\end{gather}
The function $s_1 = s_1(t,\eta)$ in \eq{3.30} is independent of
$\zeta$. It is determined in Section~\ref{S4.2} by the procedure
sketched at the end of Section~\ref{S2.2}, which we apply of course to
the boundary value problem \eq{3.29}, \eq{3.30}, \eq{3.34}, \eq{3.35}
instead of the problem \eq{2.19}, \eq{2.21} -- \eq{2.24}. The function
$s_1$, whose explicit expression is given in \eq{2.32s1} -- \eq{2.34},
forms the second term in the definition \eq{2.evolution1} of the
evolution operator ${\cal K}^{(\mu)}$.

\section{The functions $\boldsymbol{S_0,\ldots,S_2}$ from the inner expansion}\label{S4}

The functions $\tilde{S}_1,\ldots,\tilde{S}_3$ in the outer expansion
can be determined explicitly from \eq{3.11} -- \eq{3.13}, whereas the
functions $S_0,\ldots,S_2$ in the inner expansion are determined as
solutions of three coupled boundary value problems to linear and
nonlinear ordinary differential equations. It is not obvious that
these solutions exist and what properties they have. We study these
solutions in this section.

\subsection{The function $\boldsymbol{S_0}$}\label{S4.1}

The first boundary value problem determining
$S_0$ is given by \eq{2.18}, \eq{2.20},

\begin{lem}\label{L4.1} 
Assume that the double well potential $\hat{\psi}$ satisfies
\eq{2.29}. Then $S_0$ is a solution of the boundary value problem
\eq{2.18}, \eq{2.20}, if and only if $S_0$ satisfies the initial value
problem 
\begin{equation}\label{E4.1}
 S'_0(\zeta)= \sqrt{2 \hat{\psi}\big(S_0(\zeta)\big)}, \quad \zeta \in
 \R, \qquad S_0(0) = \frac{1}{2}. 
\end{equation}
\end{lem}
{\bf Proof:} Let $S_0$ be a solution of \eq{2.18}, \eq{2.20}. We
multiply \eq{2.18} by $S'_0$ and obtain
\[
\frac{d}{d\zeta} \Big(\hat{\psi}(S_0)- \frac{1}{2}(S'_0)^2\Big) = 0, 
\]
or
\begin{equation}\label{E4.2}
\hat{\psi}(S_0)- \frac{1}{2}(S'_0)^2 = C_1.
\end{equation}
By \eq{2.20} we have $\lim_{\zeta \to \infty} S_0(\zeta)=1$. From
\eq{4.2} and from \eq{2.29} we thus obtain that $\lim_{\zeta \to
  \infty}\big(S'_0(\zeta)\big)^2 = -2C_1$. Using again \eq{2.20}, we
infer from this limit relation that $\lim_{\zeta \to
  \infty}S'_0(\zeta) = 0$, hence $C_1=0$. We solve \eq{4.2} for $S'_0$
and use that because of the boundary conditions \eq{2.20} the function
$S_0$ must be increasing, hence $S'_0$ must be nonnegative. This shows
that a solution of \eq{2.18}, \eq{2.20} must satisfy the initial value
problem \eq{4.1}.
 
To prove the converse we differentiate the differential equation in
\eq{4.1} and obtain \eq{2.18}. We leave it to the reader to verify
that the solution of \eq{4.1} satisfies the boundary conditions
\eq{2.20}. \qed

\begin{theo}\label{T4.2} 
Assume that $\hat{\psi} \in C^3([0,1], \R)$ has the properties
\eq{2.29}. 
Then there is a unique solution $S_0 \in C^4(\R, (0,1))$ of the
initial value problem \eq{4.1}. This solution is strictly increasing
and satisfies \eq{2.18} and \eq{2.20}. Moreover, there are
constants $K_1, \dots , K_3 >0$ such that for $a > 0$ defined in
\eq{2.29} 
\begin{align}
\label{E4.6}
0< S_0(\zeta) \leq K_1 e^{-a |\zeta|}, & \quad -\infty < \zeta \leq
  0, 
\\
\label{E4.7} 
1-K_2 e^{- a \zeta} \leq S_0 (\zeta) < 1, &  \quad 0 \leq \zeta
  < \infty, 
\\ 
\label{E4.8}
|\pa^i S_0(\zeta)| \leq K_3 e^{-a |\zeta|}, & \quad - \infty <
  \zeta < \infty,\ i=1,\ldots,4\,. 
\end{align}  
\end{theo} 
This theorem follows immediately from the standard theory of ordinary
differential equations, and we omit the proof.

\begin{lem}\label{L4.3} 
If $\hat{\psi}$ satisfies the symmetry condition \eq{2.30}, then the
solution $S_0$ of \eq{4.1} satisfies for all $\zeta \in \R$ 
\begin{gather}
 S_0(-\zeta)=1-S_0(\zeta), \qquad S'_0(\zeta)=S'_0(-\zeta),
 \label{E4.9a} 
\\
\int_{-\infty}^\zeta S_0(\vartheta)\, d \vartheta =
\int_{-\infty}^{-|\zeta|} S_0(\vartheta)\, d\vartheta +
\zeta^+, \label{E4.9b}
\\
\Big| \int_{-\infty}^\zeta S_0(\vartheta)\, d \vartheta - \zeta^+
  \Big| \leq \frac{K_1}{a} e^{-a |\zeta|}, \label{E4.9c}
\\[1ex]
| \hat{\psi}''\big(S_0(\zeta)\big) - \hat{\psi}''\big(\hS(\zeta)
\big)| \leq K_4 e^{- a |\zeta|}.
  \label{E4.9d}
\end{gather}
\end{lem} 
{\bf Proof:} If the symmetry condition \eq{2.30} holds and if $S_0$ is
a solution of the initial value problem \eq{4.1}, then also $\zeta
\mapsto \big(1-S_0(-\zeta)\big)$ is a solution. To see this, note that
\eq{2.30} and \eq{4.1} imply
\begin{align*}
\sqrt{2\hat{\psi}\big(1-S_0(-\zeta)\big)} &= \sqrt{2
  \hat{\psi}\Big( \frac{1}{2}+\Big(\frac{1}{2}-S_0(-\zeta)\Big)\Big)} 
  = \sqrt{2 \hat{\psi}\Big(
  \frac{1}{2}-\Big(\frac{1}{2}-S_0(-\zeta)\Big)\Big)} 
\\
&= \sqrt{2
  \hat{\psi}\big(S_0(-\zeta)\big)} = 
  (\pa_\zeta S_0 )(-\zeta) = \pa_\zeta \big(1-S_0(-\zeta)\big), 
\end{align*}
whence $1-S_0(-\zeta)$ satisfies the differential equation in
\eq{4.1}. Since we obviously have $1-S_0(0)=\frac{1}{2}$, we see that
$1-S_0(-\zeta)$ is a solution of \eq{4.1}. Since the solution of this
initial value problem is unique, we infer that
$S_0(\zeta)=1-S_0(-\zeta)$ holds, which implies
$S'_0(\zeta)=S'_0(-\zeta)$. 

To prove \eq{4.9b}, note that \eq{4.9a} implies for $\zeta > 0$ 
\begin{align*}
\int_{-\infty}^\zeta S_0(\vartheta)\, d \vartheta &=
\int_{-\infty}^{-\zeta} S_0(\vartheta)\, d\vartheta +
\int_{-\zeta}^\zeta S_0(\vartheta)\, d\vartheta
\\ 
&= \int_{-\infty}^{-\zeta} S_0(\vartheta)\, d\vartheta + 
\int_0^\zeta S_0(\vartheta)+ S_0(-\vartheta)\, d\vartheta\\ 
&= \int_{-\infty}^{-\zeta} S_0(\vartheta)\, d\vartheta +
\int_0^\zeta S_0(\vartheta)+ \big(1-S_0(\vartheta)\big)\,d\vartheta =
\int_{-\infty}^{-\zeta} S_0(\vartheta)\, d\vartheta + \zeta. 
\end{align*}
From this equation we immediately obtain \eq{4.9b}. The inequality
\eq{4.9c} follows from \eq{4.9b} and from \eq{4.6}, which yield 
\[
\Big| \int_{-\infty}^\zeta S_0(\vartheta)\, d \vartheta - \zeta^+
  \Big| = \Big| \int_{-\infty}^{-|\zeta|} S_0(\vartheta)\, d\vartheta
  \Big| \leq \int_{-\infty}^{-|\zeta|} K_1 e^{-a|\vartheta|}
  d\vartheta = \frac{K_1}{a} e^{-a |\zeta|}.
\] 
For the proof of \eq{4.9d} note that $\hS(\zeta) = 1$ for $\zeta >
0$. Consequently, the mean value theorem and \eq{4.7} together imply
for $\zeta > 0$ that 
\[
| \hat{\psi}''\big(S_0(\zeta)\big) - \hat{\psi}''\big(\hat{S}(\zeta)
\big)| = | \hat{\psi}''\big(S_0(\zeta)\big) - \hat{\psi}''(1)| \leq
|\hat{\psi}'''(r^*) (S_0(\zeta) - 1)| \leq CK_2 e^{- a |\zeta|},
\]
with a suitable number $r^*$ between $S_0(\zeta)$ and $1$. For $\zeta
< 0$ an analogous estimate is obtained using \eq{4.6} and noting that
$\hS(\zeta) = 0$ if $\zeta < 0$.  
\qed

\subsection{The functions $\boldsymbol{S_1}$ and $\boldsymbol{S_2}$}\label{S4.2}

The solutions $S_1$ and $S_2$ of the second and third boundary value
problems are studied in this section. The second problem determining
$S_1$ is given by the equations \eq{2.19}, \eq{2.21} -- \eq{2.24}, the
third problem, which determines $S_2$, consists of the equations
\eq{3.29}, \eq{3.30}, \eq{3.34}, \eq{3.35}. The properties of $S_1$
and $S_2$, which we need in Section~\ref{S5}, are summarized in the
next two theorems.

To state the first theorem we need the function $\rho_1:\Gm \ti \R \ra
\R$, which is defined by   
\begin{equation}\label{E4.10}
\rho_1(t,\eta,\zeta) =  
 \ov{\ve}:\hat{T}^{(-)}(t,\eta)\, \varphi_- (\zeta)  
   + \ov{\ve}:\hat{T}^{(+)}(t,\eta)\, \varphi_+ (\zeta),  
\end{equation}
with $\varphi_\pm$ introduced in \eq{3.32}.

\begin{theo}\label{T4.4} 
Assume that $\hat{\psi}$ belongs to $C^5([0,1],\R)$ and satisfies
the assumptions \eq{2.29} and the symmetry condition \eq{2.30}.
Suppose that the function $s_0 = s_0(t,\eta)$ in \eq{2.24} is given
by \eq{2.32}. Let $S_0$ be the solution of the boundary value
problem \eq{2.18}, \eq{2.20}, which exists by \reft{4.2}.

Then for every $(t, \eta) \in \Gamma$ there is a unique solution
$\zeta \mapsto S_1 (t, \eta, \zeta):\R \ra \R$ of the boundary value
problem \eq{2.19}, \eq{2.21} -- \eq{2.24}.  The function $S_1$ belongs
to the space $C^2( \R, C^2(\Gm,\R))$. Moreover, there are constants
$K_1,\ldots, K_3$ such that for the constant $a$ defined in \eq{2.29}
and for all $(t,\eta,\zeta) \in \Gm\ti \R$ the estimates
\begin{eqnarray}
\label{E4.11}
\|D^{\alpha}_{(t, \eta)} S_1\|_{L^{\infty}(\Gamma
  \times \R)} &\leq& K_1, \qquad |\alpha| \leq 2, 
\\
\label{E4.12}
\big| \pa^j_\zeta D^\al_{(t,\eta)} \big(S_1 (t, \eta, \zeta) - 
  \rho_1(t,\eta,\zeta) \big) \big| &\leq& K_2\, e^{-a |\zeta|}, \quad
  0 \leq j,|\al| \leq 2,  
\\
\label{E4.14}
\big|  \pa^j_\zeta D^\al_{(t,\eta)} S_1(t,\eta,\zeta) \big| &\leq&
  K_3\, e^{-a |\zeta|}, \quad |\al| \leq 2,\ j=1,2,    
\end{eqnarray}
hold.

\end{theo}
We do not give the proof of this theorem, since it is almost the same
as the proof of Theorem~1.2 in \cite{CMT2011}. Morover, it is obtained
from the proof of the following theorem by simplification. The main
difference between the two proofs is that the right hand side $F_1$ of
the differential equation \eq{2.19} for $S_1$ is bounded, whereas the
right hand side $F_2$ of the differential equation \eq{3.29} for $S_2$
grows linearly for $\zeta \ra \pm \infty$.

\begin{theo}\label{T4.5} 
Assume that $\hat{\psi}$ satisfies the assumptions given in the
\reft{4.4}. Let $S_0$ be the solution of the boundary value problem
\eq{2.18}, \eq{2.20}, and let $S_1$ be the solution of the boundary
value problem \eq{2.19}, \eq{2.21} -- \eq{2.24}. Suppose that the
function $s_1 = s_1(t,\eta)$ in \eq{3.30} satisfies \eq{2.32s1} with
$s_{10}$, $s_{11}$ given in \eq{2.33}, \eq{2.34}.

(i) Then for every $(t, \eta) \in \Gamma$ there is a unique solution
$\zeta \mapsto S_2 (t, \eta, \zeta):\R \ra \R$ of the boundary value
problem \eq{3.29}, \eq{3.30}, \eq{3.34}, \eq{3.35}. The function $S_2$
belongs to $C^2( \R, C^2(\Gm,\R))$, and there are constants
$K_4,\ldots, K_6$ such that for the constant $a$ defined in \eq{2.29}
and for all $(t,\eta,\zeta) \in \Gm\ti \R$ the estimates
\begin{eqnarray}
\label{E4.11a}
|\pa^j_\zeta D^{\alpha}_{(t, \eta, \zeta)} S_2(t,\eta,\zeta)| &\leq&
K_4 (1+|\zeta|)^{1-j}, \quad |\alpha| \leq 2,\ j=0,1, 
\\
\label{E4.12a}
\big| \pa_\zeta^j D^{\alpha}_{(t, \eta)} \big( S_2 (t, \eta, \zeta) -
  \rho_2(t,\eta,\zeta) \big) \big| &\leq& K_5 (1 + |\zeta|)\, e^{-a 
    |\zeta|},  \quad 0 \leq j,|\al| \leq 2,
\\ 
\label{E4.13a}
\big| \pa^2_\zeta D^\al_{(t,\eta)} S_2(t,\eta,\zeta) \big| &\leq& K_6
 (1 + |\zeta|)\, e^{-a |\zeta|},\qquad |\al| \leq 2, 
\end{eqnarray}
hold, where $\rho_2$ is defined in \eq{3.33}.

(ii) $S_2$ is the only solution of the differential equation
\eq{3.29} with $F_2$ given by \eq{3.30}, which satisfies \eq{3.34} and
for which constants $C,\theta > 0$ exist such that 
\begin{equation}\label{E4.14a}
| S_2(t,\eta,\zeta)| \leq C e^{ (a - \theta) |\zeta| }, \qquad \zeta
\in \R, 
\end{equation}
holds.

\end{theo}

\subsection{Proof of Theorem~4.5}\label{S4.3}

In this section we give the proof of \reft{4.5}, which is divided into
five parts:

\paragraph{(I) Reduction of the boundary value problem for
  $\boldsymbol{S_2}$ to a problem in $\boldsymbol{L^2}$.}

With $\rho_2$ defined in \eq{3.33} we make the ansatz 
\begin{equation}\label{E4.16}
S_2(t,\eta,\zeta) = w(t,\eta,\zeta) + \rho_2(t,\eta,\zeta).
\end{equation}
Insertion of this ansatz into the equations \eq{3.29} and \eq{3.34},
\eq{3.35} shows that $S_2$ is a solution of the problem given by these
equations if and only if $w$ solves the equations 
\begin{gather}
\hat{\psi}''\big( S_0(\zeta)\big) w(t,\eta,\zeta) - \pa^2_\zeta
w(t,\eta,\zeta) = F_2(t,\eta,\zeta) + F_3(t,\eta,\zeta), 
\label{E4.17} \\
w(t,\eta,0) = 0, 
\label{E4.18} \\
\lim_{\zeta \ra \pm \infty} w(t,\eta,\zeta) = 0, \label{E4.19}
\end{gather}
where $F_2$ is given by \eq{3.30} and where 
\begin{equation}\label{E4.20}
F_3 = - (\hat{\psi}''(S_0) - \pa^2_\zeta ) \rho_2.
\end{equation}
To get \eq{4.18} we used that $\vp_+(0) = \vp_-(0) = 0$, which by
\eq{3.33} implies $\rho_2(t,\eta,0) = 0$. To show that the solution
$S_2$ of the problem \eq{3.29} and \eq{3.34}, \eq{3.35} exists, it
therefore suffices to prove that the reduced problem \eq{4.17} --
\eq{4.20} has a solution.  

\paragraph{(II) Spectral theory} For this proof note that
$\hat{\psi}''(S_0)-\pa^2_\zeta$ is a linear self-adjoint differential
operator in $L^2(\R)$. From the spectral theory of such operators we
know that the continuous spectrum of $\hat{\psi}''(S_0)-\pa^2_\zeta$
is contained in the interval $[a_0,\infty)$, where
\[
a_0= \min\Big\{\lim\limits_{\zeta \to
  -\infty}\hat{\psi}''\big(S_0(\zeta)\big), \lim\limits_{\zeta \to
  \infty} \hat{\psi}''\big(S_0(\zeta)\big)\Big\}, 
\]
and that the part of the spectrum in $(-\infty,a_0)$ is a pure point
spectrum. \eq{4.9d} yields $\lim\limits_{\zeta \to -\infty}
\hat{\psi}'' \big(S_0(\zeta)\big) = \hat{\psi}''(0)$,
$\lim\limits_{\zeta \to \infty} \hat{\psi}''\big(S_0(\zeta)\big)=
\hat{\psi}''(1)$, hence the assumption \eq{2.29} implies $a_0=a^2 >0$.
Therefore $0$ does not belong to the continuous spectrum. From the
spectral theory we also know that for every $\om \in \C$, which is not
in the continuous spectrum, the differential equation
$\big(\hat{\psi}''(S_0)-\pa^2_\zeta\big) w -\om w=f$ has a solution $w
\in L^2(\R)$, if and only if $f \in L^2(\R)$ is orthogonal to the
kernel of the operator $\psi''(S_0)- \pa^2_\zeta -\om$. This implies
in particular, that for every $(t,\eta)\in\Gamma$ the differential
equation \eq{4.17} has a solution $w(t,\eta,\cdot) \in L^2(\R)$, if
the right hand side $\zeta \mapsto f(\zeta)=F_2 (t,\eta,\zeta)+
F_3(t,\eta,\zeta)$ belongs to $L^2(\R)$ and is orthogonal to the
kernel of the operator $\hat{\psi}''(S_0)-\pa^2_\zeta$. To show that
the problem \eq{4.17} -- \eq{4.19} has a solution, we therefore verify
in the next two parts of the proof that $F_2+F_3$ satisfies these two
conditions.

\paragraph{(III) The asymptotic behavior of $\boldsymbol{F_2 +
    F_3}$ at infinity.}

We first show that the right hand side $F_2 + F_3$ of \eq{4.17} decays
exponentially at $\pm \infty$, which implies that $F_2 + F_3 \in
L^2(\R)$.
To simplify the notation we define 
\begin{equation}
\ov{\vp}_+(\zeta) = \vp_+(\zeta) \hat{\psi}''\big (S_0(\zeta)\big),
\qquad \ov{\vp}_-(\zeta) = \vp_-(\zeta) \hat{\psi}''\big
(S_0(\zeta)\big), \label{E4.21}  
\end{equation}
with $\vp_+$, $\vp_-$ given in \eq{3.32}. For these functions we
obtain from \eq{4.9d} that
\begin{align}
&|\ov{\vp}_- - \hat{\psi}''(0) \vp_- | = \vp_-(\zeta)\,
 |\hat{\psi}''(S_0) - \hat{\psi}''(0) | \leq C K_4 e^{-a|\zeta|},   
\label{E4.22} \\
&|\ov{\vp}_+ - \hat{\psi}''(1) \vp_+|  = \vp_+(\zeta)\, 
 |\hat{\psi}''(S_0) - \hat{\psi}''(1)| \leq C K_4 e^{-a|\zeta|}, 
\label{E4.23}
\end{align}
for all $\zeta \in \R$. Since $ \hat{\psi}''(0) \vp_-(\zeta) = 1$ for
$\zeta \leq -2$ and $\hat{\psi}''(1) \vp_+(\zeta) = 1$ for $\zeta \geq
2$, these estimates imply   
\begin{align}
&|1 - \ov{\vp}_-(\zeta) | \leq C K_4 e^{-a |\zeta|}, & -\infty <
\zeta \leq 0. \label{E4.24} 
\\
&|1 - \ov{\vp}_+(\zeta) | \leq C K_4 e^{-a |\zeta|}, & 0 \leq \zeta
< \infty, \label{E4.25} 
\\
&|1 - \ov{\vp}_-(\zeta) - \ov{\vp}_+(\zeta)| \leq C K_4 e^{-a |\zeta|},
  & \zeta \in \R. \label{E4.26}
\end{align}
To get the last estimate we combined the first two estimates and noted
that $\ov{\vp}_-(\zeta) = 0$ for $\zeta \geq -1$ and
$\ov{\vp}_+(\zeta) = 0$ for $\zeta \leq 1$.

Note that by \eq{3.30}, \eq{3.33} and \eq{4.20} the function $F_2 +
F_3$ can be decomposed in the form 
\begin{equation}\label{E4.27}
F_2 + F_3 = F_2 - \hat{\psi}''(S_0) \rho_2 + \pa^2_\zeta \rho_2 =
  \sum_{j=1}^5 I_j , 
\end{equation}
where 
\begin{eqnarray}
I_1 &=& \vsi(0) + \ov{\ve}: [\hT]\,S_1 - \ov{\ve}: \vT^{(-)}\,
  \ov{\vp}_- - \ov{\ve}: \vT^{(+)}\, \ov{\vp}_+ \,, \label{E4.28}
\\[1ex]
I_2 &=& -\frac{\hat{\psi}'''(S_0)}{2} S_1^2 + \frac{\hat{\psi}'''(0)}{2} 
\Big( 
 \frac{ \ov{\ve}: \hT^{(-)}} {\hat{\psi}''(0)} \Big)^2\, \ov{\vp}_- 
 +  \frac{\hat{\psi}'''(1)}{2} 
\Big( 
 \frac{ \ov{\ve}: \hT^{(+)}} {\hat{\psi}''(1)} \Big)^2\, \ov{\vp}_+ \,, 
\label{E4.29} \\[1ex]
I_3 &=& \la^{1/2}\, \hat{\si}'(0) \zeta\, \big(1-\ov{\vp}_- -
  \ov{\vp}_+ \big),   
\label{E4.30} \\[1ex]
I_4 &=& \la^{1/2}\,\ov{\ve}: D \ve(a^* \otimes n + \na_x u^*) \Big(
  \int_{-\infty}^\zeta S_0(\vta) \, d\vta - \zeta^+ \ov{\vp}_+ \Big),  
\label{E4.31} \\[1ex]
I_5 &=& - \frac{1}{c_1}\, \ov{\ve} : \langle \hT \rangle S_1' + \Big(
\frac{s_1} {c} - \la \ka'(0) \zeta \Big) S_0' +
\pa^2_\zeta \rho_2\,. \label{E4.32}
\end{eqnarray}
We show that everyone of these terms decays to zero for $\zeta \ra \pm
\infty$. To verify this for the first term we insert \eq{3.22a} and
\eq{3.22b} into \eq{4.28}, which results in    
\[
I_1 = \vsi(0) + \ov{\ve}:[\hT]\, S_1 - \vsi(0) (
  \ov{\vp}_+ + \ov{\vp}_- ) - \ov{\ve}:[\hT]\, \Big(
  \frac{\ov{\ve}:\hT^{(-)} } {\hat{\psi}''(0)} \ov{\vp}_- +
  \frac{\ov{\ve}:\hT^{(+)} } {\hat{\psi}''(1)} \ov{\vp}_+ \Big). 
\]
We introduce the terms $\hat{\psi}''(0) \vp_-$ and 
$\hat{\psi}''(1) \vp_+$ into this equation. Noting the definition of
$\rho_1$ in \eq{4.10}, this leads to   
\begin{align}
|I_1| & \leq \big| \vsi(0)\, (1 - \ov{\vp}_+ + \ov{\vp}_- ) +
  \ov{\ve}:[\hT]\, \big( S_1 - \rho_1 \big) \big|
\nn \\
& \mbox{} + \Big| \ov{\ve}:[\hT]\, \Big(
  \frac{\ov{\ve}:\hT^{(-)} } {\hat{\psi}''(0)} (\ov{\vp}_- -
  \hat{\psi}''(0) \vp_- ) +  
  \frac{\ov{\ve}:\hT^{(+)} } {\hat{\psi}''(1)} (\ov{\vp}_+ - 
  \hat{\psi}''(1) \vp_+ ) \Big) \Big| 
  \leq C e^{-a |\zeta|}, \label{E4.33}
\end{align}
for all $\zeta \in \R$, where we applied the estimates \eq{4.12}, 
\eq{4.22}, \eq{4.23} and \eq{4.26}.   

Next we estimate $I_2$. By definition we have $\ov{\vp}_+(\zeta) = 0$ 
for $\zeta \leq 1$. From \eq{4.29} we thus have on the half axis
$-\infty < \zeta \leq 0$ that  
\begin{align}
I_2 &= -\frac{\hat{\psi}'''(S_0)}{2} S_1^2 + \frac{\hat{\psi}'''(0)}{2} 
\Big( 
 \frac{ \ov{\ve}: \hT^{(-)}} {\hat{\psi}''(0)} \Big)^2 \ov{\vp}_- 
\nn \\
&=
 \frac{\hat{\psi}'''(0) - \hat{\psi}'''(S_0)}{2} S_1^2 -
 \frac{\hat{\psi}'''(0)}{2} S_1^2 (1 - \ov{\vp}_-) -
 \frac{\hat{\psi}'''(0)}{2} \Big( S_1^2 - \Big( 
 \frac{ \ov{\ve}: \hT^{(-)}} {\hat{\psi}''(0)} \Big)^2 \Big)
 \ov{\vp}_-
\nn \\
& = I_{21} + I_{22} + I_{23}. \label{E4.34}
\end{align}
To estimate $I_{21}$ we apply the mean value theorem to
$\hat{\psi}'''$ and use \eq{4.6} and \eq{4.11}, to estimate $I_{22}$
we use \eq{4.24} and \eq{4.11}. The result is 
\begin{equation}\label{E4.35}
|I_{21} + I_{22}| \leq C K_1 e^{-a|\zeta|}, \quad -\infty <
  \zeta \leq 0. 
\end{equation}
To estimate $I_{23}$ note that by \eq{3.31}, \eq{3.32} and \eq{4.10}
we have for $-\infty < \zeta \leq -2$ that 
\[
\rho_1(t,\eta,\zeta) = \vp_-(\zeta) \big( 
 \ov{\ve}:\hat{T}^{(-)}(t,\eta) \big) = \frac{
  \ov{\ve}: \hT^{(-)}(t,\eta) } {\hat{\psi}''(0)}.
\]  
With this equation we infer from \eq{4.11} and \eq{4.12} with
$\al,j=0$ that 
\begin{equation}\label{E4.36}
|I_{23}| =  \Big| \frac{\hat{\psi}'''(0)}{2} \Big( S_1- \frac{
  \ov{\ve}: \hT^{(-)}} {\hat{\psi}''(0)} \Big) \Big( S_1 + \frac{
  \ov{\ve}: \hT^{(-)}} {\hat{\psi}''(0)} \Big)
  \ov{\vp}_- \Big| \leq C e^{-a|\zeta|}, \quad -\infty < \zeta \leq 0. 
\end{equation}
\eq{4.34} -- \eq{4.36} together imply that $|I_2(\zeta)| \leq C
e^{-a|\zeta|}$ for $-\infty < \zeta \leq 0$. On the half axis $0 \leq
\zeta <\infty$ we estimate $I_2$ analogously. This proves that  
\begin{equation}\label{E4.37}
|I_2| \leq C e^{-a|\zeta|}, \qquad -\infty < \zeta < \infty.
\end{equation}
The estimate for $I_3$ is obtained by application of \eq{4.26} to
\eq{4.30}, which immediately yields   
\begin{equation}\label{E4.38}
| I_3 | \leq C (1+|\zeta|) e^{-a|\zeta|}, \qquad -\infty < \zeta <
  \infty. 
\end{equation}
To study the asymptotic behavior of $I_4$ note that \eq{4.9c} 
and \eq{4.25} together imply 
\begin{multline*}
\Big|\int_{-\infty}^\zeta S_0(\vartheta)\, d\vartheta -
\zeta^+ \ov{\varphi}_+(\zeta)\Big| \leq
\Big|\int_{-\infty}^\zeta S_0(\vartheta)\, d\vartheta - \zeta^+ \Big|   
 + \zeta^+ |1 - \ov{\varphi}_+(\zeta) | \\ 
 \leq \Big(\frac{1}{a} K_1+ \zeta^+ C K_4\Big) e^{-a|\zeta|}. 
\end{multline*}
Insertion of this inequality into \eq{4.31} results in 
\begin{equation}\label{E4.39}
| I_4 | \leq C (1+|\zeta|)\, e^{-a|\zeta|}. 
\end{equation}
It remains to investigate $I_5$. Note first that the third term on
the right hand side of \eq{4.32} satisfies 
\begin{equation}\label{E4.39a}
\pa^2_\zeta \rho (t,\eta,\zeta) = 0, \qquad \mbox{for } |\zeta|
\geq 2. 
\end{equation}
To show this it suffices to remark that the functions $\varphi_\pm$
are constant on the intervals $(-\infty,-2)$ and $(2,\infty)$, from
which we see by inspection of \eq{3.33} that on these intervals the
function $\zeta \mapsto \rho_2(t,\eta,\zeta)$ is a sum of constant and
linear terms, whence \eq{4.39a} follows.
If we estimate the first term on the right hand side of \eq{4.32} by
employing \eq{4.14} with $\al = 0$, $j=1$ and the second term by using
\eq{4.8}, we obtain together with \eq{4.39a} that  
\begin{equation}\label{E4.40}
| I_5 | \leq C(1 + |\zeta|) e^{-a|\zeta|}. 
\end{equation}
We combine \eq{4.27}, \eq{4.33}, \eq{4.37} -- \eq{4.39} and \eq{4.40}
to derive the estimate    
\begin{equation}\label{E4.41}
| F_2 (t,\eta,\zeta) + F_3(t,\eta,\zeta) | \leq C (1+|\zeta|)
e^{-a|\zeta|}, \qquad \mbox{for all } \zeta \in \R, 
\end{equation}
which shows in particular that the right hand side of \eq{4.17}
belongs to $L^2(\R)$.

\paragraph{(IV) The orthogonality condition determining
  $\boldsymbol{s_1}$.}

Next we must show that the right hand side of \eq{4.17} is orthogonal
to the kernel of $\hat{\psi}''(S_0)-\pa^2_\zeta$. 
This kernel is different from $\{0\}$, since $S'_0$ belongs to the
kernel. This is immediately seen by differentiation of \eq{2.18},
which yields 
\begin{equation}\label{E4.42}
\hat{\psi}'' \big(S_0(\zeta)\big) S'_0(\vartheta)- \pa^2_\zeta
S'_0(\zeta)=0.  
\end{equation}
Since by \eq{4.8} the function $S'_0$ is in the domain of definition
of $\hat{\psi}''(S_0)-\pa^2_\zeta$, it belongs to the kernel of this
operator. 

The theory of linear ordinary differential equations of second order
implies now that the kernel is one-dimensional, hence every function
from the kernel is a multiple of $S'_0$. Therefore the right hand side
$F_2+F_3$ of \eq{4.17} is orthogonal to the kernel if it is orthogonal
to $S'_0$. Note that the integrals $\int_{-\infty}^\infty F_2 S'_0
\,d\zeta$ and $\int_{-\infty}^\infty F_3 S'_0\,d\zeta$ 
both exist, since $F_2$ and $F_3$ grow at most linearly for $\zeta \to
\pm \infty$, whereas by \eq{4.8} the function $S'_0$ decays
exponentially at $\pm \infty$. Therefore we obtain from \eq{4.20} and
\eq{4.42} by partial integration that 
\begin{multline}\label{E4.43}
\int_{-\infty}^\infty (F_2+F_3)  S'_0 \, d\zeta =
  \int_{-\infty}^\infty F_2 S'_0 \, d\zeta -
  \int_{-\infty}^\infty \Big( \big(\hat{\psi}''(S_0)-\pa^2_\zeta\big)
  \rho_2 \Big) S'_0\, d\zeta
\\ 
= \int_{-\infty}^\infty F_2 S'_0 \, d\zeta-
  \int_{-\infty}^\infty \rho_2
  \big(\hat{\psi}''(S_0)-\pa^2_\zeta\big) S'_0 \, d\zeta =
  \int_{-\infty}^\infty F_2 S'_0 \, d\zeta. 
\end{multline}
To study the last integral on the right hand side note that by
\eq{4.9a} the function $S'_0$ is even, which implies that 
\begin{equation}\label{E4.44}
\int_{-\infty}^\infty \hat{\sigma}' (t,\eta,0) \zeta\,
  S'_0(\zeta)\,d\zeta=0, \quad \mbox{and} \quad  \int_{-\infty}^\infty
  \ka'(t,\eta,0) \zeta\, \big(S'_0(\zeta) \big)^2 d\zeta = 0. 
\end{equation}
Moreover, since by Lemma \ref{L4.1} the function $S'_0$ satisfies
\eq{4.1}, we obtain by substitution of $\vartheta=S_0(\zeta)$ 
\begin{equation}\label{E4.45}
\int_{-\infty}^\infty S'_0(\zeta) S'_0  (\zeta)\, d\zeta=
  \int_{-\infty}^\infty \sqrt{2 \hat{\psi}\big(S_0(\zeta)\big)}\,
  S'_0(\zeta)\, d\zeta
  = \int_0^1 \sqrt{2 \hat{\psi}(\vartheta)} \,  d\vartheta=c_1, 
\end{equation}
where the last equality sign holds by definition of $c_1$ in
\eq{2.28}. Finally, by partial integration,   
\begin{multline}\label{E4.46}
\int_{-\infty}^\infty \int_{-\infty}^\zeta S_0 
  (\vartheta) \, d\vartheta\, S'_0 (\zeta)\,d\zeta=
  \lim_{\zeta_1 \to\infty} \Big(\int_{-\infty}^{\zeta_1}
  S_0(\vartheta)\, d\vartheta\, S_0(\zeta_1)-
  \int_{-\infty}^{\zeta_1}S_0 (\zeta)^2\,\zeta\Big)
\\ 
= \lim_{\zeta_1 \to\infty} \int_{-\infty}^{\zeta_1}
  S_0(\zeta) \big(S_0(\zeta_1)-S_0(\zeta)\big)\, d\zeta
\\
= \int_{-\infty}^\infty S_0(\zeta) \big(1-S_0(\zeta)\big)\, d\zeta 
  = \int_{-\infty}^\infty S_0(\zeta) S_0(-\zeta)\, d\zeta. 
\end{multline}
In the second last step we used that $S_0$ is increasing.
The equality sign thus follows from the theorem of Beppo Levi. The
last equality sign is obtained from \eq{4.9a}. 

The equations \eq{3.30} and \eq{4.44} -- \eq{4.46} yield
\begin{multline*}
\int_{-\infty}^\infty F_2  (t,\eta,\zeta) S'_0(\zeta)\,  d\zeta
  = \int_{-\infty}^\infty\big(\check{\sigma}(0)+
  \ov\ve:[\hat{T}]S_1 \big) S'_0\, d\zeta - \frac{1}{c_1}
  \ov\ve:\langle \hat{T} \rangle \int_{-\infty}^\infty S'_1 S'_0
  \, d\zeta
\\ 
\mbox{}- \frac{1}{2} \int_{-\infty}^\infty \hat{\psi}'''
  (S_0) S_1^2 S'_0 \, d\zeta 
  \mbox{}+ \lambda^{1/2}\, \ov\ve: D\ve (a^* \otimes n+ \na_\Gamma u^*) 
  \int_{-\infty}^\infty \int_{-\infty}^\zeta S_0 
  (\vartheta) \, d\vartheta\  S_0'(\zeta) \, d\zeta  
\\
\mbox{}+ \int_{-\infty}^\infty \Big(\la^{1/2} \hat{\si}'(0)\zeta +
  \big( \frac{s_1}{c} - \la \ka'(0)\zeta\big) S'_0 \Big)
  S'_0(\zeta) d\zeta  
\\ 
= \vsi(0) + \ov\ve:[\hT] \int_{-\infty}^\infty S_1 S'_0\, d\zeta -
  \frac{1}{c_1} \ov\ve:\langle \hat{T} \rangle \int_{-\infty}^\infty
  S'_1 S'_0 \, d\zeta - \frac{1}{2} \int_{-\infty}^\infty \hat{\psi}'''
  (S_0) S_1^2 S'_0 \, d\zeta \\
\mbox{} + \la^{1/2}\, \ov\ve: D\ve (a^* \otimes n+ \na_\Gamma u^*) 
  \int_{-\infty}^\infty S_0(\zeta) S_0(-\zeta)\, d\zeta  
  + c_1 \frac{s_1}{c}\,. 
\\
= - \frac{c_1}{c} s_{10} - \la^{1/2} \frac{c_1}{c} s_{11} + \frac{c_1} {c}
  s_1 = \frac{c_1}{c} (-s_{10} - \la^{1/2} s_{11} + s_1) .  
\end{multline*}
To get the second last equality sign we inserted \eq{3.22c} for
$\vsi(0)$ and used \eq{2.33}, \eq{2.34}.  The right hand side of this
equation vanishes if and only if $s_1$ satisfies \eq{2.32s1}. From
\eq{4.43} we thus infer that $F_2+F_3$ is orthogonal to the kernel of
$\hat{\psi}''(S_0)-\pa^2_\zeta$  if and only if \eq{2.32s1} holds.

Consequently, from part (II) of the proof we conclude that the
differential equation \eq{4.17} has a solution $w$ in $L^2(\R)$ if and
only if $s_1$ satisfies \eq{2.32s1} with $s_{10}$ and $s_{11}$ given
in \eq{2.33}, \eq{2.34}. In fact, there is exactly one such $w$, which
also satisfies \eq{4.18}. To prove this assume that $\tilde{w} \in
L^2(\R)$ is a special solution of \eq{4.17}. Then we obtain every
solution contained in $L^2(\R)$ in the form $w= \tilde{w}+\beta S'_0$
with an arbitrary constant $\beta \in \R$. Since \eq{4.1} and
\eq{2.29} yield
\[
S'_0(0)= \sqrt{2 \hat{\psi}\big(S_0(0)\big)}= \sqrt{2
  \hat{\psi}\Big(\frac{1}{2}\Big)} >0, 
\]
we can choose $\beta$ such that
\[
w(t,\eta,0)= \tilde{w}(t,\eta,0)+ \beta S'_0(0)=0,
\]
which is \eq{4.18}. This equation determines $\beta$ uniquely, hence
$w$ is the unique solution of \eq{4.17} and \eq{4.18} in $L^2(\R)$.

\paragraph{(V) Existence of the solution, estimates (4.18) -- (4.20).} 
We show next that this function $w$ satisfies \eq{4.19}. To this end
we need the following   
\begin{lem}\label{L4.7}
Let $\hat{a}_- >0$, $\hat{a}_+ > 0$ and set $\hat{a} =
\sqrt{\min\{\hat{a}_-, \hat{a}_+ \}}$. Let $g:\R \ra \R$ be a smooth function
and let $f:\R \ra \R$ be a continuous function such that    
\begin{eqnarray}
| g(\zeta) - \hat{a}_-| &\leq& C e^{- \hat{a} |\zeta|}, \hspace{11ex}
  \mbox{for } \zeta < 0, \label{E4.54}
\\
| g(\zeta) - \hat{a}_+| &\leq& C e^{- \hat{a} |\zeta|}, \hspace{11ex}
  \mbox{for } \zeta > 0, \label{E4.55}
\\
| f (\zeta) | &\leq& C(1+|\zeta|)  e^{- \hat{a} |\zeta|}, \quad
  \mbox{for } \zeta \in \R. \label{E4.56}
\end{eqnarray} 
Let $\hat{w}$ be a solution of  
\begin{equation}\label{E4.57}
g(\zeta) \hat{w}(\zeta) - \pa^2 \hat{w}(\zeta) = f(\zeta), \qquad \zeta
\in \R. 
\end{equation}
(i) Then $\hat{w}$ belongs to the space $C^2(\R)$. If $\hat{w} \in
L^2(\R)$, then there is $C > 0$ such that   
\begin{equation}\label{E4.58}
|\pa^j_\zeta \hat{w}(\zeta)| \leq C(1 + |\zeta|) e^{-\hat{a} |\zeta|},
\qquad  \mbox{for all }\zeta \in \R, \mbox{ for } j=0,1,2. 
\end{equation}
(ii) If there are $C,\theta > 0$ such that 
\begin{equation}\label{E4.59}
|\hat{w}(\zeta)| \leq C e^{ (\hat{a} - \theta) |\zeta|}
\end{equation}
holds for all $\zeta \in \R$, then $\hat{w} \in L^2(\R)$.  
\end{lem}  
This is a standard result from the theory of ordinary differential
equations, and we omit the proof. 

To show that $w$ satisfies \eq{4.19}, we apply this lemma with
$\hat{a}_- = \hat{\psi}''(0)$, $\hat{a}_+ = \hat{\psi}''(1)$,
$g(\zeta) = \hat{\psi}''\big(S_0(\zeta)\big)$ and $f(\zeta) =
F_2(t,\eta,\zeta) + F_3(t,\eta,\zeta)$. Then we have $\hat{a} =
\sqrt{\min\{\hat{\psi}''(0),\hat{\psi}''(1) \}} = a$, by \eq{2.29},
and from \eq{4.9d} and \eq{4.41} we see that \eq{4.54} -- \eq{4.56}
hold for this choice of functions and constants. Moreover, with this
choice of functions the differential equation \eq{4.57} is equal to
\eq{4.17}. Since $w \in L^2(\R)$ is a solution of \eq{4.17}, we see
that all assumptions for part (i) of \refl{4.7} are satisfied, hence
\eq{4.58} holds for $w$, which means that
\begin{equation}\label{E4.60}
| \pa^j_\zeta w(\zeta)| \leq C(1+ |\zeta|) e^{- a |\zeta| }, \qquad
\zeta \in \R,\ j=0,1,2, 
\end{equation}
and this in particular implies that $w$ satisfies \eq{4.19}. 

We have now found a unique solution $w \in L^2(\R)$ of \eq{4.17} --
\eq{4.19}. By part (I) of this proof this means that $S_2$ given by
\eq{4.16} is a solution of the boundary value problem \eq{3.29},
\eq{3.30}, \eq{3.34} and \eq{3.35}. Since by \eq{4.16} we have $w =
S_2 - \rho_2$, the inequality \eq{4.60} shows that $S_2$ satisfies
\eq{4.12a} for $\al = 0$.   

To verify that $S_2$ satisfies \eq{4.12a} for $\al \neq 0$, it must
first be shown that $S_2$ is two times continuously differentiable
with respect to $(t,\eta)$. This follows if we can show that 
$w = S_2 - \rho_2$ is two times continuoulsly differentiable with
respect to $(t,\eta)$, since by our regularity assumptions the
function $\rho_2$ has this differentiability property. To prove this
differentiability of $w$, we write \eq{4.17}, \eq{4.18} as a
perturbation problem for the linear equation 
\[
A w = f(t,\eta)
\]
in $L^2(\R)$, where $A = (\hat{\psi}''(S_0) - \pa^2_\zeta)$ is the
linear differential operator on the left hand side of \eq{4.17} and
$f(t,\eta) = F_2(t,\eta,\cdot) + F_3(t,\eta,\cdot) \in L^2(\R)$ is the
function on the right hand side of \eq{4.17}, which depends two times
continuously differentiable on $(t,\eta)$ and satisfies the estimate
\eq{4.41} for every $(t,\eta) \in \Gm$. Since $0$ is an eigenvalue of
$A$ and since $f(t,\eta)$ is orthogonal to the kernel of $A$ for every
$(t,\eta)$, this linear equation has infinitely many solutions and the
solution set is affine. The condition \eq{4.18} defines a linear
subspace, which is closed in the Sobolev space $H^1(\R)$, and which
intersects the solution set in exactly one point $w$, which is the
solution of \eq{4.17}, \eq{4.18}. 

To the problem set in this way we can apply the pertubation theory of
linear operators. The theory yields that $w$ is two times continuously
differentiable with respect to $(t,\eta)$. We avoid the details but
refer to standard texts on the pertubation theory of linear operators,
for example \cite{Kato}.

With this knowledge we can derive the estimate \eq{4.12a} for $\al
\neq 0$ by applying the differential operator $D^\al_{(t,\eta)}$ with
$1 \leq |\al| \leq 2$ to the differential equation \eq{4.17} and
obtain
\begin{equation}\label{E4.61}
\hat{\psi}'' (S_0) (D^\al_{(t,\eta)} w) - \pa^2_\zeta
  (D^\al_{(t,\eta)} w) = D^\al_{(t,\eta)} (F_2 + F_3).
\end{equation}
This is a differential equation for the function $D^\al_{(t,\eta)}
w$ with right hand side satisfying the estimate    
\begin{equation}\label{E4.62}
|D^\al_{(t,\eta)} (F_2 + F_3) | \leq C (1 + |\zeta|) e^{-a |\zeta|}.
\end{equation}
The proof of this estimate proceeds in the same way as the proof of
the corresponding estimate for $\al = 0$, which we gave in part
(III). Essentially one has to replace the terms appearing in $F_2$
and $F_3$, which depend on $(t,\eta)$, by their derivatives. To avoid
repetition of many technical details, we omit this proof. 

The differential equation \eq{4.61} has the same form as the
differential equation \eq{4.17}. From \eq{4.62} we see that the
assumption \eq{4.56} holds, hence we can apply \refl{4.7} (i) to this
differential equation, from which we see that $w$ belongs to the space
$C^2(\R,C^2(\Gm,\R))$ and that the inequality \eq{4.58} holds
with $\hat{w}$ replaced by $D^\al_{(t,\eta)}w$, whence we have
\[
| \pa^j_\zeta D^\al_{(t,\eta)}w(\zeta)| \leq C(1+ |\zeta|) e^{- a
  |\zeta| }, \qquad \zeta \in \R,\ 0 \leq j \leq 2,\ 1 \leq |\al| \leq
  2. 
\] 
Since $w = S_2 - \rho_2$, this is inequality \eq{4.12a} with $\al \neq
0$. Therefore we proved that \eq{4.12a} holds for all $0\leq |\al|
\leq 2$. 

The inequality \eq{4.13a} is a consequence of \eq{4.12a} and of
\eq{4.39a}, the inequality \eq{4.11a} follows by combination of
\eq{4.12a} with the estimate
\[
| \pa^j_\zeta D^\al_{(t,\eta)} \rho_2(t,\eta,\zeta) | \leq C(1 +
|\zeta|)^{1-j}, \qquad j = 0,1,
\] 
which is seen to hold by inspection of \eq{3.33}.  

We have now proved statement (i) of \reft{4.5}, and it remains to
verify (ii). That is, we have to show that $S_2$ is the only solution
of \eq{3.29}, \eq{3.30}, \eq{3.34} satisfying \eq{4.14a}. Indeed, from
\eq{4.11a} it follows that $S_2$ satisfies \eq{4.14a}. Assume that
$S_2^*$ is a second solution satisfying \eq{4.14a}. Then 
$\hat{w} = S_2 - S_2^*$ fulfills \eq{4.59} and the equation
\[
\hat{\psi}''(S_0) \hat{w} - \pa^2 \hat{w} = 0. 
\]
\refl{4.7} (ii) thus yields $\hat{w} \in L^2(\R)$. Consequently, by
\eq{4.16} we have $S_2^* = w + \hat{w} + \rho_2$, where $w + \hat{w}
\in L^2(\R)$ is a solution of \eq{4.17}, \eq{4.18}. At the end of part
(IV) of this proof we showed that $w$ is the only solution of
\eq{4.17}, \eq{4.18} in $L^2(\R)$, whence $\hat{w} = 0$, hence $S_2^*
= S_2$.

The proof of \reft{4.5} is complete. \qed

\section{Proof of the estimates (\ref{E2.37a}) -- (\ref{E2.38b}) in \reft{2.3}}\label{S5}

The proof of \eq{2.37a} -- \eq{2.38b} is straightforward: We insert the
function $(u^{(\mu)},T^{(\mu)},S^{(\mu)})$ defined in Section~\ref{S3}
into the model equations \eq{1.1} and \eq{1.3} and compute the
residues. However, the necessary computations are long.  Therefore we
divide them into four parts:

In Section~\ref{S5.1} we compute for the functions $T^{(\mu)}_1$ and
$T^{(\mu)}_2$ from the inner and outer expansions of $T^{(\mu)}$ the
residues $\div_x T^{(\mu)}_1 + {\sf b}$ and $\div_x T^{(\mu)}_2 + {\sf
  b}$ separately. Likewise, in Section~\ref{S5.2} we insert the inner
expansion $(u_1^{(\mu)},T_1^{(\mu)},S_1^{(\mu)})$ and the outer
expansion $(u_2^{(\mu)},T_2^{(\mu)},S_2^{(\mu)})$ into \eq{1.3} and
compute and estimate the residues separately. With these residues we
can prove \eq{2.37a} -- \eq{2.38b} in the regions $Q^{(\mu\la)}_{\rm
  inn}$ and $Q^{(\mu\la)}_{\rm out}$, but in the matching region
$Q^{(\mu\la)}_{\rm match}$ we need auxiliary estimates, which are
stated in Section~\ref{S5.3}. All the estimates are put together in
Section~\ref{S5.4} to complete the proof.

\subsection{Asymptotic expansion of $\boldsymbol{\div_x T^{(\mu)} + {\sf b}}$ }\label{S5.1}

\begin{lem}\label{L5.1}
Let $(\hu,\hT)$ be the solution of the transmission problem \eq{2.8}
-- \eq{2.12}. With the splitting \eq{3.19} of $\hu$ the stress tensor
field $\hT$ satisfies  in the neighborhood ${\cal U}_\da$ of $\Gm$    
\begin{equation}\label{EhatT} 
\hat{T} = [\hT] \hS + D\ve(\na_x \hv ) + D \ve
  \big( a^* \otimes n + \na_{\Gm_\xi}  u^* \big) \xi^+ + D \ve
  (\na_{\Gm_\xi}  a^*) \frac12 (\xi^+)^2 .    
\end{equation} 
With $\hat{\si}'$ defined in \eq{3.21} and with a remainder term
$R_{\ov{\ve}:\hT} \in L^{\infty}({\cal U}_\da)$ we have for
$(t,\eta,\xi) \in {\cal U}_\da$  
\begin{align} 
\ov{\ve}:\hT(t,\eta,\xi) &= \ov{\ve}:\hT^{(+)} (t,\eta) 
  + \hat{\si}'(t,\eta,0)\, \xi 
\nn \\
\mbox{} + \ov{\ve}&:D\ve \big(a^*(t,\eta) \otimes n(t,\eta) + \na_\Gm
  u^*(t,\eta) \big) \, \xi 
 + R_{\ov{\ve}:\hT}(t,\eta,\xi)\, \xi^2, \quad
  \xi > 0,  \label{E5.epsT+}
\\
\ov{\ve}:\hT(t,\eta,\xi) &= \ov{\ve}:\hT^{(-)} (t,\eta) +
  \hat{\si}'(t,\eta,0)\, \xi + R_{\ov{\ve}:\hT}(t,\eta,\xi)\, \xi^2,
  \qquad \xi < 0. \label{E5.epsT-}
\end{align}
The functions $u^*$ and $a^*$ introduced in \eq{2.26}, \eq{2.27} and
the normal vector $n$ satisfy on the interface $\Gm$  
\begin{eqnarray}
\Big( D \big( \ve(u^* \otimes n) - \ov\ve) \big) \Big) n &=& 0, 
  \label{E5.Dn}\\
\big( D \ve (a^* \otimes n
  + \na_\Gm u^* ) \big) n + \div_\Gm D \ve(u^* \otimes n) &=& 0.  
  \label{E5.an}
\end{eqnarray}
\end{lem}
{\bf Proof:} 
With the splitting \eq{gradsplit} of the gradient operator we compute
from \eq{3.19} that 
\begin{align*}
\na_x \hat{u}(t,x) &= \pa_\xi \hat{u} \otimes n + \na_{\Gamma_\xi}
  \hat{u}  \\ 
&= \big(u^*(t,\eta) 1^+(\xi)+ a^*(t,\eta) \xi^+\big) \otimes n
    (t,\eta) + \na_{\Gamma_\xi} u^*(t,\eta) \xi^+\\ 
&\phantom{=}+\na_{\Gamma_\xi} a^*(t,\eta) \frac{1}{2} 
        (\xi^+)^2+ \na_x \hat{v}(t,x). 
\end{align*}
We insert this equation into \eq{2.9}, note that
$\hat{S}(t,x)=1^+(\xi)$, and employ that by \eq{2.42} 
\[
D \big(\ve(u^* \otimes n)- \ov\ve\big) \hat{S}=[\hat{T}] \hat{S} 
\]
to obtain \eq{hatT}.

By \eq{surfgraddeco} we have
\begin{equation}\label{E5.24}
\na_{\Gamma_\xi} u^*(t,\eta) \xi^+= \na_\Gamma u^* (t,\eta) \xi^+ +
(\na_\Gamma u^*(t,\eta)) R_A(t,\eta,\xi)(\xi^+)^2, 
\end{equation}
where we used our convention to identify $\na_\Gamma u^*$ and
$\na_\eta u^*$, since $u^*$ does not depend on $\xi$. Noting the
definition of $\hat{\sigma}$ in \eq{3.21}, we obtain from \eq{hatT}
and from \eq{5.24} that 
\begin{align}
\ov\ve : \hat{T}&(t,\eta,\xi) = \ov\ve : [\hat{T}](t,\eta) 1^+(\xi)
  + \hat{\sigma} (t,\eta,\xi)  \nn \\ 
& + \ov\ve: D \ve \big(a^*(t,\eta) \otimes n(t,\eta)+ \na_\Gamma
  u^*(t,\eta)\big) \xi^+  \nn \\ 
& + \ov\ve: D \ve \big(\na_\Gamma u^*(t,\eta) R_A (t,\eta,\xi) +
  \frac{1}{2} \na_{\Gamma_\xi} a^*(t,\eta)\big) (\xi^+)^2. \label{E5.25}
\end{align}
\eq{3.19} and \eq{2.9} together imply for $\xi<0$ that
\begin{equation}\label{E5.26}
\hat{T} (t,\eta,\xi)=D \ve \big(\na_x \hat{v} (t,\eta,\xi)\big),
\end{equation}
whence
\[
\ov\ve: \hat{T} (t,\eta,\xi)= \ov\ve: D \ve \big(\na_x
\hat{v}(t,\eta,\xi)\big)= \hat{\sigma} (t,\eta,\xi), \quad \xi <0, 
\]
and therefore
\begin{equation}\label{E5.27}
\ov\ve: \hat{T}^{(-)} (t,\eta)= \hat{\sigma} (t,\eta,0), \quad \ov\ve:
  \hat{T}^{(+)} (t,\eta)= \ov\ve: [\hat{T}] (t,\eta)+
  \hat{\sigma}(t,\eta,0). 
\end{equation}
By Taylor's formula we can express $\hat{\sigma}$ in the form
\[
\hat{\sigma}(t,\eta,\xi)= \hat{\sigma} (t,\eta,0)+ \pa_\xi
  \hat{\sigma}(t,\eta,0) \xi+ \pa^2_\xi \hat{\sigma}(t,\eta,\xi^*)
  \xi^2. 
\]
We expand $\hat{\sigma}$ in \eq{5.25} with this formula and note the
equations \eq{5.27} to obtain \eq{5.epsT+} and \eq{5.epsT-}. 

\eq{5.Dn} is an immediate consequence of \eq{2.42} and \eq{2.11}. 
To prove \eq{5.an} we apply \eq{divsplit} to calculate from  \eq{hatT}
that 
\begin{eqnarray}
  0 &=& \div_x \hat{T} + {\sf b} 
  \nn \\[1ex]
  &=& \pa_\xi ( [\hT] n \hS ) + \div_{\Gm_\xi} [\hT] \hS 
  \nn \\
  && \mbox{} +
  \pa_\xi\big( D \ve (a^* \otimes n + \na_{\Gm_\xi} u^* ) \xi^+ \big) n 
  + \pa_\xi \big( \frac12 D \ve( \na_{\Gm_\xi} a^*) (\xi^+)^2 \big)n
  \nn \\ 
  &&\mbox{} + 
  \div_{\Gm_\xi} D \ve \Big( \big(a^* \otimes n + \na_{\Gm_\xi} u^*
  \big) \xi^+  + \frac12 (\na_{\Gm_\xi} a^*) (\xi^+)^2 \Big) 
  \nn \\[1ex]   
  &&\mbox{} + \div_x D\ve(\na_x \hv) + {\sf b}. \label{EdivhatT}
\end{eqnarray}
From this equation we obtain for $\xi < 0$ that  
\begin{equation}\label{Eacdiv}
\div_x D\ve(\na_x \hv) + {\sf b} = \div_x \hT + {\sf b} = 0. 
\end{equation}
By assumption in \reft{2.3}, the function ${\sf b}$ is continuous at
$\Gm$. Moreover, by \refl{3.2} and the differentiability properties of
$\hv$ required in \reft{2.3} the function $\hv$ is two times
continuously differentiable at $\Gm$. Therefore we infer from
\eq{acdiv} that     
\begin{equation}\label{Ebcdiv}
\big(\div_x D\ve(\na_x \hv) + {\sf b} \big)^{(+)} = \big( \div_x
  D\ve(\na_x \hv) + {\sf b} \big)^{(-)} = 0, \quad \mbox{on } \Gm.
\end{equation}
With this equation and with $[\hT]n = 0$, by \eq{2.11}, we conclude
from \eq{divhatT} that   
\begin{eqnarray*}
0 = \lim_{\xi \ra 0+} (\div_x \hat{T} + {\sf b}) = \big( D \ve (a^* \otimes n
  + \na_\Gm u^* ) \big) n + \div_\Gm [\hT] .    
\end{eqnarray*}
From this relation and from $\div_\Gm [\hT] = \div_\Gm D \ve(u^*
\otimes n)$, which is a consequence of \eq{2.42}, we obtain
\eq{5.an}. \qed 
\\[1ex]
Next we study the stress field $T^{(\mu)}_1$ in the inner expansion.

\begin{lem}\label{L5.2}
Let $u_1^{(\mu)}$, $S_1^{(\mu)}$, $T_1^{(\mu)}$ be given in \eq{3.23}
-- \eq{3.25}, let $u_0$, $u_1$, $u_2$ be
defined in \eq{3.26} -- \eq{3.28}, and let $R_A$ be the remainder term
from \eq{Aexpansion}. We set $\zeta= \frac{\xi}{(\mu\lambda)^{1/2}}$. 
Then we have for $(t, \eta, \xi)$ from the neighborhood
$\cal{U}_\delta$ of $\Gamma$ 
\begin{multline}\label{E5.30}
T_1^{(\mu)}(t,\eta,\xi) = [\hat{T}] (S_0+ \mu^{1/2} S_1) +D \ve
  \big(\na_x (\hat{v}+ \mu^{1/2} \check{v})\big)
\\ 
+ (\mu\lambda)^{1/2} D \ve (a^* \otimes n + \na_\Gamma u^*)
  \int_{-\infty}^\zeta S_0(\vartheta) \, d\vartheta + \mu
  R_{T_1} (\lambda,t,\eta,\xi, \zeta), 
\end{multline}
where
\begin{equation}\label{E5.31}
R_{T_1} (\lambda, t, \eta, \xi, \zeta) = D \Big(\ve
\big(\na_{\Gamma_\xi} (\lambda^{1/2} u_1 + \lambda u_2) + \lambda\zeta
(\na_\eta u_0) R_A\big)- \ov\ve S_2\Big). 
\end{equation}
The argument of $[\hat{T}]$, $u^*$, $a^*$, $n$ is $(t,\eta)$, the
argument of $S_1$, $S_2$, $u_0$, $u_1$, $u_2$ is $(t,\eta,
\frac{\xi}{(\mu\lambda)^{1/2}})$, the argument of $\na_x \hat{v}$,
$\na_x \check{v}$, $R_A$ is $(t,\eta,\xi)$, and the argument of $S_0$
outside of the integral is $\frac{\xi}{(\mu\lambda)^{1/2}}$. 
Moreover, we have
\begin{align}
\div_x\, T_1^{(\mu)} + {\sf b} &= \div_x D\ve(\na_x
   \hat{v})+ {\sf b}+ \xi \,\div_{\Gamma,\xi} [\hT] S_0
   + \mu^{1/2} \div_{\Gamma_\xi} \big([\hT ] S_1\big)
\nn \\ 
&  + \mu^{1/2} \div_x D\ve(\na_x \check{v}) 
   + (\mu\lambda)^{1/2} \div_{\Gamma_\xi} D\ve (a^* \otimes n+
   \na_\Gm u^*) \int_{-\infty}^\zeta S_0
   (\vartheta)\, d\vartheta 
\nn \\ 
&  + \mu \,\div_x R_{T_1}\,. \label{E5.32} 
\end{align}
With $\hat{\sigma}(0)$, $\hat{\sigma}'(0)$, $\check{\sigma}(0)$
defined in \eq{3.21}, \eq{3.22} we have 
\begin{multline}
\pa_S {\sf W}\big(\ve(\na_x u_1^{(\mu)}), S_1^{(\mu)} \big) 
\\
= -\ov\ve:T_1^{(\mu)} (t,\eta,\xi) = -\ov\ve: [\hat{T}] (S_0+
  \mu^{1/2} S_1) - \hat{\sigma}(0) - (\mu\lambda)^{1/2}
  \hat{\sigma}'(0) \zeta - \mu^{1/2} \check{\sigma}(0)
\\ 
 - (\mu\lambda)^{1/2} \ov\ve :D \ve (a^* \otimes n +
    \na_\Gamma u^*) \int_{-\infty}^\zeta S_0 (\vartheta)
     d \vartheta  
 - \mu R_W (\lambda,t,\eta,\xi,\zeta), \label{E5.33}
\end{multline}
where
\begin{equation}\label{E5.34}
R_W (\la,t,\eta,\xi,\zeta)= \ov\ve  :R_{T_1}
  (\la,t,\eta,\xi,\zeta) 
  + \la \pa_\xi^2 \hat{\si} (t,\eta, \hat{\xi}) \zeta^2+
  \la^{1/2} \pa_\xi \check{\si} (t,\eta, \check{\xi}) \zeta,  
\end{equation}
with suitable $\hat{\xi}$, $\check{\xi}$ between $0$ and $\xi$ and
with $R_{T_1}$ defined in \eq{5.31}.
\end{lem}
{\bf Proof:}
With the splitting \eq{gradsplit} of the gradient operator we obtain
by definition of $u_1^{(\mu)}$ in \eq{3.23}, \eq{3.26} -- \eq{3.28}
that  
\begin{eqnarray}\label{E5.35}
\na_x u_1^{(\mu)} &=& (\mu\la)^{1/2} \na_x u_0 + \mu\la^{1/2} \na_x
  u_1+ \mu\lambda \na_x u_2 + \na_x (\hv + \mu^{1/2} \vv ) 
\nn \\ 
&=& (u^* \otimes n) S_0 +\mu^{1/2} u^* \otimes n S_1
  +(\mu\lambda)^{1/2} a^* \otimes n \int_{-\infty}^\zeta
  S_0(\vartheta) d\vartheta
\nn \\ 
&& \mbox{} + (\mu\la)^{1/2} \na_{\Gm_\xi} u_0+ \mu\la^{1/2}
  \na_{\Gm_\xi} u_1 +\mu\la \na_{\Gm_\xi}u_2 + \na_x (\hv + \mu^{1/2}
  \vv).  
\end{eqnarray}
\eq{3.26} and \eq{surfgraddeco} together yield
\begin{equation}\label{E5.36}
\na_{\Gamma_\xi} u_0 = \na_\eta u_0 (I+\xi R_A) = \na_\Gamma u^*
  \int_{-\infty}^\zeta S_0(\vartheta) 
   d\vartheta +(\mu\lambda)^{1/2} \zeta\, (\na_\eta u_0) R_A. 
\end{equation}
We insert \eq{5.35}, \eq{5.36} and \eq{3.24} into \eq{3.25}. From the
resulting equation we obtain \eq{5.30} and \eq{5.31} if we also note
that by \eq{2.42} 
\begin{multline*}
D\big(\ve(u^* \otimes n) (S_0+ \mu^{1/2} S_1) - \ov\ve (S_0+
\mu^{1/2} S_1)\big)\\ 
= D\big(\ve(u^* \otimes n)-\ov\ve\big) (S_0+ \mu^{1/2} S_1)=
[\hat{T}] (S_0+ \mu^{1/2} S_1).
\end{multline*}
To prove \eq{5.32} we employ the splitting \eq{divsplit} of the
divergence operator and \eq{surfdivmatdeco} to compute from \eq{5.30}
\begin{align}
\div_x\, T_1^{(\mu)}+ {\sf b} &= \pa_\xi ([\hat{T}] n S_0) +
   (\div_\Gm [\hat{T}]) S_0 + \xi \big(\div_{\Gm,\xi}
   [\hat{T}]\big) S_0 
\nn \\[1ex]
& + \mu^{1/2} \big( \pa_\xi ([\hat{T}] n S_1 )
   + \div_{\Gamma_\xi} ([\hat{T}] S_1 )\big) 
\nn \\[1ex] 
& + \div_x D \ve (\na_x \hat{v}) +{\sf b} +\mu^{1/2} \div_x
   D\ve(\na_x \check{v}) + \big( D\ve (a^* \otimes n + \na_\Gamma
   u^*)\big)n S_0 
\nn \\
& + (\mu\lambda)^{1/2} \div_{\Gamma_\xi} D\ve (a^* \otimes n
   +\na_\Gamma u^*) \int_{-\infty}^\zeta S_0(\vartheta)  d \vartheta +
   \mu\, \div_x R_{T_1}. \label{E5.37}  
\end{align}
By \eq{2.42} and \eq{5.an} we have
\[
\big(D\ve (a^* \otimes n + \na_\Gamma u^*)\big)n S_0 + \div_\Gamma
  [\hat{T}] S_0 = \Big( \big(D\ve(a^* \otimes n +\na_\Gamma u^*)\big)n
  + \div_\Gm D\ve(u^* \otimes n)\Big) S_0=0. 
\]
With this equation and with \eq{2.11} we obtain \eq{5.32} from \eq{5.37}.

\eq{5.33}, \eq{5.34} follow immediately from \eq{1.7}, which implies 
$\pa_S {\sf W}\big(\ve(\na_x u_1^{(\mu)}),S_1^{(\mu)} \big) = - \ov{\ve} :
T_1^{(\mu)}$, and from \eq{5.30}, \eq{5.31}, using the Taylor
expansions 
\begin{align*}
\vsi(\xi) &= \vsi(0)+ \pa_\xi \vsi (\check{\xi}) \xi = \vsi(0) +
  (\mu\la)^{1/2} \pa_\xi \vsi (\check{\xi}) \zeta,
\\ 
\hat{\si} (\xi) &= \hat{\si}(0)+ \pa_\xi \hat{\si}(0)\xi+
  \pa^2_\xi \hat{\si} (\hat{\xi}) \xi^2= \hat{\si}(0) +
  (\mu\la)^{1/2} \hat{\si}'(0)\zeta+ \mu\la \hat{\si}''(\hat{\xi}) 
   \zeta^2.   
\end{align*}
This completes the proof of \refl{5.2}. \qed
\begin{coro}\label{C5.3}
Let $Q_{\rm inn}^{(\mu\la)}$ and $Q_{\rm match}^{(\mu\la)}$ be
defined in \eq{2.25a} and let $T^{(\mu)}_1$ be given by \eq{3.25}. Then
there is a constant $C$ such that for all $0<\mu \leq \mu_0$ and all
$0 < \la \leq \la_0$  
\begin{equation}\label{E5.38}
\|\div_x\, T_1^{(\mu)}+{\sf b}\|_{L^\infty (Q_{\rm inn}^{(\mu\la)}
  \cup Q_{\rm match}^{(\mu\la)})} \leq C \Big(
\frac{\mu}{\la}\Big)^{1/2} |\ln \mu|^2.    
\end{equation}
\end{coro}
{\bf Proof:}
We estimate the terms on the right hand side of \eq{5.32}.
Note first that if $(t,\eta,\xi) \in Q_{\rm inn}^{(\mu\la)} \cup
Q_{\rm match}^{(\mu\la)}$ and $\zeta=
\frac{\xi}{(\mu\lambda)^{1/2}}$, then 
\begin{equation}\label{E5.39}
|\xi| \leq \frac{3}{a} (\mu\lambda)^{1/2} |\ln\mu|, \quad|\zeta| \leq
\frac{3}{a} |\ln \mu|. 
\end{equation}
This follows from \eq{2.25a}. With these inequalities the first two
terms on the right hand side of \eq{5.32} can be estimated as follows:
From the differentiability properties of $\hv$ and ${\sf b}$, which
in \reft{2.3} are assumed to hold, it follows by \refl{3.2} that
\begin{equation}\label{E5.regudivhu}
\div_x D \ve(\na_x \hv) + {\sf b} \in C({\cal U}_\da) \cap C^1\big( 
  (-\da,0], C(\Gm)\big) \cap C^1\big( [0,\da), C(\Gm)\big). 
\end{equation}
Because of this differentiability property we can apply the mean value
theorem to $\div_x D \ve(\na_x \hv) + {\sf b}$, which together with
\eq{bcdiv} and \eq{5.39} yields for all 
$(t,\eta,\xi) \in Q_{\rm inn}^{(\mu\la)} \cup Q_{\rm match}^{(\mu\la)}$
with $\xi \geq 0$ that  
\begin{multline}\label{E5.41}
\Big| \big(\div_x D\ve(\na_x \hv) + {\sf b}\big)(t,\eta,\xi) \Big|
  = \Big| \big(\div_x\, D\ve(\na_x \hat{v})+{\sf b}\big)^{(+)} + 
  \pa_\xi \big(\div_x D\ve(\na_x \hv)+{\sf b} \big)
  (t,\eta,\xi^*)\xi \Big|  
\\
= \big| \pa_\xi \big(\div_x D\ve(\na_x \hv)+{\sf b} \big)
  (t,\eta,\xi^*) \big| \xi
  \leq C_1 \xi \leq C_1 \frac{3}{a} (\mu\lambda)^{1/2} |\ln \mu|, 
\end{multline}
with a suitable number $\xi^*$ between $0$ and $\xi$. Since by
\eq{acdiv} the term $\div_x D\ve(\na_x \hv) + {\sf b}$ vanishes for
$\xi < 0$, the inequality \eq{5.41} holds for all $(t,\eta,\xi) \in 
Q_{\rm inn}^{(\mu\la)} \cup Q_{\rm match}^{(\mu\la)}$.   

To estimate the last term in \eq{5.32} note that 
\eq{4.9c}, \eq{4.11} and \eq{5.39} yield
\begin{align}
0 \leq \int_{-\infty}^\zeta S_0(\vartheta) d\vartheta
  &\leq \zeta^+ + C_2 \leq C_3 \frac{3}{a} |\ln \mu|,
\label{E5.intS0}\\[1ex] 
0 \leq \int_{-\infty}^\zeta \int_{-\infty}^\vartheta
  S_0(\vartheta_1) d\vartheta_1 d\vartheta &\leq
  \frac{1}{2}(\zeta^+)^2 + C_4 \leq C_5 \Big(\frac{3}{a} |\ln\mu|
  \Big)^2, 
\nn \\[1ex] 
\Big|\int_0^\zeta S_1(t,\eta,\vartheta) d \vartheta\Big| &\leq
  C_6|\zeta| \leq C_6 \frac{3}{a} |\ln\mu|. \nn
\end{align}
Using these inequalities, the definitions of $u_0$, $u_1$, $u_2$ in
\eq{3.26} -- \eq{3.28} and the inequality \eq{4.11a} we obtain from
\eq{5.31} that 
\begin{eqnarray}
|R_{T_1}| &\leq& \big(C_7 + \frac{C_8}{a^2} \big) | \ln \mu|^2,
  \label{E5.estRT1} 
\\
|\mu\, \div_x R_{T_1}| &\leq&  \mu^{1/2} \la^{-1/2} \big(C_9 +
  \frac{C_{10}} {a^2} \big) | \ln \mu|^2. \label{E5.estdivRT1} 
\end{eqnarray}
\eq{5.estRT1} is used later, \eq{5.estdivRT1} is the desired estimate
for the last term in \eq{5.32}. 

To estimate the other terms in \eq{5.32} we apply \eq{4.6}, \eq{4.7},
\eq{4.11}, \eq{5.39} and\eq{5.intS0}. Together with \eq{5.41} and
\eq{5.estdivRT1} we find for $(t,\eta,\xi)\in 
  Q_{\rm inn}^{(\mu\la)} \cup Q_{\rm match}^{(\mu\la)}$ that 
\[
\Big|(\div_x T_1^{(\mu)} + {\sf b}) (t,\eta,\xi)\Big| \leq
     \Big(\frac{C_{11}}{a^2}+ C_{12}\Big) \Big(\mu^{1/2} + 
     (\mu \la)^{1/2} |\ln\mu| + \mu^{1/2}\la^{-1/2}) |\ln\mu|^2\Big),  
\]
which implies \eq{5.38}. \qed
\\[1ex]
In the next lemma we study the outer expansion $T_2^{(\mu)}$.

\begin{lem}\label{L5.4}
Let $(\tilde{u}, \tilde{T})$ be the solution of the boundary value
problem, which consists of the elliptic system \eq{3.9}, \eq{3.10}
with $\tilde{S}_2$ given by \eq{3.18}, and of the boundary conditions
\eq{3.14} -- \eq{3.16}. Let $\tilde{S}_3$ be the solution of
\eq{3.13}. Then $T_2^{(\mu)}$ defined in \eq{3.8} satisfies on $Q
\setminus \Gamma$ 
\begin{eqnarray} 
T_2^{(\mu)} &=& \hat{T}+ \mu^{1/2} \check{T}+ \mu \tilde{T}-
  \mu^{3/2} D \ov\ve \tilde{S}_3, \label{E5.42}
\\ 
\div_x T^{(\mu)}_2 +{\sf b} &=& -\mu^{3/2} \div_x(D \ov\ve
  \tilde{S}_3).\label{E5.43} 
\end{eqnarray}
\end{lem}
{\bf Proof:}
Insertion of \eq{3.6} and \eq{3.7} into \eq{3.8} yields
\begin{align*}
T_2^{(\mu)} = D \big(\ve (\na_x \hat{u}) - \ov\ve \hat{S}\big) &+
  \mu^{1/2} D \big(\ve(\na_x \check{u})- \ov\ve \tilde{S}_1\big)
\\ 
&+ \mu D \big(\ve(\na_x \tilde{u})- \ov\ve \tilde{S}_2\big)- \mu^{3/2}
  D \ov\ve \tilde{S}_3. 
\end{align*}
Using \eq{3.17}, we see from this equation and from \eq{2.9},
\eq{2.14}, \eq{3.10} that \eq{5.42} holds. \eq{5.43} is an immediate
consequence of \eq{5.42} and \eq{2.8}, \eq{2.13}, \eq{3.9}. \qed 

\subsection{Asymptotic expansion of $\boldsymbol{S_t + c({\sf W}_S + \hat{\psi} - \Da_x S)}$ }\label{S5.2}

In this section we compute the form of the residue 
\begin{equation}\label{E5.44}
(\mu\lambda)^{1/2} \pa_t S+ c\Big(\pa_S {\sf W}\big(\ve(\na_x u),S\big)+
\frac{1}{\mu^{1/2}} \hat{\psi}' (S)- \mu^{1/2}\lambda \Delta_x S\Big), 
\end{equation}
which is obtained when we either insert for $(u,S)$ the inner
expansion $\big(u_1^{(\mu)},S_1^{(\mu)} \big)$ or the outer expansion
$\big(u_2^{(\mu)},S_2^{(\mu)} \big)$ of the asymptotic solution
$\big(u^{\mu)},S^{(\mu)}\big)$.

For functions $(t,x) \mapsto w(t,x)$ defined in a neighborhood of
$\Gm$ we write $w(t,\eta,\xi) = w(t,x)$ with $x = \eta +
n(t,\eta)\xi$, as always. However, in the following computations this
slight abuse of notation could lead to confusion when we consider
derivatives with respect to $t$. To avoid this, we introduce the
notations
\[
w_{|t}(t,x) = w_{|t}(t,\eta,\xi) = \pa_r w(r,\eta,\xi)\rain{r=t}\,,
\quad (\pa_t w)(t,\eta,\xi) = \pa_t w(t,x). 
\]
As introduced previously, for $i=0,1,2$ we write 
$S_i'(t,\eta,\zeta) = \pa_\zeta S_i(t,\eta,\zeta)$ and
$S_i''(t,\eta,\zeta) = \pa^2_\zeta S_i(t,\eta,\zeta)$.

\paragraph{Inner expansion} We first compute \eq{5.44} for $(u,S) =
\big(u_1^{(\mu)},S_1^{(\mu)} \big)$. To this end we need

\begin{lem}\label{L5.5}
Let $s^{(\mu)}(t,\eta)$ be the normal speed of the phase interface
$\Gamma(t)$ at $\eta \in \Gamma(t)$, let $\na_\eta$ be the operator
defined in \eq{naeta}, and let $w$ be a function defined in a
neighborhood of $\Gm$. Then we have   
\[
\pa_t w (t,x)= w_{|t} (t,\eta,\xi) - \xi (\pa_t n)(t,\eta) \cdot
  \na_\eta w (t,\eta,\xi) - s^{(\mu)} (t,\eta) \pa_\xi w (t,\eta,\xi).
\]  
\end{lem}
{\bf Proof:} By definition, $\na_\eta w(t,\eta,\xi)$ is a tangential 
vector to $\Gm(t)$. \refl{2.normspeed} thus yields
\begin{multline*}
\pa_t w (t,x)= \pa_t w(t,\eta,\xi) = w_{|t} (t,\eta,\xi) + \pa_t \eta
  \cdot \na_\eta w (t,\eta,\xi) +  \pa_\xi w (t,\eta,\xi) \pa_t \xi
\\
= w_{|t} (t,\eta,\xi) - \xi (\pa_t n)(t,\eta) \cdot \na_\eta w
  (t,\eta,\xi) - s^{(\mu)} (t,\eta) \pa_\xi w (t,\eta,\xi). 
\end{multline*}
This proves the lemma. \qed 
%
\\[1ex]
We apply this lemma to the function $S_1^{(\mu)}$ defined in \eq{3.24}
to obtain  
\[
\pa_t S_1^{(\mu)} (t,x)= S_{1|t}^{(\mu)} (t,\eta,\xi)- \xi (\pa_t
n)(t,\eta) \cdot \na_\eta S_1^{(\mu)} (t,\eta,\xi)- s^{(\mu)} (t,\eta)
\pa_\xi S_1^{(\mu)} (t,\eta,\xi). 
\]
From this equation and from the asymptotic expansion \eq{2.31} of
$s^{(\mu)}$ we conclude for the first term in \eq{5.44} that
\begin{align}
(\mu\lambda)^{1/2}& \pa_t S_1^{(\mu)} (t,x)
\nn \\
& = (\mu\lambda)^{1/2} \pa_t \Big(S_0 \big(
  \frac{\xi}{(\mu\lambda)^{1/2}}\big) 
  + \mu^{1/2} S_1 \big(t,\eta,\frac{\xi}{(\mu\lambda)^{1/2}}\big) 
  + \mu S_2 \big(t,\eta,\frac{\xi}{(\mu\lambda)^{1/2}}\big) \Big) 
\nn \\
&= -(s_0+\mu^{1/2} s_1) (S'_0 + \mu^{1/2} S'_1 + \mu S'_2)
+\mu\la^{1/2} \tilde{R}_{S_t} 
\nn \\
&= -s_0 S_0' - \mu^{1/2} (s_1 S_0' + s_0 S_1') + \mu R_{S_t},
   \label{E5.47} 
\end{align}
with
\begin{eqnarray}
\tilde{R}_{S_t}(\mu,\lambda,t,\eta,\xi) &=& (S_{1|t}+ \mu^{1/2}
  S_{2|t})- \xi (\pa_t n) \cdot \na_\eta (S_1+ \mu^{1/2} S_2), 
\nn \\
R_{S_t}(\mu,\lambda,t,\eta,\xi) &=& \la^{1/2} \tilde{R}_{S_t} -  
  s_1 S_1' - (s_0 + \mu^{1/2} s_1)S_2'. \label{E5.48} 
\end{eqnarray}
For the third term in \eq{5.44} we get from Taylor's formula and from
\eq{3.24} 
\begin{equation}\label{E5.49}
\frac{1}{\mu^{1/2}} \hat{\psi}'(S^{(\mu)}_1)   
= \frac{1}{\mu^{1/2}} \hat{\psi}' (S_0) + \hat{\psi}'' (S_0) S_1 
 + \mu^{1/2} \big( \hat{\psi}''(S_0) S_2 + \frac12 \hat{\psi}'''(S_0)
   S_1^2 \big) + \mu R_{\hat{\psi}'}   
\end{equation}
where
\begin{equation}\label{E5.50}
R_{\hat{\psi}'} = \frac12 \hat{\psi}'''(S_0)\big(2 S_1 S_2 +
  \mu^{1/2}S_2^2 \big) + \frac16 \hat{\psi}^{(IV)}\big(S_0 + \vartheta
  (\mu^{1/2} S_1 + \mu S_2)\big) (S_1 + \mu^{1/2} S_2)^3.
\end{equation}
with suitable $0 < \vartheta < 1$. Observe next that 
\begin{equation}\label{E5.51}
\Delta_x S_1^{(\mu)} (t,x)= \pa_\xi^2 S_1^{(\mu)} (t,\eta,\xi)-
  \ka(t,\eta,\xi) \pa_\xi S^{(\mu)}_1(t,\eta,\xi) +
  \Delta_{\Gamma_\xi} S^{(\mu)}_1 (t,\eta,\xi), 
\end{equation}
where $\Delta_{\Gamma_\xi}= \div_{\Gm_\xi} \na_{\Gm_\xi}$ denotes the
surface Laplacian and where $\ka(t,\eta,\xi)$ is twice the mean
curvature of the surface $\Gm_\xi(t)$ at $\eta \in \Gm_\xi(t)$. With
the notation $\ka'(t,\eta,0) = \pa_\xi \ka(t,\eta,0)$ we obtain from
Taylor's formula
\begin{eqnarray}
\kappa(t,\eta,\xi) &=& \kappa(t,\eta,0)+ \pa_\xi \kappa(t,\eta,0)\xi+
  \frac{1}{2} \pa^2_\xi \kappa(t,\eta,\xi^*)\xi^2  
\nn \\ 
&=& \ka_\Gm(t,\eta)+ (\mu\la)^{1/2} \ka'(t,\eta,0) \zeta + \mu\la
R_\ka(t,\eta,\xi) \zeta^2, \label{E5.52} 
\end{eqnarray}
where $R_\ka(t,\eta,\xi) = \frac{1}{2} \pa_\xi^2 \kappa(t,\eta,\xi^*)$
is the remainder term, with suitable $\xi^*$ between $0$ and $\xi$. We
insert \eq{3.24} and \eq{5.52} into \eq{5.51} and obtain for the
fourth term in \eq{5.44} that
\begin{multline}\label{E5.53}
\mu^{1/2}\la \Da_x S^{(\mu)}_1 = \mu^{-1/2} S_0'' + S_1'' + \mu^{1/2}
  S_2'' 
\\
- \la^{1/2} \Big( \ka_\Gm + (\mu\la)^{1/2} \ka'\zeta + \mu \la R_\ka
  \zeta^2 \Big) (S_0' + \mu^{1/2} S_1')  
  - \mu \la^{1/2} \ka(\xi) S_2' + \mu^{1/2}\la \Da_{\Gm_\xi}
  S^{(\mu)}_1  
\\
= \mu^{-1/2} S_0'' + \big(S_1'' - \la^{1/2} \ka_\Gm S_0'\big) 
  + \mu^{1/2} \Big( S_2'' - \la^{1/2} \ka_\Gm S_1' - \la \ka' \zeta
  S_0' \Big) + \mu R_\Da,  
\end{multline}
where
\begin{equation}\label{E5.54}
R_\Da = - \la \ka' \zeta S_1' - \la^{1/2} \ka(\xi) S_2' +
  \la \Da_{\Gm_\xi} (S_1 + \mu^{1/2} S_2) + \la^{3/2} R_\ka\zeta^2
  (S_0' + \mu^{1/2} S_1').  
\end{equation}
From \eq{5.47}, \eq{5.33}, \eq{5.49} and \eq{5.53} we obtain
\begin{align}
(\mu\la)^{1/2} \pa_t S_1^{(\mu)} &+ c\Big(\pa_S {\sf W}\big(\ve(\na_x
 u_1^{(\mu)}),S_1^{(\mu)}\big)+ 
 \frac{1}{\mu^{1/2}} \hat{\psi}' (S_1^{(\mu)})- \mu^{1/2}\la
 \Da_x S_1^{(\mu)}\Big) 
\nn \\
= \frac{c}{\mu^{1/2}}& \big(\hat{\psi}' (S_0) - S_0'' \big) 
 + c\Big( \hat{\psi}'' (S_0) S_1 - S_1'' - \ov\ve: [\hat{T}] S_0 -
 \hat{\si}(0) + (\la^{1/2} \ka_\Gm - \frac{s_0}{c}) S_0' \Big) 
\nn \\
\mbox{} + c\mu^{1/2}& \Bigg(\hat{\psi}''(S_0) S_2 - S_2'' -\ov\ve :
  [\hat{T}] S_1  - \check{\si}(0) + \big(\la^{1/2} \ka_\Gm -
  \frac{s_0}{c} \big) S_1'  
\nn \\ 
& - \la^{1/2}\Big(\hat{\si}'(0) \zeta + \ov{\ve} : D \ve (a^* \otimes n +
    \na_\Gamma u^*) \int_{-\infty}^\zeta S_0 (\vta)
     d \vta \Big) 
\nn \\
&  + \frac12 \hat{\psi}'''(S_0) S_1^2 + \big( \la
   \ka' \zeta - \frac{s_1}{c} \big) S_0' \Bigg) +
   \mu R_{S_t + c(\ldots)}\,, \label{E5.55}
\end{align}
where 
\begin{equation}\label{E5.56}
R_{S_t + c(\ldots)} = R_{S_t} + c(- R_W +
R_{\hat{\psi}'} -  R_\Da ), 
\end{equation}
with $R_{S_t}$, $R_W$, $R_{\hat{\psi}'}$ and $R_\Da$ given in \eq{5.48},
\eq{5.34}, \eq{5.50} and \eq{5.54}, respectively. 

\begin{coro}\label{C5.6}
Let $s_0$ be given by \eq{2.32} and assume that the functions $S_0$,
$S_1$ and $S_2$ satisfy the ordinary differential equations \eq{2.18},
\eq{2.19}, \eq{3.29}, with $F_1$, $F_2$ given by \eq{2.24}, \eq{3.30}.  
Assume moreover that the conditions \eq{2.20} -- \eq{2.23} and
\eq{3.34}, \eq{3.35} hold. Then there is a constant $K$ such that the
interior expansion $\big( u_1^{(\mu)},S_1^{(\mu)},T_1^{(\mu)} \big)$
defined in \eq{3.23} -- \eq{3.25} satisfies for all $(t,\eta,\xi) \in
Q_{\rm inn}^{(\mu\la)} \cup Q_{\rm match}^{(\mu\la)}$ and all $0 < \mu
\leq \mu_0$, $0 < \la \leq \la_0$ the inequality 
\begin{multline}
\Big| \pa_t S_1^{(\mu)} + \frac{c}{(\mu\la)^{1/2}}
 \Big(\pa_S {\sf W} \big(\ve(\na_x u_1^{(\mu)}),S_1^{(\mu)}\big)+ 
 \frac{1}{\mu^{1/2}} \hat{\psi}' (S_1^{(\mu)})- \mu^{1/2}\la
 \Da_x S_1^{(\mu)}\Big) \Big| 
\\
= \Big( \frac{\mu}{\la} \Big)^{1/2} | R_{S_t} + c(- R_W +
  R_{\hat{\psi}'} -  R_\Da ) | \leq K  \Big( \frac{\mu}{\la}
  \Big)^{1/2} |\ln \mu|^2. \label{E5.57} 
\end{multline}
\end{coro}
{\bf Proof:} From \eq{2.32} we obtain 
\[
\la^{1/2} \ka_\Gm - \frac{s_0}{c} = \frac{1}{c_1} \ov{\ve}:\langle
\hT\rangle, 
\]
and by \eq{5.27} we have $\ov{\ve}: \hT^{(-)} = \hat{\si}(0)$. After
insertion of these two equations into \eq{5.55}, the latter equation
takes the form
\begin{align}
(\mu\la)^{1/2} \pa_t S_1^{(\mu)}& + c\Big(\pa_S {\sf W}
 \big(\ve(\na_x u_1^{(\mu)}),S_1^{(\mu)}\big) + 
 \frac{1}{\mu^{1/2}} \hat{\psi}' (S_1^{(\mu)})- \mu^{1/2}\la
 \Da_x S_1^{(\mu)}\Big) 
\nn \\
= \frac{c}{\mu^{1/2}}& \Big(\hat{\psi}' (S_0) - S_0'' \Big) 
 + c\Big( \hat{\psi}'' (S_0) S_1 - S_1'' - F_1 \Big) 
\mbox{} + c\mu^{1/2} \Big(\hat{\psi}''(S_0) S_2 - S_2'' -  F_2
  \Big) 
\nn \\ 
&+ \mu R_{S_t + c(\ldots)} 
=  \mu \big(R_{S_t} + c(- R_W + R_{\hat{\psi}'} -  R_\Da )\big).
\label{E5.58}
\end{align}
Here we also used \eq{2.18}, \eq{2.19} and \eq{3.29}. Noting the
inequalities \eq{5.39} for $\xi$ and $\zeta$, the inequalities
\eq{4.6} -- \eq{4.8}, \eq{4.11}, \eq{4.14}, \eq{4.11a} for $S_0$,
$S_1$, $S_2$, and the inequality \eq{5.estRT1} for the term
$R_{T_1}$, which appears in $R_W$, we see by inspection of every term
in \eq{5.48}, \eq{5.34}, \eq{5.50} and \eq{5.54} that the inequality
\begin{equation}\label{E5.59}
\big| R_{S_t} + c(- R_W + R_{\hat{\psi}'} -  R_\Da )\big| \leq K |
\ln \mu |^2
\end{equation}
holds. To obtain inequality \eq{5.57} we divide \eq{5.58} by
$(\mu\la)^{1/2}$ and estimate the right hand side of the resulting
equation using \eq{5.59}. \qed 

\paragraph{Outer expansion} Next we compute \eq{5.44} for $(u,S) =
\big(u_2^{(\mu)},S_2^{(\mu)}  \big)$. Note first that \eq{1.7} and
\eq{5.42} yield
\begin{equation}\label{E5.60}
\pa_S {\sf W}\big( \ve(\na_x u_2^{(\mu)}),S_2^{(\mu)} \big) = - \ov{\ve} :
  (\hT + \mu^{1/2} \vT + \mu \tT) + \mu^{3/2} \ov{\ve} :D \ov{\ve}
  \tS_3\,.  
\end{equation}
Also, Taylor's formula and \eq{3.7} yield for a suitable $0 <
\vartheta(t,x) <1$  
\begin{align}
\hat{\psi}'(S_2^{(\mu)}) = \hat{\psi}'(\hS)& +
  \hat{\psi}''(\hS)\big(\mu^{1/2} \tS_1 + \mu \tS_2 + \mu^{3/2} \tS_3
  \big)  
\nn \\ \mbox{}
+ \frac12& \hat{\psi}'''(\hS) \big(\mu^{1/2} \tS_1 + \mu \tS_2 +
  \mu^{3/2} \tS_3 \big)^2 + \frac16 \hat{\psi}^{(IV)}(\hS)
  \big(\mu^{1/2} \tS_1 + \mu \tS_2 + \mu^{3/2} \tS_3 \big)^3 
\nn \\
&+ \frac1{24}\hat{\psi}^{(V)}\big(\hS + \vartheta( \mu^{1/2}
  \tS_1 + \mu \tS_2 + \mu^{3/2} \tS_3) \big) \big(\mu^{1/2} \tS_1 +
  \mu \tS_2 + \mu^{3/2} \tS_3 \big)^4   
\nn \\
= \mu^{1/2} \hat{\psi}''&(\hS) \tS_1 + \mu \Big(\hat{\psi}''(\hS) \tS_2
  + \frac12 \hat{\psi}'''(\hS) \tS_1^2 \Big) 
\nn \\
&+ \mu^{3/2} \Big(
  \hat{\psi}''(\hS) \tS_3 +  \hat{\psi}'''(\hS) \tS_1\tS_2 + \frac16
  \hat{\psi}^{(IV)}(\hS) \tS_1^3 \Big) + \mu^2\, \ov{R}_{\hat{\psi}'}\,.
\label{E5.61}
\end{align}  
Here we used that $\hat{\psi}'(\hS) = 0$, by \eq{2.29}. Equations
\eq{5.60} and \eq{5.61} imply that in the domain $Q \setminus \Gm$ 
\begin{align}
(\mu\la)^{1/2}& \pa_t S_2^{(\mu)} + c\Big(\pa_S {\sf W}
  \big(\ve(\na_x u_2^{(\mu)}),S_2^{(\mu)}\big) + 
  \frac{1}{\mu^{1/2}} \hat{\psi}' (S_2^{(\mu)})- \mu^{1/2}\la
  \Da_x S_2^{(\mu)}\Big) 
\nn \\
&= c \Big( - \ov{\ve} : \hT + \hat{\psi}''(\hS) \tS_1 \Big)
\nn \\
&+ c \mu^{1/2}\Big(- \ov{\ve} : \vT + \hat{\psi}''(\hS) \tS_2  + \frac12
  \hat{\psi}'''(\hS) \tS_1^2 \Big) 
\nn \\
&+ c\mu \Big(- \ov{\ve} : \tT + \hat{\psi}''(\hS) \tS_3  + 
  \hat{\psi}'''(\hS) \tS_1 \tS_2 + \frac16   \hat{\psi}^{(IV)}(\hS)
  \tS_1^3 - \la \Da_x \tS_1 + \frac{\la^{1/2}}{c} \pa_t \tS_1 \Big) 
\nn \\
& + \mu^{3/2}\, \ov{R}_{S_t + c(\ldots)}\,, \label{E5.62}
\end{align}
where
\begin{equation}\label{E5.63}
\ov{R}_{S_t + c(\ldots)} = c\,\big( \ov{\ve} :D \ov{\ve} \tS_3 +
  \ov{R}_{\hat{\psi}'} - \la \Da_x ( \tS_2 + \mu^{1/2} \tS_3 )\big) +
  \la^{1/2} \pa_t(\tS_2 + \mu^{1/2} \tS_3).
\end{equation}
Here we used that the function $\hS$ has the constant values $0$ in
$\gm$ and $1$ in $\gm'$. 
\begin{coro}\label{C5.7}
Assume that the functions $\tS_1$, $\tS_2$ and $\tS_3$ satisfy
\eq{3.11} -- \eq{3.13}. Then there is a constant $K$ such that for all
$(t,x) \in Q\setminus \Gm$ and all $0 < \mu \leq \mu_0$, $0 < \la \leq
\la_0$ 
\begin{equation}\label{E5.64} 
\Big| \pa_t S_2^{(\mu)} + \frac{c}{(\mu\la)^{1/2}} \Big(\pa_S {\sf W}
  \big(\ve(\na_x u_2^{(\mu)}),S_2^{(\mu)}\big) + 
  \frac{1}{\mu^{1/2}} \hat{\psi}' (S_2^{(\mu)})- \mu^{1/2}\la
  \Da_x S_2^{(\mu)}\Big) \Big| \leq K \frac{\mu}{\la^{1/2}}\,.  
\end{equation}
\end{coro}
{\bf Proof:} By \eq{3.11} -- \eq{3.13}, the brackets on the right hand
side of equation \eq{5.62} vanish. Therefore, if we divide the latter
equation by $(\mu\la)^{1/2}$, we obtain
\[
\pa_t S_2^{(\mu)} + \frac{c}{(\mu\la)^{1/2}} \Big(\pa_S {\sf W}
  \big(\ve(\na_x u_2^{(\mu)}),S_2^{(\mu)}\big) + 
  \frac{1}{\mu^{1/2}} \hat{\psi}' (S_2^{(\mu)})- \mu^{1/2}\la
  \Da_x S_2^{(\mu)}\Big) = \frac{\mu}{\la^{1/2}} \ov{R}_{S_t +
    c(\ldots)}. 
\]
$\tS_1$ and $\tS_2$ are given in \eq{3.17}, \eq{3.18}, and the
function $\tS_3$ is obtained by solving \eq{3.13} for this function.
From these equations we see by our general regularity assumptions that
$\| (\tS_1,\tS_2,\tS_3)\|_{L^\infty(Q\setminus \Gm)} \leq K_1$, with
the constant $K_1$ independent of $\mu$. Using this, we see by
inspection of every term in \eq{5.63} that $\| \ov{R}_{S_t +
  c(\ldots)}\|_{L^\infty(Q\setminus \Gm)} \leq K$, with $K$
independent of $\mu$ and $\la$. This inequality and the equation above
imply \eq{5.64}.  \qed


\subsection{Auxiliary estimates needed in the matching region}\label{S5.3}

The following auxiliary estimates are needed to prove \eq{2.37a} and
\eq{2.38a} in the matching region $Q_{\rm match}^{(\mu\la)}$.    
\begin{lem}\label{L5.6a}
The functions $u_2^{(\mu)}$, $T_2^{(\mu)}$, $S_2^{(\mu)}$ defined in
\eq{3.6} -- \eq{3.8} and $u_1^{(\mu)}$, $T_1^{(\mu)}$, $S_1^{(\mu)}$
defined in \eq{3.23} -- \eq{3.25} satisfy  
\begin{eqnarray}
\label{ESmu1Smu2}
\|S_1^{(\mu)} - S_2^{(\mu)}\|_{L^\infty(Q_{\rm match}^{(\mu\la)} ) }
  &\leq& K \mu^{3/2} |\ln \mu|^2, 
\\
\label{EDxSmu1Smu2}
\| D^\alpha_{x}(S_1^{(\mu)} - S_2^{(\mu)})\|_{L^\infty(
  Q_{\rm match}^{(\mu\la)}) } &\leq& K \la^{-\frac{|\al|}{2}}
  \mu^\frac{3-|\al|}{2},\quad 1 \leq |\al| \leq 2, 
\\[1ex]
\label{EDtSmu1Smu2}
\| \pa_t(S_1^{(\mu)} - S_2^{(\mu)})\|_{L^\infty(Q_{\rm match}^{(\mu\la)}
  )} &\leq& K \la^{-1/2} \mu, 
\\
\label{Eumu1umu2}
\|u_1^{(\mu)} - u_2^{(\mu)}\|_{L^\infty(Q_{\rm match}^{(\mu\la)} )}
  &\leq& K \la^{1/2} \mu^{3/2} |\ln \mu| ,  
\\
\label{Enaumu1umu2}
\| \na_x (u_1^{(\mu)} - u_2^{(\mu)}) \|_{L^\infty(
  Q_{\rm match}^{(\mu\la)} )} &\leq& K \mu,
\\
\label{ETmu1Tmu2}
\|T_1^{(\mu)} - T_2^{(\mu)}\|_{L^\infty(Q_{\rm match}^{(\mu\la)} )}
  &\leq& K \mu,  
\end{eqnarray}
for all $\mu \in (0,\mu_0]$, $\la \in (0,\la_0]$. Here $\al$ denotes a
multi-index and $K$ denotes a positive constant, which does not
necessarily have the same value in the six estimates. 
\end{lem}
{\bf Proof:} Since the proofs of these estimates are long and
technical, we present here only the proofs of \eq{Smu1Smu2} and
\eq{umu1umu2}. The proofs of the estimates \eq{DxSmu1Smu2},
\eq{DtSmu1Smu2} and \eq{naumu1umu2} run along the same lines.
\eq{Tmu1Tmu2} is an immediate consequence of the definitions \eq{3.8},
\eq{3.25} of $T_2^{(\mu)}$ and $T_1^{(\mu)}$, and of the estimates
\eq{naumu1umu2}, \eq{Smu1Smu2}.

In this proof we mostly drop the arguments $t$ and $\eta$
to simplify the notation. As usual we write $\zeta =
\frac{\xi}{(\mu\la)^{1/2}}$. We need that for $(t,\eta,\xi)\in Q_{\rm
  match}^{(\mu\la)}$ the inequalities 
\begin{equation}\label{E5.8}
\frac{3}{2} \frac{| \ln \mu|}{a} \leq
\Big|\frac{\xi}{(\mu\lambda)^{1/2}}\Big| = |\zeta| \leq 3 \frac{|\ln
  \mu|}{a},  
\end{equation}
hold, by definition of $Q_{\rm match}^{(\mu\la)}$ in \eq{2.25a}.

We begin with the proof of \eq{Smu1Smu2}. By definition of $S_1^{(\mu)}$
and $S_2^{(\mu)}$ in \eq{3.24} and \eq{3.7} we have 
\begin{eqnarray}\label{E5.9}
| S_1^{(\mu)}-S_2^{(\mu)}| &=& |S_0+ \mu^{1/2}S_1+\mu S_2- \hat{S} -
\mu^{1/2} \tilde{S}_1- \mu \tilde{S}_2- \mu^{3/2} \tilde{S}_3|
\nn \\ 
&\leq& |S_0- \hat{S}|+ \mu^{1/2} |S_1+ \mu^{1/2} S_2- \tilde{S}_1-
\mu^{1/2} \tilde{S}_2|+ \mu^{3/2} |\tilde{S}_3|. 
\end{eqnarray}
To estimate the first term on the right hand side note that since
$\hat{S}(t,x)= \hat{S}(\xi)=1^+(\xi)$, relations \eq{4.6}, 
\eq{4.7}, and \eq{5.8} imply for $(t,\eta,\xi)\in 
  Q_{\rm match}^{(\mu\la)}$ that   
\begin{equation}\label{E5.10}
\Big|S_0 \Big(\frac{\xi}{(\mu\lambda)^{1/2}}\Big)- \hat{S}(\xi)\Big| 
\leq K_1 e^{-a|\zeta|} \leq K_1 e^{- \frac{3}{2}|\ln
  \mu|} = K_1 \mu^{3/2}. 
\end{equation}
To estimate the second term on the right hand side of \eq{5.9} we
introduce the notations 
\begin{align}
\tilde{\rho}_1(\zeta)  &= 
\begin{cases}
\frac{\ov{\ve}:\hat{T}^{(-)}}{\hat{\psi}''(0)}, & \zeta < 0,
\\ 
\frac{\ov{\ve}:\hat{T}^{(+)}}{\hat{\psi}''(1)}, & \zeta > 0, 
\end{cases} \label{E5.trho1}
\\
\tilde{\rho}_2(\zeta)  &= 
\begin{cases}
\frac{1}{\hat{\psi}''(0)} \Big(
  \ov{\ve}:\vT^{(-)} -
  \frac{\hat{\psi}'''(0)}{2}  \Big( \frac{\ov{\ve} :
    \hT^{(-)}}{\hat{\psi}''(0)} \Big)^2 
  + \la^\frac12\, \hat{\si}'(0) \zeta \Big) , & \zeta < 0, 
\\
\frac{1}{\hat{\psi}''(1)} \Big(
 \ov{\ve}:\vT^{(+)} -
  \frac{\hat{\psi}'''(1)}{2}  \Big( \frac{\ov{\ve} :
  \hT^{(+)}}{\hat{\psi}''(1)} \Big)^2  
  + \la^\frac12 \hat{\si}'(0) \zeta 
\\ \hphantom{\frac{1}{\hat{\psi}''(0)} \Big( \ov{\ve}:\vT^{(+)} -}  
  + \la^\frac12 \ov{\ve} : D\ve (a^* \otimes n + \na_\Gm u^*) \zeta
  \Big),   & \zeta > 0.  
\end{cases} \label{E5.trho2}
\end{align}
By definition of the functions $\rho_1$, $\rho_2$ in \eq{4.10} and
\eq{3.33}, and by definition of $\varphi$, $\varphi_\pm$ in
\eq{3.31} and \eq{3.32}, we have 
\[
\tilde{\rho}_i(\zeta) - \rho_i(\zeta) = (1 - \varphi(-\zeta) -
  \varphi(\zeta)) \tilde{\rho}_i (\zeta) = 0, \qquad \mbox{for }
  i=1,2 \mbox{ and } |\zeta| \geq 2.
\]
We can therefore choose a suitable constant $\tilde{K}$ such that
$\big| \tilde{\rho}_i(\zeta) - \rho_i(\zeta) \big| \leq \tilde{K}
e^{-a |\zeta|}$ holds for $i=1,2$ and all $\zeta \in \R$.   
From this inequality and from the estimates \eq{4.12}, \eq{4.12a} we 
conclude that 
\begin{align}
| S_1 (\zeta) - \tilde{\rho}_1| &\leq | S_1 (\zeta) - \rho_1(\zeta)| +
  |\rho_1(\zeta) - \tilde{\rho}_1| 
  \leq (K_2 + \tilde{K}) e^{-a |\zeta|}, \label{E5.S1rho1}  
\\
| S_2 (\zeta) - \tilde{\rho}_2| &\leq | S_2 (\zeta) - \rho_2(\zeta)| +
  |\rho_2(\zeta) - \tilde{\rho}_2| 
  \leq \big( K_5 (1+|\zeta|) + \tilde{K} \big) e^{-a |\zeta|},
  \label{E5.S2rho2} 
\end{align}
for $\zeta \in \R$. 

Now we proceed to estimate the second term on the right hand side of
\eq{5.9}. We insert the functions $\tilde{\rho}_1$ and
$\tilde{\rho}_2$ into this term, use the expressions for $\tS_1$ and
$\tS_2$ given in \eq{3.17}, \eq{3.18}, and employ the triangle
inequality to obtain
\begin{multline}\label{E5.11}
| S_1 + \mu^{1/2} S_2- \tilde{S}_1- \mu^{1/2} \tilde{S}_2|
\\
\leq \left|\tilde{\rho}_1+ \mu^{1/2} \tilde{\rho}_2- \frac{\ov\ve:
    \hat{T}}{\hat{\psi}''(\hat{S})}- \mu^{1/2}
  \left(\frac{\ov\ve:\check{T}}{\hat{\psi}''(\hat{S})}-
    \frac{\hat{\psi}'''(\hat{S})}{2 \hat{\psi}''(\hat{S})}
    \Big(\frac{\ov\ve:\hat{T}}{\hat{\psi}''(\hat{S})}\Big)^2
    \right)\right|  
\\ 
 + |S_1-\tilde{\rho}_1|+ \mu^{1/2} |S_2-\tilde{\rho}_2|= |I_1|+
  |I_2|+ |I_3|. 
\end{multline}
By \eq{5.S1rho1}, \eq{5.S2rho2} and \eq{5.8} we have for
$(t,\eta,\xi)\in Q_{\rm match}^{(\mu\la)}$ 
\begin{eqnarray} 
|I_2|+ |I_3| &=&  \big| S_1 (t,\eta, \zeta ) - \tilde{\rho}_1 (t,\eta,
  \zeta )\big| + \mu^{1/2} \big|S_2 (t,\eta, \zeta) - \tilde{\rho}_2 
  (t,\eta, \zeta)\big|
\nn 
\\[1ex] 
&\leq& (K_2 + \tilde{K}) e^{-a|\zeta|} + \mu^{1/2} \big(K_5 (1 + |
 \zeta|) + \tilde{K} \big) e^{-a|\zeta|} 
\nn \\ 
&\leq& \left(C_1 + \mu^{1/2} C_2 \Big(1+3 \frac{|\ln
    \mu|}{a}\Big)\right) e^{- \frac{3}{2}|\ln \mu|} \leq C_3\,
  \mu^{3/2}. 
\label{E5.12}
\end{eqnarray}
To find an estimate for $|I_1|$ note that the definitions of
$\tilde{\rho}_1$, $\tilde{\rho}_2$ in \eq{5.trho1}, \eq{5.trho2} yield
for $(t,\eta,\xi)\in Q_{\rm match}^{(\mu\la)}$ with $\xi > 0$
\begin{align}
I_1 &=\tilde{\rho}_1 \Big(\frac{\xi}{(\mu\lambda)^{1/2}}\Big) +
  \mu^\frac12 \tilde{\rho}_2 \Big(\frac{\xi}{(\mu\lambda)^{1/2}}\Big)
  - \frac{\ov\ve: \hat{T}(\xi)}{\hat{\psi}''(1)} - \mu^\frac12 
   \left( \frac{\ov\ve: \vT (\xi)}{\hat{\psi}''(1)} -
     \frac{\hat{\psi}'''(1)}{2 \hat{\psi}''(1)} \Big( \frac{\ov\ve: 
   \hat{T}(\xi)}{\hat{\psi}''(1)} \Big)^2\right) 
\nn \\ 
&= \frac{1}{\hat{\psi}''(1)} \Bigg( \ov\ve: \hat{T}^{(+)} -
  \ov\ve: \hat{T}(\xi) + \sigma'(0) \xi + \ov\ve: D \ve (a^*
  \otimes n + \na_\Gamma u^*)\xi \label{E5.13}
\\ 
&\phantom{= \frac{1}{\hat{\psi}''(1)}} + \mu^\frac12
  \big(\ov\ve: \check{T}^{(+)} - \ov\ve: \check{T}(\xi)\big) -
  \mu^\frac12\, \frac{\hat{\psi}'''(1)}{2} \left( \Big(
  \frac{\ov\ve: \hat{T}^{(+)}}{\hat{\psi}''(1)}\Big)^2- \Big(
  \frac{\ov\ve:
    \hat{T}(\xi)}{\hat{\psi}''(1)}\Big)^2\right)\Bigg). \nn   
\end{align}
\eq{5.epsT+} and \eq{5.8} together yield  
\begin{multline}\label{E5.14}
|\ov\ve: \hat{T}^{(+)}+ \sigma'(0)\xi + \ov\ve:D \ve (a^* \otimes n 
  + \na_\Gamma u^*)\xi - \ov\ve: \hat{T}(\xi)| 
\\
= |R_{\ov{\ve}:\hT}(\xi) \xi^2| \leq C_4\, \mu\lambda |\ln \mu|^2.  
\end{multline}
Since $\vT^{(+)}= \vT(0+)$, $\hT^{(+)}= \hT(0+)$, the mean value
theorem and \eq{5.8} imply  
\begin{multline}\label{E5.15}
\mu^{1/2} \Big| \ov\ve: \big( \check{T}^{(+)}- \vT(\xi)\big) -
  \frac{\hat{\psi}'''(1)}{2\big(\hat{\psi}''(1)\big)^2} \Big((\ov\ve:
  \hat{T}^{(+)})^2- \big(\ov\ve: \hat{T}(\xi)\big)^2\Big) \Big| 
\\ 
= \mu^{1/2} |R(\xi)\xi| \leq C_5\, \mu \lambda^{1/2} |\ln \mu|,
\end{multline}
where the remainder term $R$ belongs to $L^\infty ({\cal U}_\da)$.
Combination of \eq{5.13} -- \eq{5.15} yields for $(t,\eta,\xi) \in
Q^{(\mu\la)}_{\rm match}$ with $\xi > 0$ that 
\begin{equation}\label{E5.16}
|I_1| \leq \mu \big(C_4 \lambda |\ln \mu|^2  + C_5 \lambda^{1/2} 
  |\ln \mu|\big) \leq C_6\, \lambda^{1/2} |\ln \mu|^2 \mu. 
\end{equation}
From the definitions of $\tilde{\rho}_1$, $\tilde{\rho}_2$ in
\eq{5.trho1} and \eq{5.trho2} we see that for $(t,\eta,\xi) \in
Q^{(\mu\la)}_{\rm match}$ with $\xi <0$ the term $I_1$ takes the form  
\begin{align*}
I_1 &= \frac{1}{\hat{\psi}''(0)} \Bigg( \ov\ve: \hT^{(-)}  
  - \ov\ve: \hT(\xi) + \si'(0) \xi
\\ 
& \phantom{= \frac{1}{\hat{\psi}''(1)}} +\mu^{1/2} \big(\ov\ve:
  \vT^{(-)}- \ov\ve: \vT(\xi)\big) - \mu^{1/2}
  \frac{\hat{\psi}'''(0)}{2} \left(\Big(\frac{\ov\ve :
    \hT^{(-)}}{\hat{\psi}''(0)}\Big)^2- \Big(\frac{\ov\ve:
    \vT(\xi)}{\hat{\psi}''(0)}\Big)^2\right). 
\end{align*} 
Using \eq{5.epsT-} instead of \eq{5.epsT+}, we see as above that the
estimate \eq{5.16} also holds in this case, whence the estimate
\eq{5.16} is valid for all $(t,\eta,\xi) \in Q^{(\mu\la)}_{\rm
  match}$.

To finish the proof of \eq{Smu1Smu2}, we combine \eq{5.9} with
\eq{5.10}, \eq{5.11}, \eq{5.12} and \eq{5.16} to obtain the estimate    
\[
|S_1^{(\mu)}- S_2^{(\mu)}| \leq K_1 \mu^{3/2} + \mu^{1/2} \big( C_6\, 
  \lambda^{1/2} |\ln \mu|^2 \mu + C_3\, \mu^{3/2}
  \big) + \mu^{3/2} |\tS_3|,  
\]
which implies \eq{Smu1Smu2}.

Next we prove \eq{umu1umu2}. From \eq{3.6} and \eq{3.19}, \eq{3.20} 
we conclude for $(t,x) \in \U_\delta$ that 
\begin{align*}
u_2^{(\mu)}(t,x) &= \hat{u}(t,x)+ \mu^{1/2} \check{u}(t,x)+
    \mu \tilde{u}(t,x)\\ 
\begin{split}
 = u^* \xi^+ +a^* \frac{1}{2}(\xi^+)^2+ \mu^{1/2} u^*
  \Big(\frac{\ov\ve : \hat{T}^{(+)}}{\hat{\psi}''(1)} \xi^+ +
   \frac{\ov\ve : \hat{T}^{(-)}}{\hat{\psi}''(0)} \xi^- \Big)\\  
   \mbox{} +  \hat{v}(t,x)+ \mu^{1/2} \check{v}(t,x) + \mu
   \tilde{u}(t,x). 
\end{split}
\end{align*}
Combination of this equation with \eq{3.23} and insertion of \eq{3.26}
-- \eq{3.28} yields with $\zeta= \frac{\xi}{(\mu\lambda)^{1/2}}$ that  
\begin{align}
u_1^{(\mu)} (t,x) &- u_2^{(\mu)}(t,x) 
\nn \\
= (&\mu\lambda)^{1/2} u_0 \Big(\frac{\xi}{(\mu\lambda)^{1/2}}\Big) +
  \mu\lambda^{1/2} u_1 \Big( \frac{\xi}{(\mu\lambda)^{1/2}}\Big) +
  \mu\lambda u_2 \Big(\frac{\xi}{(\mu\lambda)^{1/2}}\Big) 
\nn \\ 
& - \left(u^*\xi^+ + a^* \frac{1}{2}(\xi^+)^2+ \mu^{1/2} u^*
  \Big(\frac{\ov\ve : \hat{T}^{(+)}}{\hat{\psi}''(1)} \xi^+ +
  \frac{\ov\ve: \hat{T}^{(-)}}{\hat{\psi}''(0)} \xi^-\Big) + \mu
  \tilde{u}(\xi)\right) 
\nn \\
=(&\mu\la)^{1/2} u^* \Big(\int_{-\infty}^\zeta
  S_0(\vartheta)\,{\rm d}\vartheta-\zeta^+\Big) + \mu\lambda\, a^*
  \int_{-\infty}^\zeta 
  \Big(\int_{-\infty}^\vartheta S_0(\vartheta_1)\, d \vartheta_1 -
  \vartheta^+\Big)\,d \vartheta 
\nn \\
& \mbox{}+ \mu\lambda^{1/2} u^*\int_0^\zeta \Big(
  S_1(\vartheta) - \frac{\ov\ve: \hat{T}^{(+)}}{\hat{\psi}''(1)}
  1^+(\vartheta) - \frac{\ov\ve: \hat{T}^{(-)}}{\hat{\psi}''(0)}
  1^-(\vartheta)\Big) d \vartheta - \mu \tilde{u}(\xi) 
\nn \\[1ex]
= (&\mu\lambda)^{1/2} J_1(\zeta)+ \mu\big(J_2(\zeta)+J_3(\zeta) -
  \tilde{u}(\xi)\big). 
\label{E5.17}
\end{align}
To estimate the right hand side we use the boundary condition
\eq{3.16} for $\tilde{u}$, note that by \eq{5.trho1} the equation
$\tilde{\rho}_1(\zeta) = \frac{\ov\ve:
  \hat{T}^{(+)}}{\hat{\psi}''(1)}$ holds for $\zeta > 0$, and employ
the inequalities \eq{5.S1rho1}, \eq{4.9c} to compute for $(t,\eta,\xi)
\in Q^{(\mu\la)}_{\rm match}$ with $\xi >0$
\begin{align}
\begin{split}
|J_2(\zeta)+ J_3(\zeta)- \tilde{u}^{(+)}| 
 =\Big|J_2(\zeta)-\lambda a^* \int_{-\infty}^\infty
  \Big(\int_{-\infty}^\vartheta S_0(\vartheta_1)\, d \vartheta_1
  - \vartheta^+\Big)\, d\vartheta  
\\  
\mbox{}+ J_3(\zeta)-\lambda^{1/2} u^*\int_0^\infty
  \Big(S_1(\vartheta)- \frac{\ov\ve:
  \hat{T}^{(+)}}{\hat{\psi}''(1)}\Big)\,d\vartheta\Big|
\end{split}
\nn\\ 
&=\Big|\lambda a^*\int_\zeta^\infty
   \Big(\int_{-\infty}^\vartheta
   S_0(\vartheta_1)\vartheta_1-\vartheta^+\Big)\,d\vartheta
 + \lambda^{1/2} u^*\int_\zeta^\infty
  \big(S_1(\vartheta)- \tilde{\rho}_1(\zeta) \big)\, d\vartheta\Big| 
\nn\\[1ex] 
&\leq \lambda |a^*| \int_\zeta^\infty \frac{K_1}{a}
  e^{-a\vartheta}\, d\vartheta +\lambda^{1/2} |u^*|
  \int_\zeta^\infty (K_2 + \tilde{K}) e^{-a\vartheta}\, d\vartheta
\nn \\[1ex] 
&= \Big(\lambda |a^*| \frac{K_1}{a^2} +\lambda^{1/2} |u^*|
  \frac{K_2 + \tilde{K}}{a}\Big) e^{-a\zeta}  \leq C_1 \la^{1/2}
  \big(\|a^*\|_{L^\infty(\Gamma)} + \|u^*\|_{L^\infty(\Gamma)}\big) 
  e^{-a \zeta}. \label{E5.18} 
\end{align}
The mean value theorem implies for $\xi >0$
\[
\tilde{u}(t,x)= \tilde{u}(t,\eta,\xi)=
\tilde{u}(t,\eta,0+)+R_{\tilde{u}}(t,\eta,\xi)\xi. 
\]
Since $\tilde{u}(t,\eta,0+)= \tilde{u}^{(+)}(t,\eta)$, we infer 
from this equation and from \eq{5.18}, \eq{5.8} for all $(t,\eta,\xi)
\in Q_{\rm match}^{(\mu\la)}$ with $\xi>0$ that
\begin{align}
|J_2 (\zeta)& + J_3(\zeta) - \tilde{u}(\xi)| \leq
 |J_2(\zeta)+J_3(\zeta) - \tilde{u}^{(+)}| + |\tilde{u}(\xi) -
 \tilde{u}^{(+)}|  
\nn\\ 
&\leq C_1 \la^{1/2} \big(\|a^*\|_{L^\infty(\Gamma)}+
  \|u^*\|_{L^\infty(\Gamma)}\big) e^{-a \zeta} +
  \|R_{\tilde{u}}\|_{L^\infty (\U_\delta)} \xi
\nn\\[1ex] 
&\leq C_1 \la^{1/2}\big(\|a^*\|_{L^\infty(\Gamma)}+
  \|u^*\|_{L^\infty(\Gamma)}\big) e^{- \frac{3}{2}|\ln\mu|} +
  \|R_{\tilde{u}}\|_{L^\infty (\U_\delta)} \frac{3}{a}(\mu\lambda)^{1/2}
  |\ln\mu|
\nn\\
& \leq C_2 \la^{1/2} \mu^{1/2} |\ln\mu|. \label{E5.19}
\end{align}
Using the boundary condition \eq{3.15} for $\tilde{u}$ instead of
\eq{3.16}, we see by the analogous computation that \eq{5.19} also holds
for $(t,\eta,\xi) \in Q_{\rm match}^{(\mu\la)}$ with $\xi<0$. 

Now use \eq{5.19} to estimate the second term on the right hand side
of \eq{5.17}. The first term is estimated by \eq{4.9c}. Because
\eq{5.8} yields $e^{-a|\zeta|} \leq e^{- \frac32 |\ln \mu|} =
\mu^{3/2}$, we obtain for $(t,\eta,\xi) \in Q_{\rm match}^{(\mu\la)}$ 
\begin{multline*}
|u_1^{(\mu)}(t,x)-u_2^{(\mu)}(t,x)| 
 \leq (\mu\lambda)^{1/2} |u^*|\, \Big| \int_{-\infty}^\zeta
  S_0(\vartheta)\, d\vartheta-\zeta^+\Big|+ \mu
  \big|J_2(\zeta)+ J_3(\zeta)- \tilde{u}(\xi)\big|
\\[1ex] 
\leq (\mu\lambda)^{1/2} \big( \max_\Gm |u^*| \big)\, \frac{K_1}{a}
 e^{-a|\zeta|} + C_2 \la^{1/2} \mu^{3/2} |\ln \mu|
  \leq C_1 \lambda^{1/2} \mu^2+ C_2 \la^{1/2}\mu^{3/2} |\ln\mu|, 
\end{multline*}
which implies \eq{umu1umu2}. \qed

\subsection{End of the proof of \reft{2.3}}\label{S5.4}

To complete the proof of \reft{2.3} note first that \eq{3.3}, \eq{3.4}
imply 
\[
u^{(\mu)}\rain{\pa\Om} = u^{(\mu)}_2\rain{\pa\Om}, \qquad
\pa_{n_{\pa\Om}} S^{(\mu)}\rain{\pa\Om)} = \pa_{n_{\pa\Om}}
  S^{(\mu)}_2\rain{\pa\Om},\qquad 
S^{(\mu)}\rain{Q_{\rm inn}^{(\mu\la)}} = S^{(\mu)}_1\rain{Q_{\rm
    inn}^{(\mu\la)}}.
\]
Therefore \eq{2.39} follows from the definition of $S_1^{(\mu)}$ in
\eq{3.24}, equation \eq{2.36} follows from \eq{2.20}, \eq{2.23},
\eq{3.34}, and \eq{2.36d} is a consequence of the definition of
$u^{(\mu)}_2\rain{\pa\Om}$ in \eq{3.6} and of \eq{2.12}, \eq{2.17},
\eq{3.14}. Moreover, the estimate \eq{2.38c} for the right hand side
$f_3^{(\mu\la)}$ of \eq{2.36e} follows from the definition of
$S^{(\mu)}_2\rain{\pa\Om}$ in \eq{3.7} and from $\pa_{n_{\pa\Om}}
\hS\rain{\pa\Om} = 0$. This last equation holds, since by assumption
$\Gm(t) \subseteq \Om$, which implies that $\hS(t)$ is identically
equal to $0$ or $1$ in a neighborhood of $\pa\Om$.  

\eq{2.36b} follows from the definition of $T^{(\mu)}$ in \eq{3.5};
equation \eq{2.40} is an immediate consequence of \eq{4.11a}.

It remains to verify the estimates \eq{2.37a} -- \eq{2.38b} for the
right hand sides $f_1^{(\mu\la)}$, $f_2^{(\mu\la)}$ of the equations
\eq{2.36a} and \eq{2.36c}. To this end we put together all the
estimates derived in Sections~\ref{S5.1} -- \ref{S5.3}. We start with
the proof of \eq{2.37a} and \eq{2.37b}.

Equation \eq{3.3} yields
\[
\na_x u^{(\mu)} = \na_x u_1^{(\mu)}\phi_{\mu\la} + \na_x
  u_2^{(\mu)}(1-\phi_{\mu\la}) + \big(u_1^{(\mu)} - u_2^{(\mu)}\big) \otimes
  \na_x \phi_{\mu\la}.
\]
We insert this equation into \eq{3.5} and use \eq{3.4} and \eq{3.8},
\eq{3.25} to obtain   
\begin{equation}\label{ETmuexplicite}
T^{(\mu)} = T_1^{(\mu)} \phi_{\mu\la} + T_2^{(\mu)} (1-\phi_{\mu\la})
+ D\ve\big( (u_1^{(\mu)} -  u_2^{(\mu)}) \otimes \na_x \phi_{\mu\la} \big). 
\end{equation} 
The function $\phi_{\mu\la}$ defined in \eq{3.2} is independent of
$(t,\eta)$. The decomposition \eq{skalgradsplit} of the gradient thus
yields 
\begin{equation}\label{Enaxphi}
\na_x \phi_{\mu\la} = \frac{2a}{3(\la\mu)^{1/2} |\ln \mu |}
  \phi_{\mu\la}'n, 
\end{equation}
with the unit normal vector $n=n(t,\eta)$ to $\Gm_\xi(t)$ at $\eta \in
\Gm_\xi(t)$. We write 
\[
\phi_{\mu\la}' = \phi'\big(\frac{2a\xi}{3 (\mu\la)^{1/2}|\ln \mu|}
\big), \qquad \phi_{\mu\la}'' = \phi''\big(\frac{2a\xi}{3
  (\mu\la)^{1/2}|\ln \mu|} \big),
\]
by a slight abuse of notation. With \eq{Tmuexplicite} and \eq{naxphi}
we compute    
\begin{align}
\div_x T^{(\mu)} + {\sf b} &= (\div_x T_1^{(\mu)}+{\sf b} )
  \phi_{\mu\la} + ( \div_x T_2^{(\mu)} + {\sf b} ) (1-\phi_{\mu\la})
\nn \\[1ex] 
& + \Big( (T_1^{(\mu)} - T_2^{(\mu)}) n  + \div_x D\ve\big(
  (u_1^{(\mu)} - u_2^{(\mu)}) \otimes n \big) \Big)
  \frac{2a}{3(\la\mu)^{1/2} |\ln \mu |} \phi_{\mu\la}' \nn \\
& + \Big(D\ve\big( (u_1^{(\mu)} -  u_2^{(\mu)}) \otimes n \big)\Big) 
  n\, \Big(\frac{2a} {3 (\la\mu)^{1/2} |\ln \mu |} \Big)^2
  \phi_{\mu\la}''. \label{EdivTmu}   
\end{align}
Inequality \eq{2.37b} is an immediate consequence of this equation and
of \eq{5.43}, since $\phi_{\mu\la} = 0$ in $Q_{\rm out}^{(\mu\la)}$, and
\eq{2.37a} is obtained by estimating the right hand side of \eq{divTmu}
using  the obvious inequality 
\begin{equation}\label{Eelliest4}
| \div_x D\ve\big((u_1^{(\mu)} - u_2^{(\mu)}) \otimes n \big) | \leq
  C \big( |\na_x (u_1^{(\mu)} -  u_2^{(\mu)})| + |u_1^{(\mu)} -
  u_2^{(\mu)}| \big)   
\end{equation}
and the equation and inequalities \eq{5.38}, \eq{5.43}, \eq{umu1umu2}
-- \eq{Tmu1Tmu2}.

We next proof \eq{2.38a} and \eq{2.38b}. The inequality \eq{2.38b}
follows immediately from \eq{5.64}, since $\phi_{\mu\la} = 0$ on
$Q_{\rm out}^{(\mu\la)}$, which by \eq{3.3} and \eq{3.4} implies
$(u^{(\mu)},S^{(\mu)}) = (u_2^{(\mu)}, S_2^{(\mu)})$ on $Q_{\rm
  out}^{(\mu\la)} \subseteq Q \setminus \Gm$.

It remains to verify \eq{2.38a}. Since ${\sf W}_S (\ve,S) =
-\ov{\ve}:D(\ve - \ov{\ve} S)$, by \eq{1.7}, it follows from \eq{3.5},
  \eq{3.25}, and \eq{Tmuexplicite} that  
\begin{align*}
{\sf W}_S\big( \ve(\na_x u^{(\mu)}),&\, S^{(\mu)}\big) - {\sf W}_S
\big( \ve(\na_x  
u_1^{(\mu)}),S_1^{(\mu)}\big) = -\ov{\ve}:(T^{(\mu)}-T_1^{(\mu)}) 
\\ 
& = -\ov{\ve} : (T_2^{(\mu)}-T_1^{(\mu)})(1-\phi_{\mu\la}) -
  \ov{\ve} : D\ve \big( (u_1^{(\mu)} - u_2^{(\mu)})\otimes \na_x
  \phi_{\mu\la}\big).  
\end{align*}
The mean value theorem and \eq{3.4} imply 
\begin{align}
\hat{\psi}' (S^{(\mu)}) - \hat{\psi}' (S_1^{(\mu)}) &=
  \hat{\psi}'' \Big( S_1^{(\mu)} + \vartheta (S_2^{(\mu)} -
  S_1^{(\mu)})(1 - \phi_{\mu\la}) \Big) (S_2^{(\mu)} - S_1^{(\mu)})(1 -
  \phi_{\mu\la}),   
\nn \\ 
\intertext{for a suitable $0<\vartheta (t,x)<1$, and }
\Da_x S^{(\mu)} - \Da_x S_1^{(\mu)} &= \Da_x ( S_2^{(\mu)} -
  S_1^{(\mu)}) (1- \phi_{\mu\la}) + 2\na_x ( S_1^{(\mu)} - S_2^{(\mu)}) 
  \cdot \na_x \phi_{\mu\la} 
\nn \\ 
& + ( S_1^{(\mu)} - S_2^{(\mu)}) \Da_x
   \phi_{\mu\la}. \label{E5.DeltaS-S1} 
\end{align}
The right hand sides of the last three equations vanish on the set
$Q_{\rm inn}^{(\mu\la)}$, since $\phi_{\mu\la} = 1$ on this set. On
the set $Q_{\rm match}^{(\mu\la)}$ the right hand sides of these
equations can be estimated using \eq{Smu1Smu2}, \eq{DxSmu1Smu2},
\eq{umu1umu2}, \eq{Tmu1Tmu2}, \eq{naxphi}. In the estimation of
\eq{5.DeltaS-S1} we also note that since $\phi_{\mu\la}$ is
independent of $(t,\eta)$, analogous to \eq{5.51} the equation
\[
\Da_x \phi_{\mu\la} = - \ka \pa_\xi\phi_{\mu\la} + \pa_\xi^2
 \phi_{\mu\la} = - \ka \frac{2a}{3(\la \mu)^{1/2}|\ln \mu|}
 \phi'_{\mu\la} + \Big( \frac{2a}{3 (\la \mu)^{1/2}|\ln \mu|}\Big)^2
 \phi''_{\mu\la}  
\] 
holds, with twice the mean curvature $\ka(t,\eta,\xi)$ of the surface
$\Gm_\xi(t)$ at $\eta \in \Gm_\xi(t)$. Together we obtain that on
$Q_{\rm inn}^{(\mu\la)} \cup Q_{\rm match}^{(\mu\la)}$ the inequality
\begin{multline}
\Big | \Big ({\sf W}_S \f \ve (\na_x u^{(\mu)}),S^{(\mu)}\g +
  \frac{1}{\mu^{1/2}} \hat{\psi}'(S^{(\mu)})-\mu^{1/2} \la \Da_x
  S^{(\mu)}\Big ) 
\\ 
- \Big ({\sf W}_S \f \ve(\na_x u_1^{(\mu)}),S_1^{(\mu)}\g +
\frac{1}{\mu^{1/2}} \hat{\psi}'(S_1^{(\mu)})-\mu^{1/2} \la \Da_x
S_1^{(\mu)}\Big ) \Big |   
\\
\le K \big( \mu + \mu |\ln \mu|^2 + \mu^{1/2}\la \mu^{1/2} \la^{-1}
  \big) \le K \mu | \ln \mu|^2 \label{E5.68}
\end{multline}
holds. Similarly, \eq{3.4} implies
\[
\pa_t S^{(\mu)} - \pa_t S_1^{(\mu)} = \pa_t (S_2^{(\mu)} -
  S_1^{(\mu)})(1 - \phi_{\mu\la}) + (S_1^{(\mu)} - S_2^{(\mu)}) \pa_t
  \phi_{\mu\la}. 
\]
The right hand side of this equation vanishes on 
  $Q_{\rm inn}^{(\mu\la)}$. To estimate the right hand side on the set 
  $Q_{\rm match}^{(\mu\la)}$ we use the inequalities \eq{Smu1Smu2},
\eq{DtSmu1Smu2} and the equation  
\[
\pa_t \phi_{\mu\la} = -\frac{2as^{(\mu)}}{3(\mu \la)^{1/2} |\ln \mu|} 
\phi'_{\mu\la},  
\]
which follows from \eq{3.2} and \refl{5.5}. The result is
\begin{equation}
\label{E5.69}
|\pa_t S^{(\mu)} - \pa_t S_1^{(\mu)} | \le K\la^{-1/2} \mu  |\ln
 \mu|,  
 \qquad \text{on } Q_{\rm inn}^{(\mu\la)} \cup Q_{\rm
   match}^{(\mu\la)}.  
\end{equation}
By combination of \eq{5.57}, \eq{5.68} and \eq{5.69} we see that the
inequality 
\begin{multline*}
\Big| \pa_t S^{(\mu)} + \frac{c}{(\mu \la)^{1/2}} \Big( 
 {\sf W}_S \big( \ve(\na_x u^{(\mu)}), S^{(\mu)}\big) + \frac{1}{\mu^{1/2}} 
 \hat{\psi}' (S^{(\mu)}) - \mu^{1/2} \la \Da_x S^{(\mu)}\Big) \Big|
\\ 
\le \Big| \pa_t S_1^{(\mu)} + \frac{c}{(\mu \la)^{1/2}} \Big( {\sf W}_S
 \big(\ve(\na_x u_1^{(\mu)}), S_1^{(\mu)} \big) + \frac{1}{\mu^{1/2}}
 \hat{\psi}' (S_1^{(\mu)}) - \mu^{1/2} \la \Da_x S_1^{(\mu)}\Big) \Big| 
\\ 
+ K \la^{-1/2} \mu |\ln \mu| + \frac{c}{(\mu\la)^{1/2}} K\mu | \ln
  \mu|^2   
\\
\le K \Big(\frac{\mu}{\la}\Big)^{1/2} |\ln \mu|^2 
 + K \frac{\mu}{\la^{1/2}}|\ln \mu| + c K \Big(
 \frac{\mu}{\la}\Big)^{1/2} | 
  \ln \mu|^2 \le K_1 \Big(\frac{\mu}{\la} \Big)^{1/2} |\ln \mu|^2  
\end{multline*}
holds on the set $Q_{\rm inn}^{(\mu\la)} \cup Q_{\rm match}^{(\mu\la)}$. 
This proves \eq{2.38a} and completes the proof of \reft{2.3}.  \qed

\section{Proof of \reft{2.8}}\label{S6}

This section contains the proof of \reft{2.8}. $(u,T,S)$ denotes the
asymptotic solution constructed in \reft{2.3} and $(u_{\rm AC},T_{\rm
  AC},S_{\rm AC})$ denotes the exact solution of \eq{1.1} -- \eq{1.3},
\eq{2.BC1AC} -- \eq{2.ICAC}. For the proof we need a lemma and a
theorem, which we state first.

\begin{lem}\label{L6.1}
For all $x \in \Gm(\hat{t})$ the propagation speeds
$s_{\rm AC}$ and $s$ satisfy 
\begin{equation}\label{E6.differences}
s_{\rm AC}(\hat{t},x) - s(\hat{t},x) 
= \frac{1}{|\na_x S(\hat{t},x)|}
  \Big( f_2^{(\mu\la)}(\hat{t} ,x)
  - \frac{c}{(\mu\la)^{1/2}}\, \ov{\ve}:  \big( T_{\rm 
    AC} (\hat{t}, x) - T (\hat{t}, x) \big) \Big),
\end{equation}
where  $f_2^{(\mu\la)}$ is the right hand side of equation \eq{2.36c}.
\end{lem}
{\bf Proof:} Since the manifold $\Gm$ is a level set of $S$ and since
by \eq{2.GmAC} the manifold $\Gm_{\rm AC}$ is a level set of $S_{\rm AC}$,
it follows that $(\pa_t S(\hat{t},x), \na_x S(\hat{t},x) )$ and
$(\pa_t S_{\rm AC}(\hat{t},x), \na_x S_{\rm AC}(\hat{t},x) )$ are
normal vectors to the respective manifolds at $(\hat{t},x)$. Moreover,
\eq{2.ICAC} implies that $\na_x S_{\rm AC}(\hat{t},x) = \na_x
S(\hat{t},x)$. From \eq{2.normspeed1} we thus infer that
\begin{gather}
s(\hat{t},x) = \frac{- \pa_t
  S^(\hat{t},x)} {\na_x S(\hat{t},x) \cdot n(\hat{t},x) } = \frac{-
  \pa_t S(\hat{t},x)} {|\na_x S(\hat{t},x)|}, 
\label{E6.2} \\ 
s_{\rm AC}(\hat{t},x) = \frac{- \pa_t S_{\rm AC}(\hat{t},x)} {|\na_x 
  S_{\rm AC}(\hat{t},x)| } = \frac{- \pa_t S_{\rm AC}(\hat{t},x)}
{|\na_x S(\hat{t},x)|}.
\label{E6.3}  
\end{gather}
For brevity we do not write the argument $(\hat{t},x)$ in the
following computation. In \eq{6.2} we eliminate $\pa_t S$ with the
help of \eq{2.36c}, and in \eq{6.3} we replace $\pa_t S_{\rm AC}$
by the right hand side of \eq{1.3}. Together with another application
of \eq{2.ICAC} this results in 
\begin{multline*}
s_{\rm AC} - s = \frac{c}{(\mu\la)^{1/2} |\na_x S|} 
  \Big( \big( \pa_S {\sf W} \big( \ve(\na_x u_{\rm AC}),
  S_{\rm AC} \big) + \frac{1}{\mu^{1/2}} \hat{\psi}'(S) -
  \mu^{1/2}\la\Da_x S \big) 
\\  
- \big( \pa_S {\sf W} \big( \ve(\na_x u), S \big) +
  \frac{1}{\mu^{1/2}} \hat{\psi}'(S) - \mu^{1/2} \la \Da_x S \big) 
  \Big) + \frac{1}{ |\na_x S|} f_2^{(\mu\la)} 
\\
= \frac{1} { |\na_x S|} f_2^{(\mu\la)}  
  - \frac{c}{(\mu\la)^{1/2} |\na_x S|} \Big(\ov{\ve}: T_{\rm AC} -
  \ov{\ve}: T \Big). 
\end{multline*}
which is \eq{6.differences}. In the last step we used that by \eq{1.7}
and \eq{2.36b} we have   
\[
\pa_S {\sf W} \big( \ve(\na_x u_{\rm
  AC}), S_{\rm AC} \big) = - \ov{\ve}:  T_{\rm
  AC}, \quad \mbox{and} \quad \pa_S {\sf W} \big( \ve(\na_x
  u), S \big) = - \ov{\ve}:  T.  
\]
The proof of \refl{6.1} is complete.   \qed  

\begin{theo}\label{T6.2}
Suppose that the order of differentiability of $\hat{\psi}$, $\Gm$,
$\hu$, $\vu$, ${\sf b}$, is higher by two than required in \reft{2.3}.
Assume that the principal curvatures $\ka_1^{(\la\mu)}$,
$\ka_2^{(\la\mu)}$ of the regular $C^1$--manifold $\Gm(\hat{t}) =
\Gm^{(\mu\la)}(\hat{t})$ are bounded, uniformly with respect to $\mu
\in (0,\mu_0]$ and $\la \in (0,\la_0]$, and that there is an open
subset $\Om' \subset\subset \Om$ and $\delta > 0$ such that the
neighborhood ${\cal U}^{(\mu\la)}_\delta(\hat{t})$ of
$\Gm^{(\mu\la)}(\hat{t})$ defined in \eq{2.2} satisfies ${\cal
  U}^{(\mu\la)}_\delta(\hat{t}) \subseteq \Om'$. Then there is a
constant $K_5$ such that for all $\mu \in (0,\mu_0]$ and all $\la
\in (0,\la_0]$
\begin{equation}\label{E6.differenceT}
\| T_{\rm AC} - T \|_{L^2(\Gm(\hat{t}))} \leq K_5 | \ln \mu|^3 \mu.  
\end{equation}
\end{theo}
We postpone the proof of this theorem and first finish the proof of
\reft{2.8}. 
\\[1ex]
{\bf End of the proof of \reft{2.8}} 
By \eq{2.6} and \eq{2.39} we have for $x = (\eta,0) \in
\Gm(\hat{t})$ that 
\begin{multline}\label{E6.naxlow}
\na_x S(\hat{t},x) = n (\pa_n S)(\hat{t},x) + \na_\Gm S(\hat{t},x) 
\\
= \frac{1}{(\mu\la)^{1/2}}  \Big( S_0'(\zeta) + \mu^{1/2}
 \pa_\zeta S_1(t,\eta,\zeta) + \mu \pa_\zeta S_2(t,\eta,\zeta)
 \Big)\rain{\zeta = 0}n(\hat{t},\eta) 
\\ 
+ \mu^{1/2} \na_\Gm S_1(\hat{t},\eta,0) + \mu \na_\Gm
  S_2(\hat{t},\eta,0).       
\end{multline}
\eq{4.1} implies   
\[
S_0'(0) = \sqrt{2 \hat{\psi}(S_0(0))} = \sqrt{2 \hat{\psi}(1/2)} > 0,
\]
whence, from \eq{6.naxlow} for $\mu \in (0,\mu_0]$ and $\la \in
(0,\la_0]$ with $\mu_0$ sufficiently small,   
\begin{align}
| \na_x S(\hat{t},x)| &\geq \frac{1}{(\mu\la)^{1/2}}
  \Big(\sqrt{2 \hat{\psi}(1/2)} - \mu^{1/2}
  |\pa_\zeta S_1(t,\eta,0)| - \mu |\pa_\zeta S_2(t,\eta,0)| \Big) 
\nn \\
 & - \mu^{1/2} |\na_\Gm S_1(\hat{t},\eta,0)| - \mu |\na_\Gm 
  S_2(\hat{t},\eta,0)| \geq \frac{1}{2(\mu\la)^{1/2}} \sqrt{2
  \hat{\psi}(1/2)}. \label{E6.naxlow2}  
\end{align}
Combination of \eq{6.differences} with the inequalities \eq{2.38a},
\eq{6.differenceT} and \eq{6.naxlow2} yields
\begin{multline*}
\|s_{\rm AC}(\hat{t}) - s(\hat{t})\|_{L^2(\Gm(\hat{t}))}  
\\
\leq \frac{1}{\min_{\Gm(\hat{t})} |\na_x S(\hat{t})|} \Big( 
  \| f_2^{(\mu\la)}(\hat{t})\|_{L^2(\Gm(\hat{t}))}  
  + \frac{c|\ov{\ve}|}{(\mu\la)^{1/2}} \big\| T_{\rm AC} (\hat{t})-T 
    (\hat{t}) \big\|_{L^2(\Gm(\hat{t}))} \Big) 
\\
\leq \frac{2(\mu\la)^{1/2}}{\sqrt{2\hat{\psi}(1/2)}} \Big(
  |\ln \mu|^2
  \Big(\frac{\mu}{\la}\Big)^{1/2} K_3\,  {\rm meas}
  (\Gm(\hat{t}))^{1/2} +
  \frac{c|\ov{\ve}|}{(\mu\la)^{1/2}} K_5 |\ln \mu|^3 \mu \Big) 
\\
\leq K_6 |\ln \mu|^2 \mu + K_7 |\ln \mu |^3 \mu. 
\end{multline*}
\eq{2.assumpA} follows from this estimate. The proof of \reft{2.8} is
complete.  \qed
\\[1ex]
{\bf Proof of \reft{6.2}:} 
Note that the function $\big( u_{\rm AC}(\hat{t}),T_{\rm
  AC}(\hat{t})\big)$ solves the equations \eq{1.1}, \eq{1.2} in $\Om$
with $S_{\rm AC}(\hat{t}) = S(\hat{t})$, by the initial condition
\eq{2.ICAC}. Moreover, \eq{2.BC1AC} holds for $u_{\rm AC}(\hat{t})$.
From the equations \eq{2.36a}, \eq{2.36b}, \eq{2.36d} we thus conclude
that the difference $(u_{\rm AC} - u, T_{\rm AC} - T)$ satisfies
\begin{eqnarray}
- \div_x ( T_{\rm AC} - T) (\hat{t} )
   &=& - f_1^{(\mu\la)}(\hat{t}),  \label{E6.linelast1}
\\[1ex]
(T_{\rm AC} - T)(\hat{t}) &=& D \ve \big( \na_x (u_{\rm AC} - u)  
  (\hat{t}) \big), \label{E6.linelast2}
\\[1ex]
(u_{\rm AC} - u) (\hat{t},x) &=& 0, \qquad x
  \in \pa \Om. \label{E6.linelast3}
\end{eqnarray}
This is the Dirichlet boundary value problem for the elliptic system
of elasticity theory in the domain $\Om$. It suggests itself to derive
the inequality \eq{6.differenceT} by using the $L^2$--regularity
theory of elliptic systems, which allows to estimate the norm
$\|T_{\rm AC}-T\|_{L^2(\Gm(\hat{t}))}$ by the $L^2$--norm of the right
hand side $-f^{(\mu\la)}_1(\hat{t})$ of \eq{6.linelast1}. To apply this
theory directly we would need that the $L^2$--norm of
$f^{(\mu\la)}_1(\hat{t})$ decays to zero for $\mu \ra 0$ uniformly
with respect to $\la$. However, the relation ${\rm meas}
\big(Q^{(\mu\la)}_{\rm inn} (\hat{t}) \cup Q^{(\mu\la)}_{\rm match}
(\hat{t})\big) \leq C_1 (\mu\la)^{1/2} |\ln \mu|$, which follows from
\eq{2.25a}, and the estimates \eq{2.37a}, \eq{2.37b} yield
\[
\| f^{(\mu\la)}_1 (\hat{t})\|_{L^2(\Om)} \leq |\ln \mu|^{5/2}
  \frac{\mu^{3/4}}{\lambda^{1/4}} K_1 C_1^{1/2}. 
\] 
The right hand side does not decay to zero for $\mu \ra \infty$
uniformly with respect to $\la$, but blows up for $\la \ra 0$.
Therefore direct application of the $L^2$--regularity theory is not
possible. Before giving the detailed proof of \eq{6.differenceT} we
sketch how to circumvent this difficulty.

Set
\begin{equation}\label{E6.4n}
{\sf A}(\mu)= \frac{3}{a} \mu^{1/2} |\ln \mu|,
\end{equation}
where $a>0$ is the constant defined in \eq{2.29}. By \eq{2.25a} we have
\begin{equation}\label{E6.3n}
Q^{(\mu\lambda)}_{\rm inn} (\hat{t}) \cup Q^{(\mu\lambda)}_{\rm match}
 (\hat{t})= \left\{(\eta,\xi) \in \mathcal{U}_\delta (\hat{t})
  \,\big\vert\, |\xi| \leq {\sf A}(\mu)\lambda^{1/2} \right\}.
\end{equation}
Define
\begin{equation}\label{E6.1n}
\begin{split}
f_{11}^{(\mu\lambda)}(x) &= 
\begin{cases}
f_1^{(\mu\lambda)}(\hat{t},x), & x \in Q^{(\mu\lambda)}_{\rm inn}
(\hat{t}) \cup Q^{(\mu\lambda)}_{\rm match} (\hat{t})\\ 
0, & x \in Q^{(\mu\lambda)}_{\rm out} (\hat{t}),
\end{cases}\\
f_{12}^{(\mu\lambda)}(x) &= 
\begin{cases}
0, & x \in Q^{(\mu\lambda)}_{\rm inn} (\hat{t}) \cup
Q^{(\mu\lambda)}_{\rm match} (\hat{t})\\ 
f_1^{(\mu\lambda)}(\hat{t},x), & x \in Q^{(\mu\lambda)}_{\rm out}
(\hat{t}), 
\end{cases}
\end{split}
\end{equation}
hence $f_1^{(\mu\la)}(\hat{t})=f_{11}^{(\mu\la)}+
f_{12}^{(\mu\la)}$. For $x = x(\hat{t},\eta,\xi) \in {\cal
  U}_\da(\hat{t})$ we write as usual $f^{(\mu\la)}_{11}(x) = 
f^{(\mu\la)}_{11}(\eta,\xi)$.  
From \eq{2.37a} and \eq{6.3n} we obtain for $\eta \in \Gm(\hat{t})$ and
$-{\sf A}(\mu)\la^{1/2} \leq \xi \leq {\sf A}(\mu) \la^{1/2}$ that 
\begin{equation}\label{E6.2n}
|f_{11}^{(\mu\la)}(\eta,\xi)| \leq |\ln \mu|^2
\left(\frac{\mu}{\lambda}\right)^{1/2} K_1. 
\end{equation}
Define $\da_*^{(\mu\la)}:\Gm(\hat{t}) \to \R$ by
\begin{equation}\label{E6.5n}
\da_*^{(\mu\la)} (\eta) = \int_{-{\sf A}(\mu)}^{{\sf A}(\mu)}
 \la^{1/2} f_{11}^{(\mu\la)} \big(\eta,\la^{1/2} \zeta \big)\,
 d\zeta. 
\end{equation}
\eq{6.2n} and \eq{6.4n} together imply that
\begin{equation}\label{E6.6n}
|\da_*^{(\mu\la)}(\eta)| \leq 2 {\sf A}(\mu) \la^{1/2} |\ln
 \mu|^2 \left(\frac{\mu}{\la}\right)^{1/2} K_1 = \frac{6}{a} K_1
 \mu |\ln \mu|^3, 
\end{equation}
for all $\eta \in \Gm(\hat{t})$. Examination of the boundary value
problem \eq{6.linelast1} -- \eq{6.linelast3} suggests that for $\mu$
fixed and $\la \ra 0$ the solution $(u_{\rm AC}-u, T_{\rm AC}-T)$
converges to the solution $(u_*, T_*):\Omega \to \R^3 \times {\cal
  S}^3$ of the transmission problem
\begin{eqnarray}
-\div_x T_* &=& 0, \label{E6.7n}\\
T_* &=& D \varepsilon (\na_x u_*), \label{E6.8n}\\
{[T_*]} n &=& \da^{(\mu)}_*, \qquad \mbox{on } 
 \Gm(\hat{t}), \label{E6.9n}\\ 
{[u_*]} &=& 0, \qquad \mbox{on } \Gm(\hat{t}), \label{E6.10n}\\
u_* (x) &=& 0, \qquad  x \in \pa \Om, \label{E6.11n}
\end{eqnarray}
where $\da_*^{(\mu)} (\eta)= \lim_{\la \to 0}
\da_*^{(\mu\la)}(\eta)$ for $\eta \in \Gm(\hat{t})$. If
this limit exists, it follows from \eq{6.6n} that 
\[
|\da_*^{(\mu)} (\eta)| \leq \frac{6}{a} K_1 \mu |\ln \mu|^3.
\]
This implies that the solution $(u_*, T_*)$ will be bounded by $C \mu
|\ln \mu|^3$ with a suitable constant $C$, and this limit behavior
suggests that though the $L^2$--norm of $f_1^{(\mu\la)}(\hat{t})$
blows up for $\la \ra 0$, the solution $(u_{\rm AC}-u, T_{\rm
  AC}-T)(\hat{t})$ of \eq{6.linelast1} -- \eq{6.linelast3} is bounded
by $C \mu |\ln \mu|^3$ with $C$ independent of $\la$. The reason for
the blow up of $\| f_1^{(\mu\la)}(\hat{t})\|_{L^2(\Om)}$ for $\la \ra
0$ is therefore not that the norm of the solution $(u_{\rm AC}-u,
T_{\rm AC}-T)(\hat{t})$ would blow up, but that the solution looses
regularity in a neighborhood of the surface $\Gm(\hat{t})$, which is
shown by the equation \eq{6.9n} for the limit solution. This equation
implies that $T_*$ does not belong to the Sobolev space $W^{1,2}(\Om)$.

In the following proof we do not study the limit $(u_*,
T_*)$. Instead, based on the idea of the behavior of the regularity of
$(u_{\rm AC}-u, T_{\rm AC}-T)(\hat{t})$, we decompose this function in
the form 
\[
(u_{\rm AC}-u, T_{\rm AC}-T)(\hat{t}) = \big(u^{(\la)}, T^{(\la)}
\big) + \big(u^{(\la)}_*, T^{(\la)}_* \big), 
\]
where $(u^{(\la)}, T^{(\la)}) = (u^{(\la)}_\mu, T^{(\la)}_\mu)$ is
bounded by $C\mu |\ln \mu|^3$, uniformly with respect to $\la$, and
for $\la \ra 0$ has the same regularity behavior as
$(u_{\rm AC}-u, T_{\rm AC}-T)$, but otherwise does not approximate
$(u_{\rm AC}-u, T_{\rm AC}-T)$. The construction is such that the
difference $(u^{(\la)}_*, T^{(\la)}_*) = (u^{(\la)}_{*\mu},
T^{(\la)}_{*\mu}) = (u_{\rm AC}-u, T_{\rm AC}-T) - (u^{(\la)},
T^{(\la)})$ does not loose its regularity for $\la \ra 0$. Hence, we
can use the 
standard $L^2$--theory for elliptic equations to show that also $(u^{(\la)}_*,
T^{(\la)}_*)$ is bounded by $C \mu |\ln \mu|^3$ independently of
$\la$.  

To construct $(u^{(\la)}, T^{(\la)})$ let
${\cal U}_\da(\hat{t})$ be the neighborhood of $\Gm(\hat{t})$
defined in \eq{2.2} and let $\phi_* \in C_0^\infty
\big((-\delta,\delta)\big)$ be a function satisfying 
\begin{equation}\label{E6.12n}
\phi_* (\xi)=1, \quad -\delta/2 \leq \xi \leq \delta/2.
\end{equation}
We set
\begin{eqnarray}
u^{(\la)} (x) &=&
\begin{cases}
\displaystyle \la^{1/2} V \Big(\la,\eta,\frac{\xi}{\la^{1/2}}\Big)
  \phi_* (\xi), 
   &  x=x(\hat{t},\eta,\xi) \in {\cal U}_\da(\hat{t}),\\ 
0, & x \in \Om \setminus {\cal U}_\da(\hat{t}),
\end{cases}\label{E6.13n}
\\
T^{(\la)} (x) &=& D \ve \big(\na_x u^{(\la)} (x)\big), \hspace{2cm} x
 \in \Omega, \label{E6.14n} 
\end{eqnarray}
where the function $\zeta \mapsto
V(\lambda,\eta,\zeta):[-\frac{\delta}{\lambda^{1/2}},
\frac{\delta}{\lambda^{1/2}}] \to \R^3$ is constructed as follows: 
We use the notations $V'=\partial_\zeta V$, $V''=\partial^2_\zeta
V$. In the interval $[-{\sf A}(\mu), {\sf A}(\mu)]$ the function $V$
is the solution of the boundary value problem 
\begin{eqnarray}
\Big( D\ve \big( V''(\la,\eta,\zeta) \otimes n\big) \Big)
 n &=& \la^{1/2} f_{11}^{(\mu\la)} (\eta,\la^{1/2} \zeta
 ), \qquad -{\sf A}(\mu) \leq \zeta \leq {\sf A}(\mu), \label{E6.15n}\\ 
V \big(\la,\eta, \pm {\sf A}(\mu)\big) &=& 0, \label{E6.16n}
\end{eqnarray}
where $n = n(\eta)$ is the unit normal vector to $\Gm(\hat{t})$ at
$\eta \in \Gm(\hat{t})$. The equation \eq{6.15n} is a second order
linear system of ordinary differential equations for the three
components of $V$, which can be written in the form
\begin{equation}\label{E6.16nn}
B V'' = \lambda^{1/2} f_{11}^{(\mu\lambda)},
\end{equation}
with a $3 \ti 3$--matrix $B = B(\eta)$ defined by the equation
\begin{equation}\label{E6.16nnn}
B \om =  \big( D \ve(\om \otimes n) \big)n ,
\end{equation}
which must hold for all $\om \in \R^3$. The matrix $B$ is symmetric
and positive definite uniformly with respect to $\eta$. To see this,
note that since the elasticity tensor $D: {\cal S}^3 \to {\cal S}^3$
is a linear, symmetric, positive definite mapping, we compute for
$\om_1, \om_2 \in \R^3$
\begin{multline*}
(B \om_1) \cdot \om_2 = \Big(\big(D \ve(\om_1 \otimes n)\big)
  n\Big) \cdot \om_2\\
= (\om_2 \otimes n): D \ve(\om_1 \otimes n) = \ve (\om_2 \otimes n):
  D \ve (\om_1 \otimes n)\\ 
 = \big(D\ve (\om_2 \otimes n)\big) : \ve (\om_1 \otimes n) =
 \Big(\big(D \ve (\om_2 \otimes n)\big)n\Big) \cdot \om_1 =  
  (B \om_2) \cdot \om_1.
\end{multline*}
This shows that $B$ is symmetric. For $\om \in \R^3$ we have with a
suitable constant $C_0 > 0$, which only depends on $D$ but is
independent of $\eta$, that 
\[
(B \om) \cdot \om= \ve (\om \otimes n):D \ve (\om \otimes n) \geq C_0
|\ve(\om \otimes n)|^2 \geq \frac{C_0}{2} |\om|^2 ,
\]
hence $B$ is positive definite uniformly with respect to
$\eta \in \Gm(\hat{t})$. 

Therefore the boundary value problem \eq{6.15n}, \eq{6.16n} has a
unique solution $V$ on $[-{\sf A}(\mu), {\sf A}(\mu)]$. 
To extend $\zeta \mapsto V(\lambda,\eta,\zeta)$ to all of $[-
\frac{\delta}{\lambda^{1/2}}, \frac{\delta}{\lambda^{1/2}}]$, we
continue $V$ to the intervals $\big(- \frac{\delta}{\lambda^{1/2}},
-{\sf A}(\mu)\big)$ and $\big({\sf A}(\mu),
\frac{\delta}{\lambda^{1/2}}\big)$ by affine functions: 
\begin{equation}\label{E6.17n}
V(\lambda,\eta,\zeta)=
\begin{cases}
\big(\zeta+{\sf A}(\mu)\big) V'\big(\lambda,\eta,-{\sf A}(\mu)\big), &
-\frac{\delta}{\lambda^{1/2}} \leq \zeta \leq -{\sf A}(\mu),\\ 
\big(\zeta-{\sf A}(\mu)\big) V'\big(\lambda,\eta,{\sf A}(\mu)\big), &
{\sf A}(\mu) \leq 
\zeta \leq \frac{\delta}{\lambda^{1/2}}\,. 
\end{cases}
\end{equation}
By this extension, $\zeta \mapsto V(\lambda,\eta,\zeta)$ is
continuously differentiable at $\zeta= \pm {\sf A}(\mu)$. For $x =
x(\hat{t},\eta,\xi) \in {\cal U}_\da$ we use the notation 
\[
V(\la,x) = V\big(\la,\eta,\frac{\xi}{\la^{1/2}}\big).
\]
In the remaining part of the proof of \reft{6.2} we need the following
lemma, which we prove first.

\begin{lem}\label{L6.3}
There are constants $C_1, \dots, C_4$ such that for all $\mu \in
(0,\mu_0]$, $\la \in (0,\la_0]$, $(\eta,\zeta) \in
\Gm(\hat{t}) \times \big(- \frac{\da}{\la^{1/2}},
\frac{\da}{\la^{1/2}}\big)$ and $x \in {\cal U}_\da(\hat{t})$ the
estimates  
\begin{eqnarray}
|\na_\eta^j V'(\la,\eta,\zeta)| &\leq& C_1 |\ln \mu|^3 \mu, \qquad
 j=0,1,2, \label{E6.18n} \\ 
|\la^{1/2} \na_\eta^j V(\la,\eta,\zeta)| &\leq& C_2 |\ln \mu|^3 \mu,
\qquad j=0,1,2, \label{E6.19n} \\
|\la^{1/2} \na_x V(\la,x) | &\leq& C_3 | \ln \mu |^3 \mu,
 \label{E6.21n} \\
|\la^{1/2} \pa_{x_k} \na_{\Gm_\xi} V(\la,x) | &\leq& C_4 
 | \ln \mu |^3  \mu, \qquad k = 1,\ldots ,3,  \label{E6.22n}
\end{eqnarray}
hold. Moreover, there is a function $g^{(\mu\la)}:\Om \ra \R^3$ and a
constant $C_5$ such that $T^{(\la)}$ defined in \eq{6.14n} satisfies 
\begin{equation}\label{E6.23n}
\div_x T^{(\la)} = f^{(\mu\la)}_{11} +
g^{(\mu\la)},  
\end{equation}
with
\begin{equation}\label{E6.24n}
| g^{(\mu\la)}(x)| \leq C_5 | \ln \mu |^3 \mu,
\end{equation}
for all $x \in \Om$ and all $\mu \in (0,\mu_0]$, $\la \in (0,\la_0]$.  
\end{lem}
{\bf Proof:} In the following computations we drop the arguments $\la$
and $\eta$. Integration of \eq{6.16nn} yields
\begin{align}
B V' (\zeta) &= \int_{-{\sf A}(\mu)}^\zeta \la^{1/2} f^{(\mu\la)}_{11}
(\la^{1/2}\vta) d\vta + BV'(-{\sf A}(\mu)), \label{E6.25n}\\
BV (\zeta) &= \int_{-{\sf A}(\mu)}^\zeta \int_{-{\sf A}(\mu)}^{\vta_1}
  \la^{1/2} f^{(\mu\la)}_{11} (\la^{1/2}\vta) d\vta d\vta_1 
  + \big( \zeta + {\sf A}(\mu) \big) BV'(-{\sf A}(\mu)), \label{E6.26n}
\end{align}
where we used the boundary condition \eq{6.16n} to get the second
equation. Since $V({\sf A}(\mu)) = 0$, the relations \eq{6.26n} and
\eq{6.2n} together yield   
\begin{multline*}
2 {\sf A}(\mu) | BV'(-{\sf A}(\mu))| = \Big| - \int_{-{\sf A}(\mu)}^{{\sf A}(\mu)}
 \int_{-{\sf A}(\mu)}^{\vartheta_1} \la^{1/2} f^{(\mu\la)}_{11} d\vartheta
 d\vartheta_1 \Big| \\ 
\leq \int_{-{\sf A}(\mu)}^{{\sf A}(\mu)}
 \int_{-{\sf A}(\mu)}^{\vartheta_1} |\ln \mu|^2 \mu^{1/2} K_1 d\vartheta
 d\vartheta_1 = 2 {\sf A}(\mu)^2 |\ln \mu|^2 \mu^{1/2} K_1\,, 
\end{multline*}
hence, by \eq{6.4n}, 
\[
| BV'(-{\sf A}(\mu))| \leq {\sf A}(\mu) |\ln \mu|^2 \mu^{1/2} K_1 = \frac{3}{a}
K_1 |\ln \mu|^3 \mu. 
\]
Since $B = B(\eta)$ is positive definite uniformly with respect to
$\eta$, this inequality implies the estimate \eq{6.18n} for $j=0$ and
$-{\sf A}(\mu) \leq \zeta \leq {\sf A}(\mu)$. Since by definition
\eq{6.17n} we have $V'(\zeta) = V'(-{\sf A}(\mu))$ for $\zeta \leq
-{\sf A}(\mu)$ and  $V'(\zeta) = V'({\sf A}(\mu))$ for ${\sf A}(\mu)
\leq \zeta$, the estimate \eq{6.18n} with $j=0$ holds 
also for the values of $\zeta$ outside of the interval
$[-{\sf A}(\mu),{\sf A}(\mu)]$.  

To prove \eq{6.19n} for $j=0$ we use that $V(-{\sf A}(\mu))=0$. By
integration we thus obtain from \eq{6.18n} for $\zeta \in
[\frac{-\da}{\la^{1/2}},\frac{\da} {\la^{1/2}}]$ that 
\[
|V(\zeta) | = \Big| \int_{-{\sf A}(\mu)}^\zeta V'(\vartheta) d\vartheta
 \Big| \leq |\zeta + {\sf A}(\mu)| C_1 |\ln \mu|^3 \mu 
 \leq \big( \la^{-1/2} \da + {\sf A}(\mu)\big) C_1 |\ln \mu|^3 \mu,
\]
which implies \eq{6.19n} for $j=0$.  

To verify \eq{6.18n} and \eq{6.19n} for $j=1,2$ we differentiate the
differential equation \eq{6.16nn} and the boundary condition \eq{6.16n}
with respect to $\eta$. For $j=1$ we obtain the differential equation 
\[
B(\eta) (\pa_{\eta_k} V)'' = \la^{1/2} \big(\pa_{\eta_k} f^{(\mu\la)}_{11} -
\pa_{\eta_k} B(\eta) B(\eta)^{-1} f^{(\mu\la)}_{11} \big),
\]
and a similar equation for $j=2$. We then use the estimate
\[
|\na_\eta^j f^{(\mu\la)}_{11}(\eta,\xi)| = |\na_\eta^j
f^{(\mu\la)}_1(\eta,\xi)| \leq  |\ln \mu|^2
\Big(\frac{\mu}{\la}\Big)^{1/2} K,  \qquad j =1,2.
\]
This estimate is obtained by differentiation with respect to $\eta$ of
the asymptotic expansions in Section~\ref{S5.1} leading to \refc{5.3}.
Under the regularity assumptions in \reft{6.2} these derivatives
exist. With this estimate we can employ the same arguments as above
for the case $j=0$ to derive \eq{6.18n} and \eq{6.19n} for $j=1,2$.

To prove \eq{6.21n} we use the decomposition \eq{gradsplit} of the
gradient and \eq{surfgradtrans} to compute
\[
\na_x V(\la,x) = \pa_\xi V\big(\la,\eta, \frac{\xi}{\la^{1/2}} \big)
\otimes n + \na_{\Gm_\xi} V\big(\la,\eta, \frac{\xi}{\la^{1/2}} \big)
=  
\la^{-\frac12} V'\otimes n + (\na_\eta V) A(\hat{t},\eta,\xi).
\] 
The right hand side is estimated by \eq{6.18n} and \eq{6.19n} to obtain
\eq{6.21n}. The estimate \eq{6.22n} is obtained from \eq{6.18n} and
\eq{6.19n} by similar decompositions. 

To prove \eq{6.23n}, \eq{6.24n} note that by \eq{6.13n}, \eq{6.14n} we
have $T^{(\la)} = D \ve (\na_x u^{(\la)}) = D \ve \big(\na_x (\la^{1/2}
V \phi_* )\big)$. Using \eq{gradsplit} and \eq{divsplit} we therefore
obtain by a similar computation as in \eq{divTmu} that  
\begin{eqnarray}
\div_x T^{(\la)}  \nn
&=& \div_x D\ve \Big(\na_x \big( \la^{1/2}
  V(\la,\eta,\frac{\xi}{\la^{1/2}}) \phi_*(\xi)\big) \Big) \nn\\
&=& \Big( \la^{-1/2} \big( D\ve (V'' \otimes n)\big) n + \div_{\Gm_\xi} 
  D\ve (V' \otimes n ) \Big) \phi_* \nn\\ 
&& \mbox{} + \la^{1/2} \Big( D\ve ( \pa_\xi \na_{\Gm_\xi} V)\big)n + 
  \div_{\Gm_\xi} D\ve (\na_{\Gm_\xi} V) \Big) \phi_* \nn\\
&& \mbox{} + \Big( \big( D\ve( \la^{1/2} \na_x V)\big) n + \div_x
  D\ve(\la^{1/2} V \otimes n) \Big) \phi_*' \nn\\
&&\mbox{} + \Big( D\ve(\la^{1/2} V \otimes n) \Big)n\, \phi_*'' \nn\\ 
&=& f^{(\mu\la)}_{11} + g^{(\mu\la)}.  \label{E6.28n}
\end{eqnarray}
In the last step we used the differential equation \eq{6.15n} and
noted that for $\xi \in ([-\da,-{\sf A}(\mu)\la^{1/2}] \cup [{\sf
  A}(\mu)\la^{1/2},\da])$ we have
$V''\big(\la,\eta,\frac{\xi}{\la^{1/2}}\big) = 0$, by definition of
$V$ for such values of $\xi$ in \eq{6.17n}. We also used that
$\phi_*(\xi) = 1$ for $\xi \in [-{\sf A}(\mu)\la^{1/2},{\sf
  A}(\mu)\la^{1/2}]$, which follows from \eq{6.12n} and \eq{6.4n},
since we have chosen $\mu_0$ and $\la_0$ small enough such that ${\sf
  A}(\mu)\la^{1/2} < \da/2$ for all $0 < \mu\leq \mu_0$ and $0< \la
\leq \la_0$.

The function $g^{(\mu\la)}$ is the sum of terms number $2$ to $7$ in
the middle expression of equation \eq{6.28n}. If we examine everyone
of these six terms and apply \eq{6.18n} -- \eq{6.22n} and also note
that the functions $\phi_*$, $\phi_*'$ and $\phi_*''$ are
bounded independently of $\mu$ and $\la$ and vanish outside of ${\cal
  U}_\da(\hat{t})$, which follows from $\phi_* \in C_0^\infty((-\da,
\da))$, we see that \eq{6.24n} holds for $g^{(\mu\la)}$. This
completes the proof of \refl{6.3}. 
\qed
\\[1ex]
To conclude the proof of \reft{6.2} let $( u^{(\la)}_*,T^{(\la)}_*)$
be the solution of the boundary value problem
\begin{eqnarray}
-\div_x T^{(\la)}_* &=& g^{(\mu\la)} - f^{(\mu\la)}_{12},
\label{E6.30n}\\ 
T^{(\la)}_* &=& D\ve(\na_x u^{(\la)}_*), \label{E6.31n}\\
u^{(\la)}_*(x) &=& 0, \qquad x \in \pa \Om. \label{E6.32n} 
\end{eqnarray}
From these equations and from \eq{6.13n}, \eq{6.14n}, \eq{6.23n} we
see that the function $( u^{(\la)} + u^{(\la)}_*,T^{(\la)} +
T^{(\la)}_*)$ satisfies 
\begin{eqnarray*}
-\div_x (T^{(\la)} + T^{(\la)}_*) &=& - f^{(\mu\la)}_{11} -
  g^{(\mu\la)} + g^{(\mu\la)} - f^{(\mu\la)}_{12} = -
  f^{(\mu\la)}_1(\hat{t}), 
  \\ 
(T^{(\la)} + T^{(\la)}_*) &=& D\ve\big(\na_x (u^{(\la)} + u^{(\la)}_*)
 \big), \\
(u^{(\la)} + u^{(\la)}_*)(x) &=& 0, \qquad x \in \pa \Om,
\end{eqnarray*}
hence $( u^{(\la)} + u^{(\la)}_*,T^{(\la)} + T^{(\la)}_*)$ is equal to
the unique solution of the boundary value problem \eq{6.linelast1} --
\eq{6.linelast3}, which means that $( u^{(\la)} + u^{(\la)}_*,T^{(\la)} +
T^{(\la)}_*)  = (u_{\rm AC} - u,T_{\rm AC} -
u)(\hat{t})$. Consequently, we have
\begin{equation}\label{E6.35n}
\| T_{\rm AC} - T\|_{L^2(\Gm(\hat{t}))} \leq \| T^{(\la)}
\|_{L^2(\Gm(\hat{t}))} + \| T^{(\la)}_*\|_{L^2(\Gm(\hat{t}))}. 
\end{equation}
To estimate $\| T^{(\la)}_*\|_{L^2(\Gm(\hat{t}))}$ we can use the
theory of interior regularity for the elliptic boundary value problem
\eq{6.30n} -- \eq{6.32n}. By this theory there is a constant $C$ such
that $\| u^{(\la)}_*\|_{W^{2,2}(\Om')} \leq C \| g^{(\mu\la)} -
f^{(\mu\la)}_{12} \|_{L^2(\Om)}$, where $\Om'$ is the subdomain of
$\Om$ introduced in \reft{6.2}, hence by the Sobolev embedding theorem
and by \eq{6.31n},
\begin{equation}\label{E6.regularityest}
\|T^{(\la)}_*\|_{L^2(\Gm(\hat{t}))} \leq C_1 
 \|T^{(\la)}_*\|_{W^{1,2}(\Om')} \leq C_2 \| g^{(\mu\la)} -
 f^{(\mu\la)}_{12} \|_{L^2(\Om)}, 
\end{equation}
where by our assumptions on $\Gm^{(\mu\la)}(\hat{t})$ in \reft{6.2}
the constants $C_1,C_2$ can be chosen independently of $\mu$ and
$\la$. By definition of $f^{(\mu\la)}_{12}$ in \eq{6.1n} and by
\eq{2.37b} we have $|f^{(\mu\la)}_{12}(x)| \leq \mu^{3/2} K_2$ for all
$x \in \Om$. From this inequality, from \eq{6.24n} and from
\eq{6.regularityest} we conclude that
\begin{multline}\label{E6.36n}
\|T^{(\la)}_*\|_{L^2(\Gm(\hat{t}))} \leq C_2 \Big(\int_\Om
\big(|g^{(\mu\la)}(x)| + |f^{(\mu\la)}_{12}(x)|\big)^2 dx\Big)^{1/2}
\\
\leq C_2 (C_5 | \ln \mu |^3 \mu + \mu^{3/2} K_2) \Big(\int_\Om
 dx\Big)^{1/2} \leq K |\ln \mu|^3 \mu. 
\end{multline}
From \eq{6.14n}, \eq{6.13n} and from the inequalities \eq{6.19n},
\eq{6.21n} we infer that 
\begin{multline*}
| T^{(\la)}(x)| \leq C | \na_x u^{(\la)}(x)| 
 = C |\na_x \big(\la^{1/2} V(\la,x) \phi_*(\xi) \big)| 
\\ 
=  C | \big( \la^{1/2} \na_x V(\la,x)\big) \phi_*(\xi) + \la^{1/2}
 V(\la,x) \otimes  \big(n \phi_*'(\xi)\big) | \leq K' |\ln \mu|^3 \mu,    
\end{multline*}
whence
\[
\| T^{(\la)} \|_{L^2(\Gm(\hat{t}))} \leq K'' |\ln \mu|^3 \mu.
\]
Combination of this inequality with \eq{6.35n} and \eq{6.36n} yields
\eq{6.differenceT}. The proof of \reft{6.2} is complete.  
\qed



\end{document}